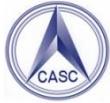

中国航天

# 甚高精度星敏感器星点定位和校正技术研究

## Research on the Starspot Centroid Estimation and Calibration Technologies for the Super-high Accuracy Star Tracker

（申请中国空间技术研究院工学博士学位论文）

| | |
|---|---|
| 培 养 单 位 ： | 中国空间技术研究院 |
| | 北京控制工程研究所 |
| 学　　　科 ： | 导航、制导与控制 |
| 研 究 方 向： | 航天器姿态测量 |
| 研 究 生 ： | 张 俊 |
| 指 导 教 师 ： | 郝云彩 研究员 |

二〇一六年三月

# Research on the Starspot Centroid Estimation and Calibration Technologies for the Super-high Accuracy Star Tracker

The dissertation submitted in partial fulfillment of
the requirements for
the Doctor's degree of Engineering in
Navigation,Guidance and Control
at the
Beijing Institute of Control Engineering, Chinese
Academy of Space Technology

Candidate:           Jun Zhang

Supervisor:         Prof. Yuncai Hao

March 2016

# 摘 要

甚高精度星敏感器一般是指精度优于 1"(3σ)的星敏感器，在星敏感器家族中是精度最高的一类，在超高分辨率卫星遥感、天文观测、深空探测等高级航天任务中发挥着不可替代的重要作用。国外甚高精度星敏感器技术已经十分成熟，在轨产品在 1°/s 角速率下的指向精度可达 0.2"(1σ)。国内该类技术处于探索研发阶段，目前已经完成了第一阶段的产品开发，取得了可喜进展，相关产品静态指向精度地面验证达到了亚角秒(1σ)量级，然而在动态条件下精度会下降，在技术指标和功能方面照比国外仍然存在不小的差距。在诸多的精度影响因素中，星点的定位和校正技术占有非常大的比重，直接决定了输出精度。现阶段主要在星点提取稳定算法、运动畸变识别动态补偿算法、实时在轨自动校正等关键技术方面存在瓶颈，需要理论上和方法上提出新的思路加以解决。

本论文以自然基金项目"高精度星敏感器星像分布对精度的影响研究"(批准号 61174004)、国家重大仪器发展专项等为背景，围绕星点定位和校正环节，对定位算法、噪声、运动、焦距漂移、恒星光谱中心变化、模型误差等误差源展开研究，主要工作如下：

针对现有星敏感器测量精度改进思路存在的问题，揭示频域滤波不是降低敏感器自身误差主要方法，基于单星定位误差公式提出对各误差源单独建模再降低误差的办法。

针对现有质心算法无法实现甚高精度星点定位问题，展开了对迭代加权质心(IWCOG)算法精度和效率特性的研究。推出了实用的优化算法和差分式 IWCOG 算法，以 Cramér Rao Lower Bound (CRLB)理论证明了 IWCOG 类算法在高斯噪声情况下的最优特性和泊松噪声下的近优特性。以 CRLB 导出的定位精度为基础，给出了 0.01pixel 静态定位精度的基本策略。地面试验和仿真验证了上述若干结论。针对实际星点偏离高斯星点形貌导致的模型误差，推导了 PSF(点扩散函数)高采样匹配算法，获得了模型误差、高采样数据噪声对定位精度的影响。

针对现有动态性能提升困难的问题，研究了加速度、速度导致的时域运动畸变，获得了能够补偿时域运动畸变的一般性动态补偿公式。以上述公式为基础，

提出多帧解析补偿技术实现高动态、高精度和高姿态输出率技术指标的解耦,三个指标可独立提升,最终获得并保持全精度。

针对在轨校正存在的问题,在判断星点数据是否适合在轨校正上,提出 CRLB 指标约束误差方法,推出 CRLB 全约束公式实现在轨参数全约束,使得收集的星场数据能够用于在轨校正,或者使得数据能够用于姿态输出或跟踪流程。首次得到各在轨参数约束条件和整体误差公式,由此得到三个重要推论。

针对现有姿态确定技术未能有效空域滤波的问题,根据星点星等及其所处视场不同给予不同加权改进 QUEST 算法,仿真验证该方法可提高现有姿态确定精度 10-20%。

论文理论和推论可直接应用于各种类型星敏感器,如甚高精度星敏感器、高动态星敏感器、CMOS 型星敏感器等,相关成果可为下一代产品研发和技术攻关提供重要的设计参考。

**关键词:星敏感器;甚高精度;单星定位误差;全精度;解析补偿;动态性能;在轨校正;Cramér-Rao Lower Bound;迭代加权质心算法;差分迭代加权质心算法;CRLB 约束方法**

# Abstract

As the superior product in the family of star sensors, normally with a pointing accuracy better than 1"(3σ), the super-high accuracy star tracker plays an irreplaceably significant role for the advanced scientific missions, such as super-high resolution satellite remote sensing, astrometry and deep space survey. For the foreign developers, the associated technology is very mature and the pointing accuracy of the in-flight products can reach 0.2"(1σ) pointing accuracy at an angular rate up to 1°/s. For the domestic counterpart, some techniques are still in progress. Now the first phase of product development is finished, with an encouraging progress in static pointing accuracy, which is validated reaching down to sub-arcsecond (1σ) steadily on ground test. However, it degrades rapidly under dynamic conditions, showing a big gap to the abroad in terms of the technical indicators and functions. Among the factors affecting the pointing accuracy, the starspot centroid estimation and calibration techniques are more critical, which directly determine the attitude precision. At present the star centroiding algorithm, the motion distortion recognition and compensation method and the real-time in-orbit calibration techniques are the bottlenecks. Theories and a new train of thoughts are needed to solve the issues in the design process.

Founded by the National Natural Science Foundation "Studies of the influence of star pattern distribution on the accuracy for the high accuracy star tracker"(Grant No. 61174004), and the National Key Scientific Instrument and Equipment Development Projects, this thesis makes a series of studies on the typical error sources associated with the positioning and calibration processes, such as the centroiding algorithm, detector noise, motion, focal length drift, variation of the center wavelength of stellar spectrum, the model error, etc., some works of which are listed as follows:

For the problem put forward by the current accuracy improvement strategies, it is revealed that the frequency filtering will not be the solution to reduce error a star tracker produces. And a new method is addressed, where each error source is modeled independently based on the measurement model for individual star positioning error.

To seek a solution to attain super-high centroiding accuracy, the iteratively weighted center of gravity (IWCOG) algorithm is studied for the characteristics of both the accuracy and efficiency. A practically optimized IWCOG algorithm and a differential IWCOG algorithm are derived. Using Cramér-Rao Lower Bound (CRLB) theorem, they are centified to have an approximate optimum property for signal with Possion noise and the optimum feature in case of Gaussian noise. On the basis of the locating accuracy derived, an useful strategy to achieve 0.01pixel positioning accuracy under static condition is proposed. The ground test and simulation results validate some conclusions we obtain. To determine the centroid deviation resulting from the model error casued by the star pattern distortion, the PSF (point spread function) high sampling matching algorithm is developed, with which the influence of high sampling noise and model error to positioning accuracy is found.

In order to enhance the dynamic performance, the time-domain motion distortion arising from acceleration and velocity is investigated. The generallized dynamic compensation formula is derived to compensate out the error from motion. Based on this motion compensation, a scheme of multi-frame analytical compensation is presented, dedicated to realizing the decoupling of three technical indicators including dynamic performance, pointing accuracy and attitude output rate. Thus three indicators can enhance their values independently and the full performance of a star tracker is achieved and preserved.

To solve the problems exhibited in the in-orbit correction procedure, a brilliant idea, known as the CRLB constraint method, is presented to minimize some small error sources, or to be used to judge whether the collected star data effective for the in-orbit calibration or to output the attitude information or not. To fulfill this goal, the full CRLB constaint formulas are derived to build a budget to limit the in-flight parameters. Using this method the effective value scope of the in-flight parameters and the overall error are obtained, which in turn results in three significant inferences.

The attitude determination process is further optimized, to avoid the limitations of filters like Kalman, which is confirmed not an effective spatial filter. By giving different weights to each star according to their magnitudes and the position over different star field, the QUEST (QUaternion ESTimator) algorithm is thus upgraded. The simulation results show that this method can improve the attitude measurement accuracy statistically by 10-20%.

The theory and deductions can be directly applied to various types of star trackers, whether it is a super-high accuracy one, or a high dynamic one, or a star tracker which uses CMOS device as the detector, and will pave a road for the development of next generation products.

# 目 录

# 第一章 绪论

## 1.1 研究背景和意义

星敏感器(又称星跟踪器)作为航天器位姿系统精度最高的角姿态测量部件，是以恒星为信号源的天文导航敏感器，具有自主导航、测量姿态精度高、无累积误差、抗干扰能力强、可靠性高等优点，除为卫星本体提供空间指向外，也为有效载荷或其他系统如太阳帆板或通信天线提供指向，是卫星常备的关键姿态测量部件之一。通常来讲，星敏感器产品的指向精度和动态性能是 2 个最为关键的指标。经过四十多年的发展，国外星敏感器技术已经十分成熟，高精度产品指向精度普遍做到了1"(1σ)量级，0.3º/s 下保持全精度(full performance)，国内在高精度产品研制技术上也日趋成熟，指向精度在 1"(1σ)量级，0.2º/s 下保持全精度，其它指标接近甚至超越国外同类产品，已在轨运行在多颗卫星和其他平台上。随着先进航天任务的深入开展，如 0.3 米级对地遥感观测、斯皮策类空间天文台的天文观测、火星探测、深空探测等，对甚高精度星敏感器的需求与日俱增，研制甚高精度级别星敏感器已经是当务之急。

目前欧美等厂商研制的高端星敏感器产品均能达到甚高精度，有多款产品指向精度做到了亚角秒(0.2", 1σ)量级，1º/s 下保持全精度，相关技术也日趋成熟。国内最近几年开展了甚高精度星敏感器研制工作。北京控制工程研究所预研立项研制了甚高精度星敏感器原理样机并开发了相关产品，同期获得相关项目支持，如自然基金项目"高精度星敏感器星像分布对精度的影响研究"(批准号 61174004)、2013 年度国家重大仪器发展专项等，旨在弄清其中的关键技术，提升产品性能，但从相关过程和结果看，国内此类产品在核心指标上要达到国外水平仍然有一定难度。

从研发思路上看，国内在研制 0.2"(1σ)高稳定甚高精度星敏感器时多依赖工程经验和参考国外公开的相关技术，如优化质心算法、采用区域标定技术、提高像元分辨率、优化电路性能和结构设计、在轨校正与卡尔曼滤波结合等方法，有些方法从本质上讲属增量技术，增量技术在高精度星敏感器产品研制中虽显示一定价值，能否将 1 角秒精度提升至 0.1 角秒量级，是值得商榷的。国外甚高精度产品则报道了多项较大改进技术，如星点定位方面采用新的质心提取算法提高星点提取精度，

动态方面采用图像运动自适应技术或可变帧率技术提高动态性能，校正方面采用在轨焦距校正和空域畸变校正修正星点定位误差等，将高精度产品提升至甚高精度级别，全精度动态性能大幅提升，然而，与此相关的技术细节和基础理论公开甚少。因此，研究可用于甚高精度星点定位和在轨校正的理论和技术对从根本上解决静态精度和动态性能问题，突破现有产品技术瓶颈，具有十分重要的意义。

## 1.2 国内外研究现状

### 1.2.1 甚高精度星敏感器研制现状

自首个星敏感器问世至今，经历了第一代非自主星敏感器阶段，第二代自主星敏感器阶段两代产品多种类型。

第一代星敏感器为非自主CCD星敏感器阶段(1970-1990)：70年代CCD器件发展使得星敏感器进入固态成像阶段，JPL率先开发了STELLAR星敏感器，视场为5°，可探测6等星，精度在10"等级。比较典型的产品还有Ball公司的PST系列、HDOS公司的ASTRA星敏感器、TRW公司的MADAN星敏感器等。这一代星敏感器主要是CCD型，主要特点是视场小、跟踪星等较高、星数较少、中等精度、无自主星图识别能力。

第二代星敏感器是自主星敏感器阶段(1990-现在)。这一代星敏感器可存储导航星表实现全天覆盖，具有全天星图识别能力，可直接输出姿态测量信息。第二代产品包括中高精度和甚高精度，目前高精度星敏产品X/Y轴指向精度普遍到1"(1σ)量级。

第二代星敏感器产品和高端应用场合仍然以CCD型为主，近几年来，APS CMOS型技术的成熟(Bigas M. et al. 2006)，如sCMOS型产品的问世，基于APS CMOS型的星敏感器产品不断涌现(Liebe C. C. et al. 1998)，在高精度和甚高精度星敏感器领域也已经大显身手。以下首先分别介绍国外CCD型和CMOS型星敏感器，分析各类型相关的甚高精度产品，再介绍国内情况。

(1) CCD 型甚高精度星敏感器

比较典型的产品有 Jena-Optronik 公司的 ASTRO-XX 系列、法国 SODERN 公司的 SEDx6 系列、Ball 公司的 CT-XXX 系列、Lockheed Martin 公司的 AST 系列、Denmark DTU 的 ASC 系列、意大利 Galileo Avionica 公司的 STR 系列等。主要特点采是中大视场、较低探测星等、较高数据更新率、中高精度、采用一体化和模块化设计或分体式结构。

本世纪以来，欧美等国家对星敏感器若干关键技术重大改进，基于 CCD 器件的产品大幅提高了精度和动态性能，典型的型号大部分出自前述公司和单位，主要体现在后期研制的甚高精度产品上。

表 1.1 典型的 CCD 型高精度和甚高精度星跟踪器型号及技术指标

| 技术指标 | AST-301 | HAST | SED36 | ALOS ST | Herschel STR |
|---|---|---|---|---|---|
| 精度（X/Y 轴，1σ） | 0.18" | 0.18" | $0.33^{"}(LFE)$ | $0.25^{"}(LFE)$ | 0.84" |
| FOV | 5° × 5° | 8.8° × 8.8° | 25°×25° | 8 × 8° | 16.4° × 16.4° |
| 像元数 | 512 × 512 | 2048 × 2048 | 1024×1024 | | 512 × 512 |
| 像元分辨率/尺寸 | 36", 15 $\mu m$ | 15.46", 15 $\mu m$ | 87.89" | | 116", 15 $\mu m$ |
| 动态性能 | *> 3.1°/s, X | 4°/s, 4°/s² | 10°/s | | |
| (*, 全精度) | * 0.1°/s, Y | * 1°/s | *>1°/s | | |
| 可跟踪星数 | 50 | 16,6.9$M_i$ | 12(8Hz) | 10 | 9-15 |
| /星等 | 9.5Mv | 11,5.5$M_i$ | | 6.5Mv | |
| 全天捕获模式 | ≤ 3s(99.98%) | | ≤ 4s | ≤ 3s | |
| 更新速率 | 2Hz | 2Hz | 1-10Hz | 1Hz | 4Hz |
| 重量 | 7.1kg | <25kg | ≤ 3.7kg | | |
| 功耗 | 18W | <120W(min) | 8.4W | | |
| 寿命 | >10 年(LEO) | >10 年(LEO) | 5 年(LEO) | | 4 年 |

表 1.1 为目前文献公开的几款 CCD 型高精度和甚高精度星敏感器。

从表1.1所述型号特点及其公开的文献资料来看，国外多款CCD型甚高精度星敏感器精度X/Y轴指向精度在亚角秒量级(1σ)、全精度动态性能不低于1°/s、动态性能在5°/s以上、探测星等优于6等星、视场较小、姿态更新率为2-10Hz、寿命较长。

(2)APS CMOS 型甚高精度星敏感器

相比 CCD 型，APS 型星敏感器具有明显优势，主要表现在：

(a) 较宽的视场(20 × 20°)，更多较亮的导航星，星等阈值可降低；

(b) CMOS 图像传感器高度数字集成，实现了单芯数字成像系统，省去了 CCD 复杂的外围设计、驱动设计；

(c)随机存取能力，更新速率快，避免转移效率限制，具有更好的抗辐射和抗饱和能力；

(d)体积小，功耗低，噪声小。

但 APS 的仍然存在较低 QE、响应不均性、较低的填充率等问题。

表 1.2 典型的 APS CMOS 型高精度和甚高精度星跟踪器型号及技术指标

| 技术指标 | ASTRO APS | HYDRA | Colorado Daystar | IST |
|---|---|---|---|---|
| 精度(X/Y 轴,1σ) | $< 1"$ | $0.27"(LFE)$ | $0.05"(night)$ | 1" |
| | | $0.23"(FOV)$ | $0.1"(day)$ | |
| FOV | $20 \times 20°$ | | $5.28° \times 6.26°$ | |
| 像元数 | 1024×1024 | | $2160 \times 2560$ | |
| 像元分辨率/尺寸 | $72", 15\ \mu m$ | | $8.8", 6.5\ \mu m$ | |
| 动态性能 | 3°/s | 10°/s,10°/s² | 6°/s | 5°/s |
| (*,全精度) | * 0.3°/s | * 1°/s, 0.05°/s² | * 1.4°/s | |
| 可跟踪星数 | 5.8M$_i$ | $\geq$ 15/OH | 45(night),8Mv | |
| /星等 | | | 5(day),7Mv | |
| 全天捕获模式 | $\leq$ 2s | $\leq$ 1s(< 1°/s) | | |
| 更新速率 | 1-30Hz | 1-30Hz | 10Hz | |
| 重量 | <2kg | 3.22kg/OH+EU | 40kg | 0.2kg |
| 功耗 | <6W | 2.6W/OH | 150W(84.4 W) | |
| 低轨寿命 | >18 年 | 7 年 | <1(Balloon-borne)年 | |

表 1.2 为几款典型的 APS CMOS 型高精度和甚高精度星敏感器产品。

APS 型星敏感器已经实现多个产品在轨飞行，高精度产品精度在1"量级，甚高精度产品指向精度低至0.1"以下，可匹敌 CCD 型产品。国外基于近空间观测平台的需求，研制了若干近空间星敏感器并成功在轨运行，这种类型星敏感器可用于白天

观测。如 ST5000、DayStar 系统等，其性能指标优越，部分指标甚至超越了目前在轨运行的 CCD 型甚高精度星敏感器。

我国从上个世纪 80 年代开始星敏感器的研究工作，参与的单位有北京控制工程研究所、中科院长春光机所、西安光机所、光电技术研究所、北京航空航天大学、哈尔滨工业大学、清华大学、国防科技大学、航天八院 812 所、国家天文台等。产品如上个世纪长春光机所研制的 XG-1 星敏感器；北京航空航天大学研制的 YK010、SS2K 型敏感器；清华大学研制的微小型高精度 APS 星敏感器等。目前北京控制工程研究所研制的星敏感器 X/Y 轴指向精度已经从10"提高到0.3"(1σ)，产品批量化用于多颗卫星上。

表 1.3 北京控制工程研究所研制的高精度和甚高精度星跟踪器及技术指标

| 技术指标 | 4th ST | 高动态 SST | CCD SST | SST |
|---|---|---|---|---|
| 精度(X/Y 轴,1σ) | ≤ 1" | <0.67" | 0.31" | 0.31" |
| FOV | 20°×20° | 20°×20° | 5°×5° | 7.1°×7.1° |
| 动态性能 | 1°/s | 20°/s | 0.5°/s | 0.5°/s |
| (*, 全精度) | | *0.6°/s | *0.1°/s | |
| 可跟踪星数 | >5.5Mv | | 35/ | 16/ |
| /星等 | | | 8.1Mv | 9Mv |
| 全天捕获模式 | ≤ 8s | ≤ 1s | < 15s(95%) | ≤ 4s |
| 更新速率 | ≤8Hz | ≤30Hz | 2Hz | |
| 重量 | 3.5kg | 3.5kg | 7.5kg | 2.2kg |
| 功耗 | <11W | <16W | <13W | 6.8W |
| 低轨寿命 | 8 年 | 8 年 | 8 年 | 8 年 |

表1.3为北京控制工程研究所研制的高精度和甚高精度星敏感器产品。

将表1.3研制的星敏感器同国外同期产品的对比看，北京控制工程研究所研制的星敏感器在光、机、电、算法设计方面已经取得了重大突破毋庸置疑，掌握了若干关键技术，在中高精度星敏感器研制能力和技术指标上，已经可以和国外主流产品同台竞技。在甚高精度星敏感器细分指标上，该类产品与国际同类产品尚有差距，主要表现在亚角秒精度(稳定性)、高动态范围、初次捕获时间等，具体来说，提高

单星提取精度的技术、提高全天星图识别速率和成功率的技术以及提高动态范围的技术是甚高精度星敏感器几项典型的关键技术，同时研制甚高精度敏感器离不开在轨标定技术，因而在轨标定技术也是甚高精度及高动态星敏感器应该解决的问题。

## 1.2.2 关键技术研究现状

图 1.1 为星敏感器工程流程示意图。由图 1.1 可知，星敏感器工作流程分为以下几步：第 1 步，恒星经过光学系统成像于探测器焦平面上，通过探测器的光电转换效应，将光子信号转换为电子信号，并转化为电压信号，再通过 A/D 转换输出数字信号；第 2 步，经过星场背景估计去除，星点提取环节得到恒星在像平面的位置、亮度等信息，经过校正环节得到更为精确的参数；第 3 步，同内置的星载星表做比对，完成星图识别；第 4 步，根据姿态确定方程计算光轴指向，通过星敏感器与其他载荷或部件的安装矩阵，确定航天器或其他载荷或部件在惯性坐标系下的三轴指向并输出。

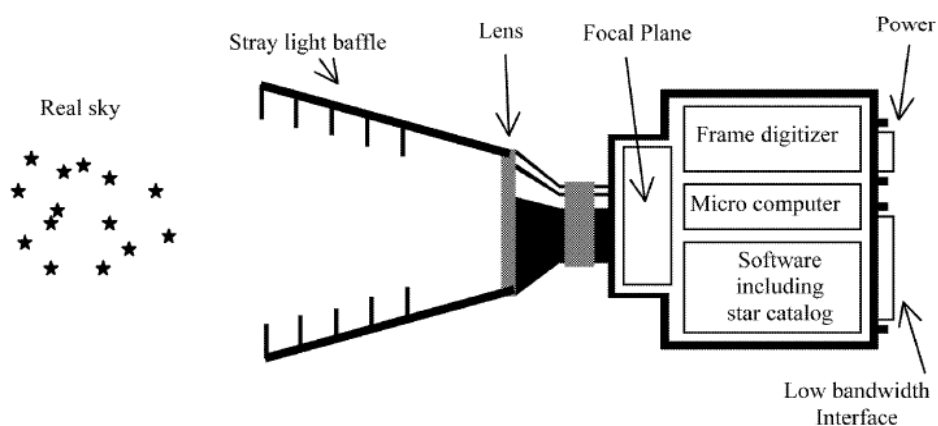

图 1.1 星敏感器工作流程示意图(Liebe C. C. 2002)

四个流程中，星点提取的精度直接决定了星敏感器姿态测量精度，间接影响星图识别成功率和速度(张俊 2014)。由于空域和时域的畸变特性，提取的星点会偏离真实位置，需要通过畸变校正将其校正到更精确的位置，这也是提高姿态确定精度必须的环节(Anderson J. et al. 2003)。星图识别的实质是寻找观测星图中观测星在星表中对应的导航星(van Bezooijen R. W. H. 1998)。一般而言，星敏感器至少包含两个工作模式：初始姿态捕获模式和跟踪模式。一般当星敏感器进入

跟踪模式后，有先验信息，可以利用前一帧的信息对下一幅星图进行预测，因而跟踪模式星图识别速率和成功率要高(Samaan M. A. 2003; Samaan M. A. et al. 2005)。姿态确定是根据测量、校正和匹配的星点信息，利用姿态确定算法计算星敏感器光轴指向的过程，以及根据星敏感器相对于航天器和其他载荷或部件的安装方位角，计算航天器及其他部件姿态的过程(Markley F. L. 2000; Crassidis J. L. 2002)。

这四个方面涉及多种技术，如对杂光抑制、噪声滤波、除去"热"像素和伪星等星敏感器通常面临的问题研究相对完善，而星点提取下的背景阈值预测、星点分离、图像识别下的跟踪模式算法和姿态确定技术较为成熟，在此不再赘述。在全天星图识别环节将提一下北京控制工程研究所周建涛的研究进展，其他几个方面，如甚高精度星点提取、在轨校正是关系甚高精度能否实现的关键问题，将在本文后续章节重点讨论。

### 1.2.2.1 单星定位技术

随着星敏感器精度要求日益提高，需要对星点提取精度的影响机理深入研究，从而得到更好的星点提取方法。对于理想光学系统，恒星在焦面上成像点可以用点扩散函数(PSF)来描述，尺度在亚微米量级，而目前探测器像元尺度均在 10μm 量级，这种情况下像点完全落在单个像元内，定位误差可达 1pixel，为达到亚像元定位目的，Liebe 从工程角度出发，提出使用离焦技术，将星点弥散在若干个像元上，通过亚像元定位算法确定质心，从而得到亚像元的定位精度(Liebe C. C. 1995)。这种离焦方法，可以等效为将艾里斑函数乘以一个倍数，并不改变 PSF 形貌(Zhang J. et al. 2016)，这种形貌可近似为高斯分布。目前已经研究的较为成熟的亚像元定位算法主要有以下 2 类。

第一类定心算法直接使用星点的像素灰度直方图分布研究单星定位误差。如最早研究的一阶矩法及带阈值的一阶矩算法(Auer L. H. et al. 1978; Stanton R. H. et al. 1987)，优点是利用了每像元的灰度信息，计算简单，带阈值的一阶矩法一定程度上去除了噪声对提取精度的影响，缺点是对噪声敏感，二者均存在难以去除的周期性像素相位误差(Anderson J. et al. 2000)，带阈值的一阶矩法存在阈值难确定的问题(潘波等 2008)。改进的灰度平方加权法突出了中心像素的影响，但当计算像元数较少时，质心偏离较大(Trinder J. C. 1989; 李朋等 2011)。采用窗口灰度法突出了窗口内

像素的影响，但窗口难以确定，质心提取精度比一阶矩法略高(El-Hakim S. F. 1986)。Alexander 等从频域出发，给出了周期性 S 型相位误差的理论解释(Alexander B. F. 1991; 曹西平等 2011)。国内研究人员在其基础上，推出了周期性误差表达式(贾辉 2010)。第一类技术计算简单，但误差较大，考虑噪声的存在，得到较好的精度在 0.1pixel 左右。这类技术已经被国内外学者多次证明在高斯信号不含噪声情况下存在 S 误差，在存在噪声情况下误差以 S 误差曲线为中心分布(Anderson J. et al. 2000)，波音公司获得的 S 误差补偿国际专利也印证了这一点(Wu Y. W. et al. 2003)。但是实际光学系统存在 PSF 畸变、噪声大等干扰因素导致理想情况下推出的 S 误差补偿技术并不凑效。

第二类定心算法以星点像素灰度直方图为起点，采用形貌拟合方法提取质心。如研究的高斯曲面和抛物面拟合法(Auer L. H. et al. 1978; Stone R. C. 1989)，精度较一阶矩法高，但计算量大，当星点分布偏离高斯形貌时，所得误差较大；多项式插值质心法通过阶次增加来获得更高精度，但阶次过高会导致计算复杂，且该算法并未消除系统误差(Hou H. et al. 1978)。其他类似算法有，亚像素插值算法(Quine B. M. 2007)，基于线性内插和最小二乘的高斯曲面拟合算法(王广君等 2005)，信息论信息加权法(Flewelling B. R. et al. 2011)等。信息加权估计质心，算法鲁棒，去除了星点位置噪声、背景噪声、量化噪声对质心估计的影响。第二类算法在既定星点成像形貌假设下，以形貌拟合的精度比质心法要高，但计算较复杂，运算量大。

哈勃(HST)等太空望远镜为对天体定位和测光进行研究，指出获取精确的点扩散函数(PSF)是提高单星定位精度的基本方法。PSF 也具有渐变特性，非理想的成像各子系统和外界因素导致 PSF 具有空变、时变、色变性，PSF 建模、插值等技术可用于解决这些问题，星点定位精度已经做到优于 0.01pixel (Anderson J. et al. 2000, 2003, 2006)。基于此，国内研究人员提出基于迭代方案的亚像素插值定心算法(张志渊等 2010; 张艳 2012)。但从研究过程看，PSF 技术路径较为复杂，不太适合星敏感器实时星点提取过程。

其他方法如中值法适用范围窄，Alexander 及贾辉等频率补偿法均存在假设过于理想、抗噪性差、技术窗口窄的问题。实际星点成像过程受光学像差、焦距变化、CCD 像素不均匀响应、CTI 效应、航天器运动和振动、宇宙射线、杂光、噪声等影

响，星点成像多偏离高斯分布状态(唐勇等 2004; 郝云彩等 2012)，这也是低频误差产生的原因之一，星点提取算法应能处理非高斯分布。

### 1.2.2.2 全天星图识别技术

星图识别技术是星敏感器实现自主定位最基本的技术。如前所述，星敏感器进入跟踪模式后，由于有先验信息，星图匹配速率快、成功率高，不再是整个工程流程的瓶颈。而当星敏感器由于故障姿态丢失或进入工作初始时刻("太空迷失")时，星敏感器处于初始姿态捕获模式，这个阶段完全没有先验的姿态信息，需要进行全天星图识别，需要高效的匹配算法保证成功率和效率。所以，星图识别的瓶颈在于全天识别模式的速率和成功率，随着星敏感器精度要求的提高，需要星敏感器加强对弱星响应能力，识别的星点数增加，导航星数量也成指数增加，降低了识别速率和成功率，寻找合适的方法是全天识别模式成功的关键，相应技术由此成为星敏感器关键技术。目前星图识别算法根据特征提取方式的不同分为两类：子图同构类算法和模式识别类算法(张广军 2011)。

**(1)子图同构类算法**

子图同构类算法以星与星之间的角距为变量，直接或间接利用角距，以线段、三角形、四边形为基本匹配元素，并按照一定方式组织导航特征数据库，根据这些基本匹配元素的组合，在全天星图中找到唯一符合匹配条件的导航星的过程。传统的星图识别算法包括多边形算法、三角形算法及改进算法、金字塔算法、匹配组算法等。

Gottlieb 最先提出了多边形角距算法，其实现过程即是计算观测星图中两颗星的角距，然后将这个角距与存储在导航星库中的所有导航星进行角距比较，如果匹配，则选取第 3 颗星继续类似操作，直到剩下唯一匹配为止。多边形角距算法简单直观，易于实现，但当导航星数目较大时，算法复杂度增加，匹配时间很长，且所需存储容量较大。Liebe 提出了"太空迷失"算法，又称三角形识别算法，是目前用得最多最为成熟的方法。其基本思想与多边形角距匹配算法类似，只是以 3 颗星的角距为匹配特征。由于三角形算法需存储导航三角，导航数据库一般较大。Quine 等人改进了三角形算法(Quine B. M. et al. 1996)，降低了导航星库容量和搜索时间，但这些方法同样需要比较准确的星等信息。Mortari 在三角形算法基础上，提

出了金字塔算法(Mortari D. et al. 2004)，其思想是在三角形算法外找第 4 颗星，用于判断三角形算法的正确性，减少冗余匹配，金字塔算法已经在 Draper 的星惯导航敏感器和 MIT 的 HERE 和 HERE-2 卫星上成功做过测试，在 StarNavI、II、III 上成功应用。

Kosik 和 van Bezooijen 提出匹配组算法，又称为主星识别法，在 AST-301 星跟踪器上得到成功应用(van Bezooijen R. W. H. 2003)。其思想是在观测星图中选择一个主星，除主星外，其余星称为伴星，每一个伴星与主星构成星对，按照三角形算法的原则寻找星对的匹配，与主星对应的导航星应该在上述星对匹配的交集中。匹配组算法以角距组织特征模式，导航数据库容量大，在许多星等相近的亮星区域内，其识别成功率也受干扰星和噪声的影响。

**(2)模式识别类算法**

模式识别类算法为每颗星构造一个独一无二的特征"星模式"，通常以一定邻域内其他星的几何分布特征来构成。这样星图识别实质上就是在星表中寻找与观测"星模式"最相近的导航星。最具代表性的为栅格算法。其他方法还有 SLA 算法，采用径向和环向特征的星图识别法。

Padgett 提出栅格算法，将观测星邻域的伴星的几何分布特征用栅格的形式来表示，并作为星的特征模式，栅格算法采用查找表来存储特征，加快匹配的速度(Padgett C. et al. 1997)。与传统算法相比，栅格算法有较高的识别率、较快的识别速度、导航星库容量也较小，但在近邻星选择错误情况下，识别失败率高。在栅格算法基础上，魏新国等提出基于径向和环向特征的识别算法，利用径向特征的旋转不变性和一维特性，成功率达到 97%以上，该算法对星点噪声、干扰星、缺失星有很强抗干扰能力(魏新国等 2004)。魏新国使用 log-polar 变换用于星图识别，每颗星的特征以字符串编码形式存储起来，并用字符串模糊匹配的方式实现特征模式的识别(魏新国等 2006)。

Mortari 提出了 SLA 算法以加快数据库搜索效率，并建议使用"K-向量"来搜索星库，该算法已经成功运用于印度的卫星上(Mortari D. et al. 2014)。由于 SLA 大量时间耗费在交叉检验阶段，Kolomenkin 推出了改进的 SLA 算法，即增加投票算法使得不正确匹配的星对得到的票数最少，这个算法能够处理伪星，降低错误匹配概率(Kolomenkin M. et al. 2008)。Tichy 等提出了两星投票算法，对于能够和星库匹配

的星对给予一定的票数，并认为票数最多那个匹配即是识别结果。投票算法不仅可以用在有关模式识别类算法中，也可以与子图同构类算法相结合增加验证过程加快搜索(Tichy V. et al. 2011)。

若假定星库中参考的星体数目为$n$，相机视场内平均星体数目为$f$，匹配模式要求的星数为$b$，$k$为$K$-向量数目。上述若干星图识别算法特征参见表 1.4。

表 1.4 几种典型星图识别算法指标特征(Tichy V. et al. 2011)

| 作者 | 年份 | 特征提取 | 星库容量 | 搜索时间 | 验证时间 |
|------|------|----------|----------|----------|----------|
| Junkins | 1981 | $O(b)$ | $O(nf^3)$ | $O(f^3)$ | N/A |
| Liebe | 1992 | $O(flgb)$ | $O(n)$ | $O(n)$ | N/A |
| Scholl | 1995 | $O(blgb)$ | $O(nf^2)$ | $O(nf^2)$ | $O(k)$ |
| Quine | 1996 | $O(flgb)$ | $O(n)$ | $O(lgn)$ | N/A |
| Padgett | 1997 | $O(f)$ | $O(n)$ | $O(n)$ | N/A |
| Mortari | 1997 | $O(b)$ | $O(nf)$ | $O(k)$ | $O(bk^2)$ |
| Kolomenkin | 2008 | $O(b)$ | $O(nf)$ | $O(k)$ | $O(kf^2)$ |

从表 1.4 中看到，Mortari 的金子塔算法和 Kolomenkin 改进的 SLA 算法搜索时间最快，特征提取时间最短，有希望成为未来星敏感器的主流识别算法。北京控制工程研究所周建涛最近改进了 Kolomenkin 和 Tichy 投票算法，误识别率大大降低，效率提高至少 2 倍，可将全天识别时间降低到 1 秒以内。

### 1.2.2.3 高动态技术

为完成高级空间任务，遥感卫星，科学探索和国防任务等需要频繁的机动操作，要求有较高机动能力。但是，一旦星敏感器速率超过 0.1º/s，在未采取任何措施下，精度会很快下降到预期目标以下，动态性能已经是限制一些先进空间任务能否完成的核心指标(Pasetti A. et al. 1999; Samaan M. A. 2002; Zhang W. et al. 2012; Sun T.et al. 2013; Liao Y.et al. 2014; Liu C. S. et al. 2015; Zhang J. et al. 2015)。研究提高动态性能的技术已经是星敏感器的热点技术之一。

由于航天器角速率的影响，图像帧捕获的星点在积分时间内会运动若干像元，形成明显的拖尾，星点形貌总体不再是高斯形貌，导致运动下的定位过程存在许多问题。第一，信噪比指标会大幅下降，影响星等探测能力，降低定位精度。第二，

为捕获足够的能量，较低星等的恒星被切断而不能用于定位。第三，并不是所有像素均用于计算质心，这也会引起误差。第四，如果有加速度，星点形貌则时时刻刻发生变化，速度畸变较为严重，难以确定质心位置。调节积分时间是个减轻运动影响的可行方法(Liu C. S. et al. 2015; Zhang J. et al. 2015)。延迟积分时间(TDI)技术(van Bezooijen R. W. H. 2002; van Bezooijen R. W. H. et al. 2003)、可变帧率技术(Michaels D. L. et al. 2004)、运动自适应补偿(IMA)技术(van Bezooijen R. W. H. et al. 2003)、dynamical binning algorithm(DBA)(Pasetti A. et al. 1999)、Richardson Lucy(RL)图像复原技术(Sun T.et al. 2013; Liu C. S. et al. 2015)等，就是几种典型的技术，致力于提高星敏感器的动态性能。

延迟积分时间技术，就是根据焦平面内星点的实时速度一帧帧同步移动星点，据报道，该技术能够获得几乎与静态一样的精度。该技术缺点是：(1)需要精确的速度和时钟同步信息，定位性能依赖于硬件实施情况，若时钟出现问题，则单帧数据失效；(2)只能处理单轴，垂直轴仍然受运动干扰，需与 IMA 技术搭配使用，运动补偿才能达到较高精度。可变帧率技术就是通过分析实时速度信息调整积分时间来减弱像移效应。该技术缺点是：(1)高帧率导致可用星点数下降，这需要更高敏感能力和更低噪声的探测器；(2)由于信噪比的限制，帧速不能无限制提高；(3)仍需做运动补偿。运动自适应补偿技术通过将每帧曝光时间的末端与同步信号同步，若曝光时间精确至 1 毫秒，运动增量项可精确计算，从而实现运动补偿。但运动自适应补偿技术面临如下问题：(1)计算运动增量需要精确的速度和时钟同步信息；(2)速度存在畸变、抖动、加速效应时误差不易补偿；(3)需计算补偿后定位精度，为在轨校正提供基准。Pasetti 等发明了 DBA，将单帧再细分为多个小帧，用软件模拟 TDI 技术补偿像移误差，该算法效率较高、鲁棒。但在速度畸变场合，硬件实施上有一定缺陷(在像移量为 binning 指数整数倍值表现较好，在非整数倍时会累积像素取整误差)。且多帧数据参与定位，探测器读出噪声不可忽略，存在星点强背景干扰，导致延长积分时间对提高信噪比贡献十分有限。可变帧率技术和 DBA 技术均要求背景和读出噪声非常小，适合较高运动速率情形，同 TDI 技术一样，在速度畸变情况下，无法保证相平面所有星点保持静态精度。孙婷、刘朝山等采用了 RL 算法复原星点，可应用于速度畸变(SVPSF)场合，能够有效利用先验速度信息，恢复星点真实位置，但在多个星点参与姿态确定场合显得较为复杂,RL 算法在效率上需要商榷，

仅适用于低速情形。国防科技大学等研究人员研究帧间相关技术降低像移影响，该技术北京控制工程研究所研究人员亦有提及，未有应用的报道。

另外一种特殊的补偿技术，在星敏感器前端安置折叠镜面(Denver T. et al. 2004)。通过调整镜片倾斜角度，辐射光路能够在预定方向反射。根据运动调整固定倾斜角，能够获得转动平台给定方向长时间的积分图像。由于星敏感器能够决定并输出运动物体指向，该仪器适合对运动物体进行闭环追踪。不过目前该技术主要用于望远镜上。

从表 1.1-1.2 几款典型的甚高精度星跟踪器型号可以看出，甚高精度星敏感器动态范围普遍较高。AST-301 星跟踪器在X轴 3.1°/s，Y轴 0.1°/s时保持全精度，HAST、HYDRA 等星跟踪器(Blarre L. et al. 2008)在1°/s时仍然能保持准静态精度。从公开文献可知，AST-301 能够达到这样的性能与以下因素有关：(1)速度方向采用 TDI 技术；(2)垂直速度方向使用图像运动补偿技术。Ball 公司 HAST 产品在以下方面做出改进：(1)采用可变积分时间技术，实现二轴不受运动像移影响或影响较小。SODERN 公司的 HYDRA 星跟踪器在以下方面做出改进：(1)多探头分离和信息融合技术，提高探头探测能力；(2)APS 技术，能够根据速度改变积分时间，类似于可变帧率技术。ST5000、DayStar 装载在近空间气浮平台(35km)上，该产品采用可变帧率、长焦距和增益设计等实现1.4°/s的"静态"精度(Young E. F. et al. 2012; Truesdale N. A. et al. 2013)。归结起来，TDI、图像运动自适应、可变帧率和多探头等技术是提高动态性能或实现全精度的主要方法。从上述改进可以看出，解决高动态的根本途径是提高弱星等探测能力和降低运动像移影响，即要求：(1)能够采用类似 TDI 技术提高对弱星的响应能力，提高星点的信噪比，防止信号被噪声淹没；(2)能够进行运动图像补偿。北京控制工程研究所对此做了大量工作，如使用 APS 和 EMCCD 器件提高星等探测能力，采用多帧图像叠加和自卷积技术技术降低像移的影响，自适应调节积分时间以得到最佳信噪比，对误差进行辨识校正，采用快速星图跟踪算法等。

## 1.2.2.4 在轨误差校正技术

由于光学系统像差、光轴指向误差、器件的不均匀性、CCD 相面不平整、CCD 制造工艺缺陷、焦距误差等，会造成成像星点位置与实际位置存在一定偏差或抖动，

星敏感器内部参数校正是星敏感器另一项关键技术，包括几何畸变校正、焦距和主点位置校正等(钟红军 2011)。

镜头畸变包括径向畸变、离心畸变和薄棱镜畸变(张靖等 2007)。设 $x, y$ 为恒星在相平面面测量坐标，$x_0, y_0$ 为主点坐标。径向畸变含桶形和枕形畸变，数学模型为：

$$\Delta x_r = x_d(a_1 r^2 + a_2 r^4 + a_3 r^6 + ...), x_d = x - x_0, y_d = y - y_0 \tag{1.1}$$
$$\Delta y_r = y_d(b_1 r^2 + b_2 r^4 + b_3 r^6 + ...), r^2 = x_d^2 + y_d^2.$$

离心畸变是棱镜系统光轴不同面所致，数学模型为：

$$\Delta x_r = c_1(3x_d^2 + y_d^2) + 2c_2 x_d y_d + ... \tag{1.2}$$
$$\Delta y_r = 2d_1 x_d y_d + d_2(x_d^2 + 3y_d^2) + ...$$

薄棱镜畸变源于设计、装配不完善，数学模型为：

$$\Delta x_p = s_1(x_d^2 + y_d^2) \tag{1.3}$$
$$\Delta y_p = s_2(x_d^2 + y_d^2)$$

这些畸变合成起来，如 Herschel 星敏感器采用 5 阶多项式描述几何形变 (Feuchtgruber H. 2012)：

$$x_{d'} = -k_0 + k_1 x_d + k_2 y_d + k_3 x_d(x_d^2 + y_d^2) + k_4 x_d(x_d^2 + y_d^2)^2 - k_5 x_d^2 - k_6 x_d y_d - k_7 y_d^2 \tag{1.4}$$
$$y_{d'} = -h_0 + h_1 y_d + h_2 x_d + h_3 y_d(x_d^2 + y_d^2) + h_4 y_d(x_d^2 + y_d^2)^2 - h_5 y_d^2 - h_6 x_d y_d - h_7 x_d^2$$

其中，$x_{d'}, y_{d'}$ 表示真实位置与主点的距离。

对上述误差的校正技术包括地面标定或在轨标定技术。地面标定只相当于粗校正。针对主点校正，地面采用单星模拟器和转台标定，将星敏感器转动 360 度，每个转动位置模拟星坐标均可得到，通过圆拟合即得到主点位置，以上过程重复多次即能得到较高精度的主点坐标(钟红军等 2010；乔培玉等 2012)。对于焦距和光学畸变标定，利用天文观测的方法，将地球作为匀速转动的转台，并固定星敏感器的方法地面标定星敏感器参数，此方法即容易受大气因素(如不同的反射、图像运动、不同的色散)干扰和重力效应的影响(李春艳等 2006)。

航天器发射后的引力释放效应、复杂多变轨道的环境、宇宙射线、球形天体坐标系像面阵的非线性投影、温度交变等，导致地面校正的偏离，为能反映星点位置

偏离的真实情况，必须实施高精度的在轨校正，这也是任何高精度甚至所有星敏感器必须具备的(Liebe C. C. 2002; Anderson J. et al. 2003)。

目前在轨标定方法有两类，一类是根据外部姿态信息的校准，另一类是根据星内角距原理的校正(Samaan M. A. et al. 2002)。前者要求事先提供一个已知的精确姿态，如果姿态存在误差，则误差会引入校准过程，如采用扩展卡尔曼滤波对在轨星图进行处理，较最小二乘估计法收敛速度快、稳定性好且精度较高(刘一

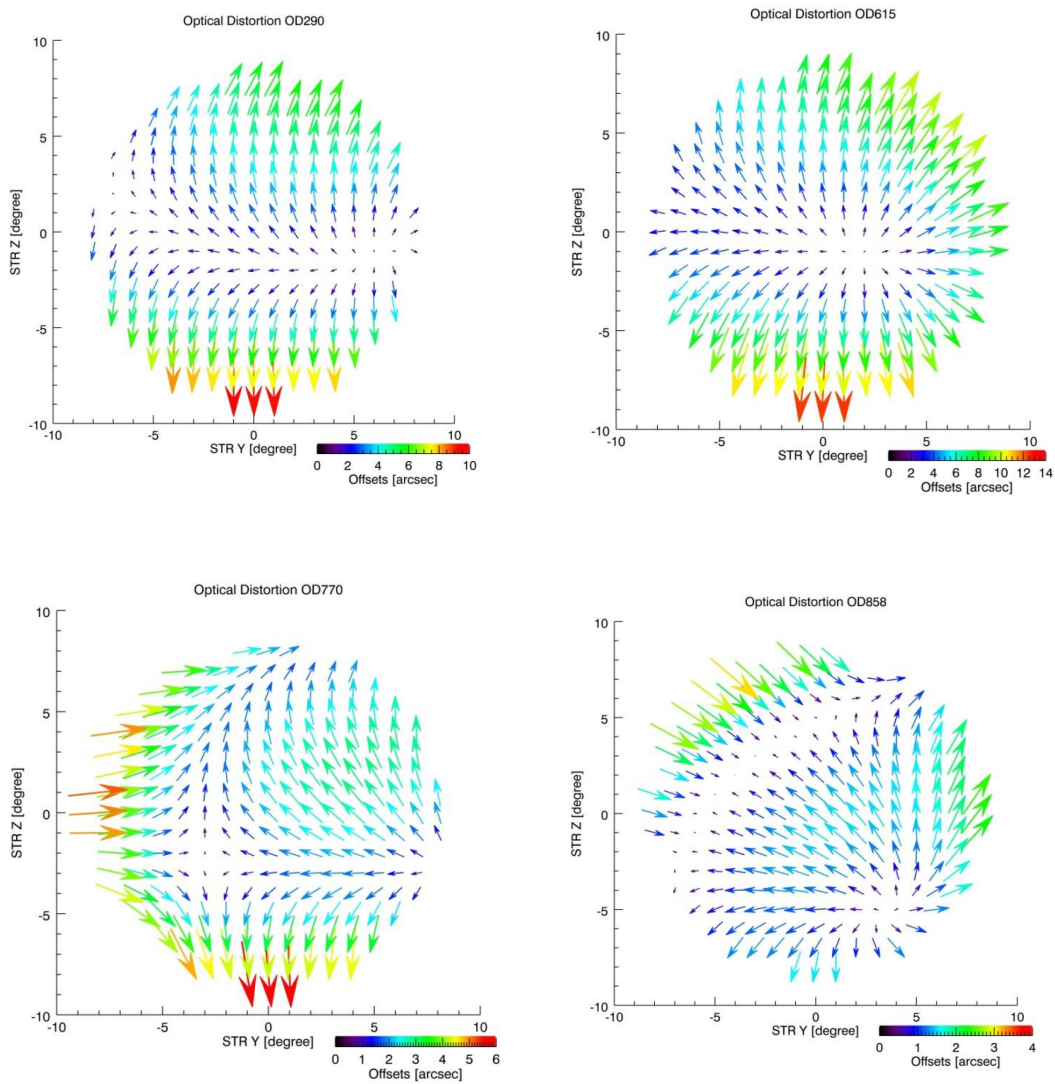

图 1.2 Herschel 星敏感器不同时段光学畸变分布

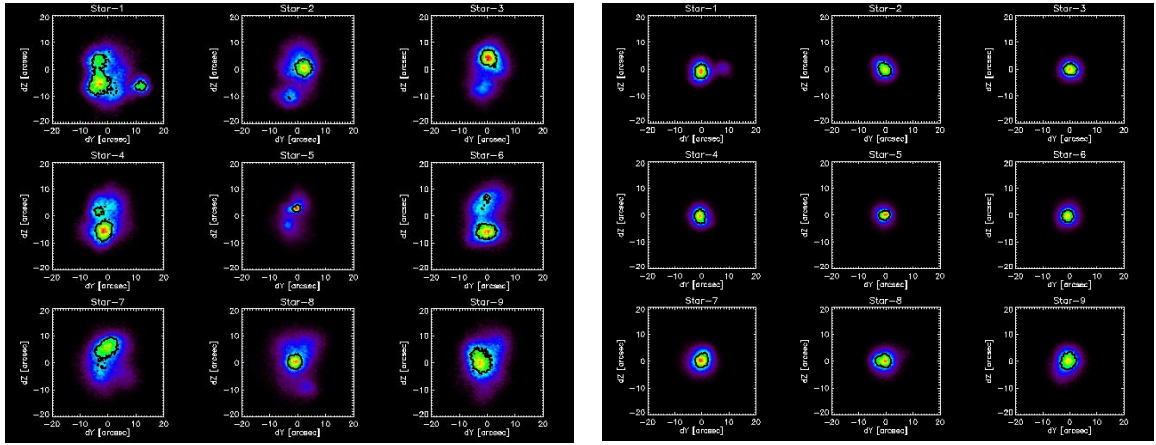

图 1.3 Herschel 星敏感器光学畸变校正前(左)后(右)的单星误差直方图

武等 2003)。另外，基于卡尔曼滤波的在轨校准方法，采用一阶径向畸变模型，利用摄像机标定中的径向排列约束对外部姿态和内参数进行在轨校准(王涛等 2005)，据报道误差在 0.05pixel，这根据外部姿态信息的校准的方法，其精度改进有一定限制。

后一类方法基于星内角距测量值不变的原理，检测在轨飞行期间星内角距测量值和真实值之间的偏差，这是最常用的方法。Samaan 使用非线性最小二乘优化估计校正光学镜头焦距误差，用序列递归在轨校正方法校正焦平面的畸变，主要思想就是利用正交变换与角度无关特性将测量星的内角距与星库做比较，实现在轨校正(Samaan M. A. et al. 2002)。SIRTF 利用 PCRS 系统测量已知主导星的精确位置校准望远镜的内部参数(Mainzer A. K. et al. 2003)。SED36 通过将探头与电子单元分离，避免了温度场的耦合，并设计了恒温系统，用钛/碳化硅结构替代铝合金架构，使得安装更牢固，在热和机械方面降低了温度交变带来的低频误差；在光学扭曲校正方面，采用多项式校正+多区域再次校正的方法大幅降低了空域低频误差，最后光轴指向精度达到0.33"。HAST 通过改变 CCD 帧速率改变星点成像信噪比并降低运动像移的影响，并校正了来自由于发射和温度交变导致的视轴误差以及光学扭曲或 CCD 响应老化导致的误差。图 1.2 为 Herschel 星敏感器不同时期几何畸变效果图。图 1.3 为修正前后的单星定位误差直方图。由图 1.2-1.3 可知，畸变是时空域变化的量，空域变化较大，时域会漂移。在星点提取完成后，需要建立模型，对不同星场位置、不同时期的几何畸变进行校正，进而得到更精确的位置。STR 首先通过星载软件自适应修正焦距，再通过 CCD 像平面二维校正(5 阶多项式)、三维校正及 CCD 像元结构校正，最终将定位精度从 1.57arcsec 降到 0.84arcsec (Schmidt M. et al. 2012)。国内

的在轨标定研究中，哈尔滨工业大学提出类似于 Samaan 的在轨标定方法用于标定星像点偏差和光学透镜焦距变化(杨博等 2001)。北京控制工程研究所高动态敏感器采用的校正方式类似于 SED-36。

## 1.3 课题问题描述

对星敏感器精度起决定性作用的是单星定位精度，总结起来前述几个关键技术均与降低单星定位误差息息相关，而降低单星定位误差目前存在多个问题。

首先，目前定心算法精度或效率不高。常用的质心法或其改进算法，存在 S 误差，对噪声、光扰敏感，较好的结果在 1/13pixel 左右，虽然高斯曲面拟合法精度很高，耗时却长，离在轨应用较远。在有限积分时间内，由于航天器运动或机动，星点信号会产生像移并可能被噪声切断，在白天或杂光等强干扰下信噪比可能低于最小值，直接降低了星点提取精度。其他多种因素如 CCD 像素响应不一致、飞行器环境温度交变导致的焦距变化、CTI 效应、星点对高斯形貌得偏离等同样导致星点提取存在偏差。国外产品报道的先进质心提取算法对噪声、运动误差等抑制较好，效率高，在一定动态情况下精度优于 1/50pixel。

其次，动态情况下航天器运动和振动会导致运动畸变难以识别校正，动态下单星定位误差较大。国内主要通过调节积分时间、帧相关等技术，保持静态精度的角速率小于 0.6°/s(200ms)。国外产品如采用的 IMA 算法、可变帧率技术等均能较好的在极限星等和高动态等极限环境下保证星点提取精度，1°/s 速率下精度可保持(500ms)。相比国外产品，国内产品在实时精确的运动补偿上仍然欠缺。

再次，关于光学畸变导致的空域畸变误差，焦距漂移导致的质心偏差等在轨校正工作尚不完善，单星定位偏差较难纠正。无论是高精度产品还是甚高精度产品，国内均采用在轨标定技术，技术方案成熟，标定精度高达 1/100pixel，国内仍处于研究和实证阶段，仅对部分参数实施在轨标定技术，精度提升有待验证。

最后，在甚高精度星敏感器若干核心指标上，如动态性能、姿态输出速率等，与目前的精度指标似有矛盾，提高了其中一项，另外两项就可能存在问题。

基于上述问题可知，解决在弱星、像移、强干扰、空域畸变等干扰因素下甚高精度单星定位和校正问题，对甚高精度星敏感器研制具有重要意义。

## 1.4 论文内容和结构安排

本论文的研究内容包括三个部分：如何进一步提高静态星点提取精度；如何进一步提高动态星点定位精度；如何有效实施在轨校正再次提高星点定位精度。论文的主要研究框架概述如下：

第一章：绪论。包括甚高精度星敏感器部件的研制现状，与敏感器相关关键技术的研发现状，课题问题描述，论文内容和组织架构。

第二章：CRLB(Cramér-Rao Lower Bound)理论和探测器噪声模型。本章属基础知识，介绍现代统计数学著名的 CRLB 理论，以及与 CRLB 计算相关的探测器件噪声模型，将用于后续章节计算星点定位误差。

第三章：单星定位误差降低方法。根据目前姿态确定算法如 QUEST(QUaternion ESTimator)算法特点，给出本文的误差模型和后面的降低路线。

第四章：甚高精度星点定位技术研究。研究迭代加权质心算法(IWCOG)优化理论及可能存在的问题，给出差分 IWCOG 技术。使用 CRLB 理论导出器件的理论精度，论证各种算法的最优性。分析与 Meanshift 算法的内在关联性。针对模型误差建模，分析模型误差导致的定位误差，并对高采样技术进行误差分析。给出其他 PSF 形貌的 2 种等效模型。分析提高星敏感器静态精度的方法。

第五章：全精度高动态技术研究。首先研究单帧解析补偿动态技术，研究导出运动状态下的定位精度，给出速度畸变机制，并导出速度畸变补偿公式和定位精度。其次，给出多帧解析补偿实现全精度原理，分析帧差分法不能单独实现恒星陀螺机理并给出可能解决办法。

第六章：在轨校正技术研究。分析在轨校正存在的问题，针对一些不易校正的误差，提出 CRLB 误差约束思路。导出 CRLB 全约束公式，分别研究几大因素导致的误差，对总体误差分析，给出 3 个重要推论及相关验证。

第七章：姿态确定技术优化。将第六章 2 个推论引入姿态确定方程并对其进行仿真验证。

第八章：总结和展望。论文成果总结，创新点概括和下一步工作建议。

为与已经发表的论文内容相对接，各章节部分变量的定义有可能并不一致，该变量的具体含义以上下文解释为准。并且，为保持论文简洁，省略了一些公式推导细节，仅给出重要的推导步骤及结果。

# 第二章 CRLB 理论和探测器噪声模型

星点定位精度是评价定位算法和校正技术好坏的关键指标，为精确获得探测器噪声等因素对其影响，本章介绍现代数理统计中非常著名的 Cramér-Rao Lower Bound 理论，以及用于 CRLB 相关计算的探测器信号和噪声模型。

## 2.1 CRLB 理论

作为数理统计领域诞生的里程碑式理论之一， CRLB 是有关参数统计在理论和工程上最基本的界(Rao C. R. 1992; Cramér H. 1999; Jutten C. et al. 2004)。CRLB，反比于费希尔(Fisher)信息量，是任何确定样本空间下参数估计的下限，任何该参数的无偏估计器不能获得低于 CRLB 的方差。它的创建者，如 Rao C. R.、Cramér H.等几乎在同一时期(1945 前后)建立了这个理论，Frechet M. (非参估计)和 Darmois G. (多参估计)等研究者对该项工作有重要贡献，因此，该下界有时也称为 CRFD 下界。该工作有几个方面的影响：回答了一些统计学家们关心的基本的问题，如一个统计估计器到底有多好，当估计某参数时是否有下界等。

除此外，非参统计领域(Hardle W. 1990)还有诸如 Ziv-Zakai 界、Barankin 界、GBhattacharyya 界等，是否使用这些均方误差下界取决于模型和可能的性能测量，但 CRLB 依然是最有影响的界，计算简单，有理论支撑。对于给定的模型，它是一个无偏估计器是否具备最优性能的判据，可避免不必要的在大量估计器中寻优过程，节省设计者在建模方面的时间。这个理论与数学统计中的许多分支都有联系，如识别、线性模型、最大似然、假定测试等。在通信等领域已经得到大规模应用，在星敏感器领域方面少见报道，但后文研究表明，CRLB 涉及星敏感器方方面面。

下面给出 CRLB 理论的定义、性质、证明及推论(Streit R. L. 2010)。该理论已经十分成熟，详细的内容请参考参数估计等书籍，不再赘述。

(1) 无偏估计(unbiased estimate)

假定样本观测量 $\mathbf{x} \in \mathrm{X} \subset \mathrm{R}^n$，服从 pdf $p(\mathbf{x};\boldsymbol{\theta})$ 分布，其中 $\boldsymbol{\theta}$ 是要确定的参数

矢量，$\boldsymbol{\theta} \in \Theta \subset R^{n_\theta}$，令 $\boldsymbol{\theta}_0 \in \Theta$ 为某未知参数 $\boldsymbol{\theta}$ 真实估计量，$\hat{\boldsymbol{\theta}}(\mathbf{x})$ 为 $\mathbf{x}$ 关于 $\boldsymbol{\theta}$ 的估计，对于 $\boldsymbol{\theta}$ 的期望定义为：$E_\theta(\bullet) = \int_X (\bullet) p(\mathbf{x}; \boldsymbol{\theta}) d\mathbf{x}$，则当且仅当

$$E_\theta(\hat{\boldsymbol{\theta}}(\mathbf{x})) = \boldsymbol{\theta}_0 \tag{2.1}$$

成立时称之为无偏估计。

(2)有效率(efficient)

定义该模型的 Fisher 信息矩阵(FIM)为：

$$I(\boldsymbol{\theta}) \triangleq E_\theta \left\{ s(\mathbf{x}; \boldsymbol{\theta}) s^T(\mathbf{x}; \boldsymbol{\theta}) \right\}, s(\mathbf{x}; \boldsymbol{\theta}) \triangleq \left. \frac{\partial \log p(\mathbf{x}; \boldsymbol{\theta}')}{\partial \boldsymbol{\theta}'} \right|_{\boldsymbol{\theta}' = \boldsymbol{\theta}} \tag{2.2}$$

对于正则条件，如 $h(\mathbf{x}) \equiv 1$ 或 $h(\mathbf{x}) \equiv \hat{\boldsymbol{\theta}}(\mathbf{x})$，由公式(2.1)知，pdf 对 $\boldsymbol{\theta}$ 的偏微分满足：

$$\frac{\partial}{\partial \boldsymbol{\theta}^T} E_\theta(h(\mathbf{x})) = E_\theta(h(\mathbf{x}) s^T(\mathbf{x}; \boldsymbol{\theta})) \tag{2.3}$$

特别地，当 $h(\mathbf{x}) \equiv \hat{\boldsymbol{\theta}}(\mathbf{x})$，公式(2.3)写成分量形式为：

$$\frac{\partial}{\partial \boldsymbol{\theta}_j^T} E_\theta(\hat{\boldsymbol{\theta}}_i(\mathbf{x})) = E_\theta(\hat{\boldsymbol{\theta}}_i(\mathbf{x}) s^T(\mathbf{x}; \boldsymbol{\theta}_j)) = \delta_{ij} \tag{2.4}$$

定义该模型的协方差矩阵为：

$$Var(\hat{\boldsymbol{\theta}}(\mathbf{x})) = E \left\{ (\hat{\boldsymbol{\theta}}(\mathbf{x}) - \boldsymbol{\theta}_0)(\hat{\boldsymbol{\theta}}(\mathbf{x}) - \boldsymbol{\theta}_0)^T \right\} \tag{2.5}$$

假定 $p(\mathbf{x}; \boldsymbol{\theta})$ 的 Jacobian 和 Hessian 矩阵关于 $\mathbf{x}$ 和 $\boldsymbol{\theta}$ 是而二阶可积可微的，则有如下性质成立。

**性质 1：** FIM 的反比即为 CRLB 界，$CRLB(\boldsymbol{\theta}) = I^{-1}(\boldsymbol{\theta})$。如果存在上述下界，则任何无偏估计器满足以下不等式：

$$Var(\hat{\boldsymbol{\theta}}(\mathbf{x})) \geq CRLB(\boldsymbol{\theta})$$

等式成立当且仅当均方意义上 $\hat{\boldsymbol{\theta}}(\mathbf{x}) - \boldsymbol{\theta} = CRLB(\boldsymbol{\theta}) s(\mathbf{x}; \boldsymbol{\theta})$ 成立。满足 CRLB 界的无偏估计器称有效率的，又名最小无偏方差估计器(MUVE)。

**证明：**

利用柯西-施瓦茨不等式可以证明上述定理。

假定公式(2.1)成立，考虑以下随机变量，$U(\mathbf{x}) = a^T \hat{\boldsymbol{\theta}}(\mathbf{x}), V(\mathbf{x}) = b^T s(\mathbf{x}; \boldsymbol{\theta})$，由前述无偏定义可知，$U(\mathbf{x})$ 均值为 $a^T \boldsymbol{\theta}_0$，$V(\mathbf{x})$ 均值为 0，由公式(2.1)和(2.5)，二随机变量的协方差矩阵分量为：

$$Cov(U, V) = E \left\{ (U(\mathbf{x}) - a^T \boldsymbol{\theta}_0)(V(\mathbf{x}) - 0)^T \right\} = a^T b$$

$$Var(U) = a^T Var(\boldsymbol{\theta})a, Var(V) = b^T I(\boldsymbol{\theta})b \tag{2.6}$$

由柯西-施瓦茨不等式，

$$Cov(U,V) \leq \sqrt{Var(U)}\sqrt{Var(V)} \tag{2.7}$$

可得，

$$\frac{(a^T b)^2}{b^T I(\boldsymbol{\theta}_0)b} \leq a^T Var(\boldsymbol{\theta})a \tag{2.8}$$

上式对所有情况成立，即 $\max\limits_{b \neq 0} \dfrac{(a^T b)^2}{b^T I(\boldsymbol{\theta}_0)b} \leq a^T Var(\boldsymbol{\theta})a$，求解过程可等价为：

$$\max_{b \neq 0}(a^T b)^2, b^T I(\boldsymbol{\theta}_0)b = 1 \tag{2.9}$$

由拉格朗日乘子方法可得 $b^* = \dfrac{I^{-1}(\boldsymbol{\theta}_0)a}{a^T I(\boldsymbol{\theta}_0)a}$，带入公式(2.8)得，

$$a^T(Var(\boldsymbol{\theta}) - I^{-1}(\boldsymbol{\theta}_0))a \geq 0 \tag{2.10}$$

可验证当 $\hat{\boldsymbol{\theta}}(\mathbf{x}) - \boldsymbol{\theta} = CRLB(\boldsymbol{\theta})s(\mathbf{x};\boldsymbol{\theta})$ 时，等式成立。得证。

**性质 2：** $p(\mathbf{x};\boldsymbol{\theta})$ 二次可微，则可知 $I_{ij}(\boldsymbol{\theta}_0) = -E\left\{\dfrac{\partial^2 \log p(\mathbf{x};\boldsymbol{\theta}')}{\partial \theta_j \partial \theta_i}\right\}$。

**证明：** 
$$\frac{\partial^2 \log p(\mathbf{x};\boldsymbol{\theta}')}{\partial \theta_j \partial \theta_i} = \frac{\partial}{\partial \theta_j}\left\{\frac{1}{p(\mathbf{x};\boldsymbol{\theta}')}\frac{\partial}{\partial \theta_j}p(\mathbf{x};\boldsymbol{\theta}')\right\}$$

$$= -\frac{1}{p^2(\mathbf{x};\boldsymbol{\theta}')}\left[\frac{\partial}{\partial \theta_j}p(\mathbf{x};\boldsymbol{\theta}')\right]^2 + \frac{1}{p(\mathbf{x};\boldsymbol{\theta}')}\frac{\partial^2}{\partial \theta_j \partial \theta_i}p(\mathbf{x};\boldsymbol{\theta}') \tag{2.11}$$

$$= -s^2(\mathbf{x};\boldsymbol{\theta}) + \frac{1}{p(\mathbf{x};\boldsymbol{\theta}')}\frac{\partial^2}{\partial \theta_j \partial \theta_i}p(\mathbf{x};\boldsymbol{\theta}')$$

$$I_{ij}(\boldsymbol{\theta}_0) = -E\left\{\frac{\partial^2 \log p(\mathbf{x};\boldsymbol{\theta}')}{\partial \theta_j \partial \theta_i}\right\} = E\left\{s^2(\mathbf{x};\boldsymbol{\theta}) - \frac{1}{p(\mathbf{x};\boldsymbol{\theta}')}\frac{\partial^2}{\partial \theta_j \partial \theta_i}p(\mathbf{x};\boldsymbol{\theta}')\right\} = E\left\{s^2(\mathbf{x};\boldsymbol{\theta})\right\} \tag{2.12}$$

**性质 3：** 如果 $E_o(\hat{\boldsymbol{\theta}}(\mathbf{x})) = \boldsymbol{\theta}_0 + r(\boldsymbol{\theta}_0)$，$r$ 可微，则称估计是有偏的，且有偏估计的方差下界为：

$$Var(\boldsymbol{\theta}) = r(\boldsymbol{\theta}_0)r^T(\boldsymbol{\theta}_0) + (I + \nabla r^T(\boldsymbol{\theta}_0))I^{-1}(\boldsymbol{\theta}_0)(I + \nabla r^T(\boldsymbol{\theta}_0))^T \tag{2.13}$$

证明略(Streit R. L. 2010)。

　　CRLB 至少有以下三种特点:(1)该下界针对无偏估计器，不针对有偏估计器;(2)工程上可能不存在达到该下界的算法；(3)当样本数达到一定规模时，最大似然估计能够渐进达到这个界。由性质 1 可知，比较协方差可得各种估计器的性能，满足 CRLB 的即是最优估计器。性质 2 提供了另外一种计算 CRLB 方法。性质 3

表明存在有偏估计器，不满足 CRLB 定理。CRLB 在本文中主要用于证明高精度定位算法的最优性(第四章)，给出定位技术在静态、动态情况下的理论下限(第四、五章)，约束其他误差源(第六章)，以及姿态优化(第七章)。

## 2.2 探测器噪声模型及信噪比指标

CCD 的典型噪声源(Magnan P. 2003; Basden A. G. 2015)，如下图 2.1 所示，有 3 类：(1)与信号和背景相关，光子噪声；(2)与信号响应、传输、转换相关，包括暗电流噪声、光响应不一致噪声(PRNU)、暗电流不一致噪声(DCNU)、复位噪声；(3)与信号输出相关，包括源跟随器噪声的热噪声、1/f 噪声和读出噪声、量化噪声。光子噪声源于信号和背景光子计数。暗电流噪声与积分时间的开方成正比，温度每上升 8K 暗电流增长一倍，可通过降低电流密度和降温来降低，由于埋入通道的使用,暗电流已经可以忽略。PRNU 是像素间光电响应不一致造成，DCNU 也是像素暗电流响应不一致造成,无法消除。前述几种噪声均服从泊松分布。复位噪声又称 KTC 噪声，源于电容采样过程中的复位过程。源跟随器会产生热噪声和 1/f 噪声。这三种噪声可通过相关双采样(CDS)技术得到一定消除。量化噪声源于 A/D 转换过程，更高位数的 A/D 转换器，量化噪声也较小。由于这几种噪声与信号输出相关，又称为读出噪声，在经过 CDS 技术处理后，表现为高斯分布。

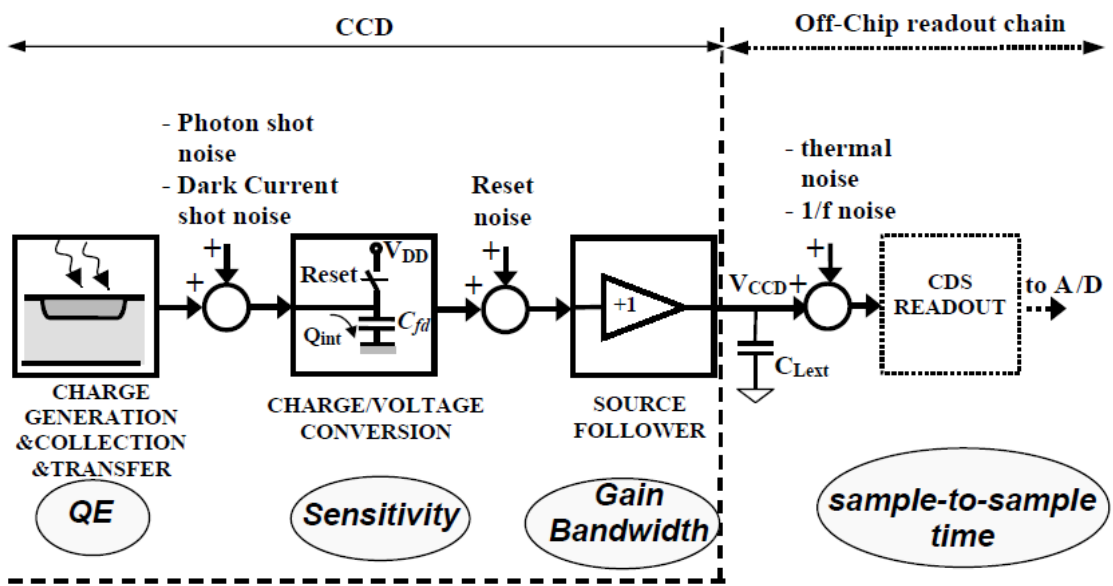

图 2.1 CCD 光电转换流程图(Magnan P. 2003)

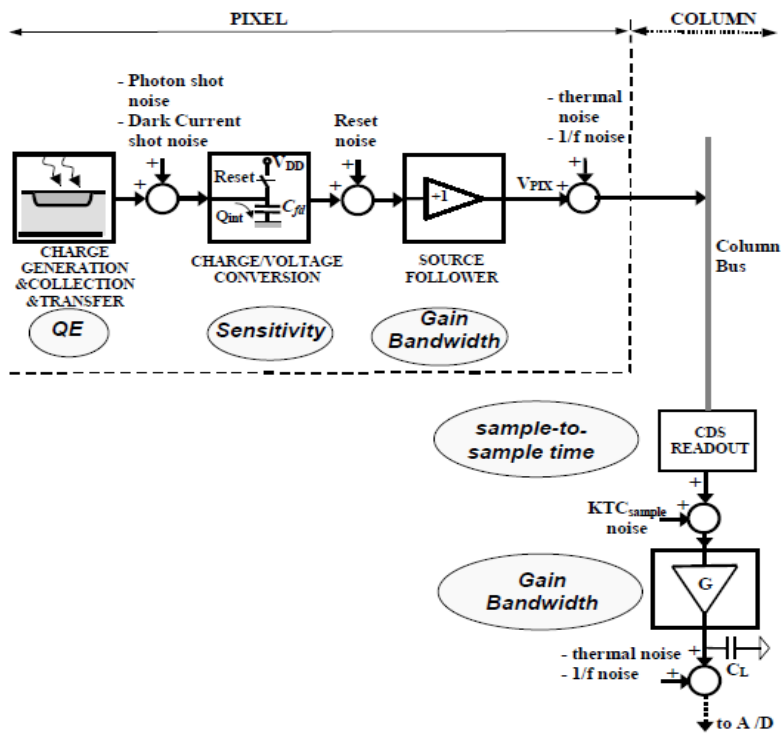

图 2.2 CMOS 光电转换流程图(Magnan P. 2003)

采用 CDS 技术的好处是：降低扩散复位噪声(KTC)并对源跟随器噪声滤波。CDS 对于 1/f 噪声表现为高通滤波模式，限制噪声频率最低为像素读出速率。剩余的噪声即为源跟随器的白噪声，与输出信号帧率和总像素成正比。

CMOS 和 CCD 一个明显的区别是读出噪声的差异，如图 2.2 所示。CMOS 列并行组织架构的读出带宽为底部列采样电路的行读出带宽。由于使用 CDS，内像素源跟随器晶体管的热噪声是主要的噪声源。因而，不同于 CCD 的噪声与帧频率成正比，CMOS 与帧频率几乎没有关系，但是填充因子和像素尺度约束了内置源跟随器的优化范围。通过 CDS 去除 1/f 噪声和 KTC 感应噪声，CMOS 的读出噪声可能会比 CCD 更低。

由于 CMOS 的优势，目前星敏感器正使用 CMOS 探测器代替 CCD 探测器(Liebe C. et al. 1998)。这些器件信号模型与 CCD 类似。因此，关于信号模型的一些推导同样适用于 CMOS 型。但是，由于 CMOS 型每个像元均有独立的读出单元，读出噪声是空域变化的(PRNU 或不同的读出策略等影响)。因此，CMOS 像元的读出噪声水平随像素不同而有差异，不仅如此，相同像素的读出噪声随每帧也有所不同(Bigas M. et al. 2006; Basden A. G. 2015)。如果信号模型使用单一的读出噪声值用于模拟或计算，就会出现性能过估计现象。为解决此问题，建议：根据每个像元单独对其进行计算，或先用一个平均读出噪声进行计算，然后再计算

噪声浮动对参数估计的影响。典型的器件如 sCMOS，由于其独特组织架构，其读出噪声随不同像元和读出模式而不同，是个变化量，其 RMS 值为 0.8e，如果使用单一的读出噪声模式模拟 sCMOS 工作模式，则会过高估计其性能指标。本节主要的噪声性质和符号表示可参见表 2.1。

表 2.1 噪声符号和性质

| 符号 | 源 | 噪声类型 | N=噪声量 |
|------|------|----------|----------|
| o | 目标信号 | 泊松分布 | $N_o^2 = S_o$ |
| d | 暗电流 | 泊松分布 | $N_d^2 = S_d$ |
| b | 信号偏置 | 泊松分布 | $N_b^2 = S_b$ |
| s | 天空背景 | 泊松分布 | $N_s^2 = S_s$ |
| r | 读出噪声 | 高斯分布 | $N_r$ |

以 S 表示信号，N 表示噪声。则每个像元收集的信号和噪声为：

$$S = S_o + S_d + S_b + S_s$$
$$N^2 = N_o^2 + N_d^2 + N_b^2 + N_b^2 + N_r^2 = S + N_r^2 \tag{2.14}$$

单个像元的信噪比可表示为：

$$R = S/N = S/(S + N_r^2)^{0.5} \tag{2.15}$$

设 $C = \sum S_o + S_d + S_b + S_s = C_o + C_d + C_b + C_s$，则整个星点的信噪比可表述为：

$$R = (\sum C^2/(C + N_r^2))^{0.5} \tag{2.16}$$

通常来讲，器件偏置和暗电流噪声比较小，像元内的主要噪声来自于天空背景和读出噪声，以下将其分为背景主导型和读出噪声主导型两种。

(1) 读出噪声主导型 (短曝光时间)

此时信噪比为：

$$R = (\sum C^2/(C_o + N_r^2))^{0.5} \tag{2.17}$$

S/N 随读出噪声增大而快速下降，为提高信噪比，需采用一个低读出噪声的探测器，或者采用合成技术，使得窗口内数据通过一个读出端口输出，此时亚采样对定位有一定影响。

(2) 天空背景主导型 (长时间曝光或背景较亮)

此时信噪比为：

$$R = (\sum C_o^2 / (C_o + C_s))^{0.5} \tag{2.18}$$

R 随天空背景噪声增大而快速下降，为提高信噪比，需降低窗口内像素数量，或者在更好的视宁度下观测。后文 CRLB 理论将以公式(2.17)-(2.18)计算信噪比，将最终的定位精度表示为信噪比的函数。

# 第三章 单星定位误差降低方法

在轨航天器要正常飞行或完成任务指标，需要姿态测量部件提供一定精度姿态信息。作为常备的角姿态测量部件，星敏感器的指向精度受多种因素影响，如何降低误差一直是设计者关注的重点。本章对星敏感器现有误差分类及其特点进行分析，通过研究相关解决方案优劣之处，基于单星定位误差公式提出一种为误差源建模并逐个降低的方法，为甚高精度星点定位技术、高动态技术、在轨误差校正技术误差改进建立统一评价指标。

## 3.1 姿态确定和误差

设系统状态为 $x(t) = [q(t), \omega(t)]$ ，则一个自由的刚性航天器其姿态运动和动力学方程可表述为：

$$\dot{\hat{x}}(t) = \mathbf{F}(\hat{x}(t), t)$$
$$\hat{x}(t)\,|_{t=0} = x_0 \tag{3.1}$$

假设知道航天器精确模型和初始状态，牛顿机械论的观点认为未来任何时刻航天器的运行状态均可以推算出来，这在理想情况下是成立的。由于轨道环境多变性(光压、热环境)、航天器存在大量的柔性机构等因素，为完成预定的空间任务，又需要航天器处于机动或执行状态，这种情况下的姿态运动和动力学方程可表述为：

$$\dot{\hat{x}}(t) = \mathbf{F}(\hat{x}(t), t) + \mathbf{G}(t)d(t)$$
$$\hat{x}(t)\,|_{t=0} = x_0, d(t)\,|_{t=0} = d_0 \tag{3.2}$$

可见，即便精确了解初始状态 $(x_0, d_0)$ 和初始模型( $\dot{\hat{x}}(t) = F(\hat{x}(t), t)$ )，由于控制律和执行器的模型( $G(t), d(t)$ )难以确定，未来姿态肯定会偏离预定曲线且难以预测。以遥感卫星等成像仪器为例，高的指向精度和稳定度是高质量成像和地面几何配准的必然要求。为完成特定任务，提供实时高精度卫星指向是必须的环节。

基于太阳敏感器、地平仪、磁强计等单点解析法姿态确定技术，其测量器件本身精度低，加上仅有少数矢量参与提供姿态，所得指向误差较大。由于星敏感器采用多星定位，姿态角测量方程通常为迭代姿态估计法(如QUEST算法)，而

陀螺的带宽较大，高精度的陀螺器件其姿态角速率测量精度很高。相比单点姿态确定法，基于星敏感器和陀螺的姿态确定系统能满足大部分航天器要求。所以，现在航天器主要依靠星敏感器和陀螺仪提供姿态信息。这种情况下，航天器姿态运动和动力学方程可表示为：

$$\dot{\hat{x}}(t) = \mathbf{F}(\hat{x}(t), t) + \mathbf{G}(t)d(t)$$
$$\hat{y}(t) = \mathbf{H}(\hat{x}(t), t) + v(t) \tag{3.3}$$
$$\mathrm{E}\left\{v(t_k)\right\} = 0, \mathrm{E}\left\{v(t_k)v^T(t_k)\right\} = \mathbf{R}\delta_{kk'}$$

上式表明，航天器处于稳态模式下，需要敏感器提供优异的指向精度和稳定度，如星敏感器，测量误差需为白噪声。由于星敏感器实际成像过程受光学畸变、焦距变化、探测器单元非均匀性响应、填充因子、航天器运动和振动、宇宙射线、杂光、噪声等影响，其测量误差很难满足上述定义，而呈现出一定的低频和高频误差特性。因此，从技术层面，必须研究这些误差特性以及降低误差的关键技术。

## 3.2 主流误差理论分析

研究误差如何产生和降低定位误差相关技术一直是热点问题。对于误差如何产生和降低方案，流行的有两种思路。一种是根据姿态输出结果分析误差特点反溯各种可能误差源及其解决办法，有大量国内外文献报道，也是星敏研究人员的主要参考；另外一种是从误差源头分析误差特点，找出解决办法，文献多集中在星点定位算法和处理像移上。两种方法均有优缺点，本文将在后续章节探讨后一种思路。为比较，首先介绍第一种思路。

按照 ESA 标准，根据误差变化周期的不同，星敏感器输出误差可划分为常值偏差、低频误差和高频误差(武延鹏 2011; 卢欣等 2014; 边志强等)。

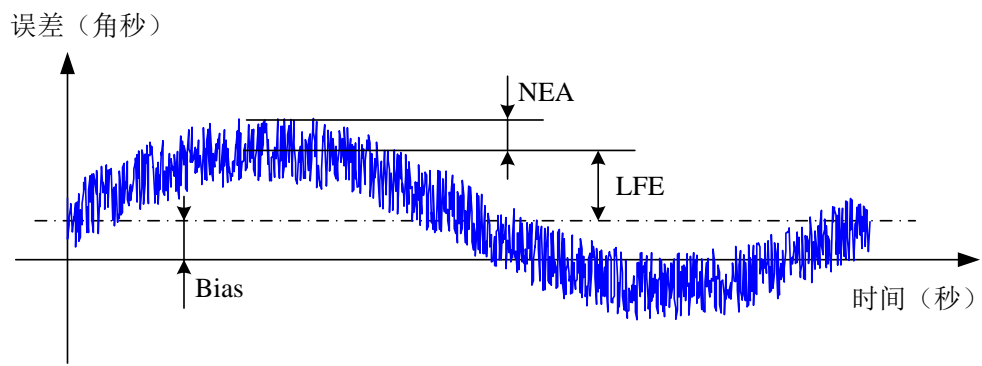

图 3.1 星敏感器测量误差示意图(边志强等)

图 3.1 为星敏感器测量误差示意图。从图 3.1 中可以看出，星敏感器总的测量误差由这三部分叠加而成。星敏感器的误差模型可以用傅里叶级数形式表示：

$$\Delta\theta(t) = \Delta\theta_o + \sum_{i=1}^{k} A_i \sin(p_i t + \varphi_i) + n_{ST} \tag{3.4}$$

其中，$\Delta\theta_o$ 表示星敏感器的视轴指向偏差，$n_{ST}$ 表示星敏感器噪声误差，右边第二项代表了星敏感器各个谐波频率的误差，主要表现为低频慢变误差。

(1)视轴指向偏差：指星敏感器视轴指向相对于相平面坐标系中心指向的偏差，主要与发射段振动和冲击引起的机械结构的变化、安装矩阵的变化以及引力释放等相关，是星敏感器工作平均温度的函数。该误差在空域在轨标定后较为稳定。

(2)高频误差 NEA(也称噪声等效角)：受离焦影响，星像斑一般覆盖若干个像元，受成像器件噪声、电路噪声以及亚像元质心算法等缺陷因素影响，一段时间内对同一位置星像斑所求出的中心坐标总是不尽相同的，总会围绕着某一数值在一定范围内波动，从而导致星敏感器输出有所波动，就产生了星敏感器高频随机误差。主要包括：

(a)像素空间误差：指由于探测器空间不均匀性及星点定位算法带来的测量误差。探测器不均匀性一般表现为高频白噪声；星点定位算法指采用亚像元定位算法带来的误差，表现为高频误差。

(b)时域误差：由于硬件噪声产生的测量误差，与空间无关，属于时域白噪声随机误差，该项可通过姿态滤波大部分去除。

需要指出的是，探测器空间不均匀性实际上是偏差而非白噪声(Piterman A. et al. 2002)，是可以校正的，详细可见 Herschel 星跟踪器的报道。星点定位算法主要指 S 误差(Liebe C. C. 2002；贾辉 2010)，由于噪声引入误差校大，实际上是难以校正的。另外，硬件噪声并不完全是通过姿态滤波抑制，主要降低方式仍然是高精度定位算法和空域滤波。这些结论，详细可参见第四章、第七章分析。

(3)低频误差(LFE)：其变化周期一般受像素周期、轨道周期影响。一是与观测条件有关，如瞬时视场中的目标星数量、亮度、光谱特性以及在视场中的位置等；二是在轨飞行阶段，星敏感器所处热环境周期性变化等因素引起的光轴变形。主要包括：

(a)视场空间误差(短周期项)：指单星校正残差，主要来自镜头畸变残差，镜头材料老化，CCD 响应电子漂移，星表误差、恒星残余标定系数等，是空域变化的，覆盖整个视场。

(b)热变形误差(长周期项)：在轨热变形导致星敏感器测量坐标系相对于机械系发生的慢变漂移误差。热变形模型十分复杂，很难用数学模型进行描述。一般表现为轨道周期性变化，是低频非随机误差。视场空间误差项和热变误差项使星敏感器输出呈现周期性波动，称为低频误差。

图 3.2 和 3.3 为 STECE 卫星的 Vondrak 滤波后的 3 轴低频误差和不同轨道周期的误差。由图中可见，相比于低于 1 角秒的随机噪声误差，低频误差是星敏感器的主要误差源，且随轨道周期呈现一定的规律性。SODERN 星敏感器专家认为低频误差(残差)是不可校正的(Blarre L. et al. 2008)。虽然轨道环境、视场均有周期性，热变形导致的误差、空域校正残差并不一定有强周期性，目前未发现国外有关周期性的数据报道。

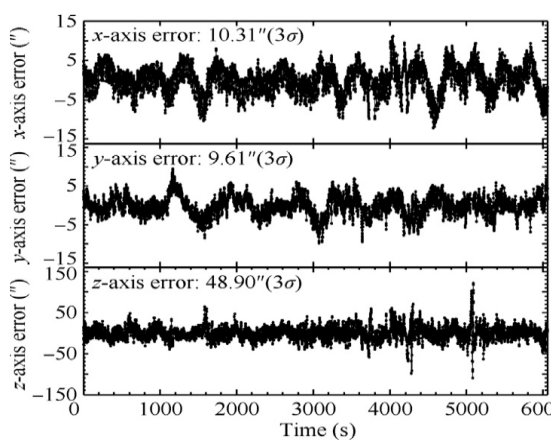 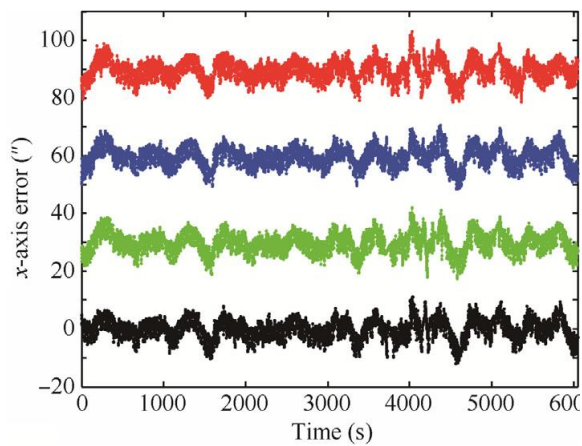

图 3.2 STECE 卫星的 Vondrak 滤波后的 3 轴低频误差(Lai Y. et al. 2014)　图 3.3 STECE 卫星不同轨道周期的 Vondrak 滤波后的 x 轴低频误差(Lai Y. et al. 2014)

综上所述，低频误差和高频误差是星敏感器输出姿态具有的主要误差特性。低频误差很难修正，很大程度上决定了卫星姿态基准，在选型时应将 LFE 指标作为重点指标(卢欣等 2014; 边志强等)。

Liebe 通过研究星跟踪器技术(Liebe C. C. 2002)，给出单星定位精度公式：

$$\varepsilon = \frac{A_{FOV}}{N_{pixel}} \times \delta \qquad (3.5)$$

其中，$\varepsilon$ 为星敏感器单星定位精度，$A_{FOV}$ 为全视场角，$N_{pixel}$ 为探测器阵列长度(或者宽度)像元个数，$\delta$ 为星点质心提取误差，包括上述 3 种误差。从公式(3.5)可以看出，$A_{FOV}$、$N_{pixel}$、$\delta$ 均会影响单星定位最终精度。

从公式(3.5)可知，为提高单星提取精度，减小像元等效角，增加参与姿态确定的恒星数量是提高最终姿态确定精度常规手段。这个结论有一定局限性，详细参见第六章分析。

从姿态输出的误差特征，很容易让人想到使用在线滤波技术和控制技术来纠正由于设计和器件缺陷等引入的误差。如将低频误差进行傅立叶分解，然后使用识别技术将低频分量提取出来再补偿回去的方案，试图使得误差降下来。但该法要取得成功，需要满足一些假设和条件：频率有限且稳定，幅值在各频率段比较稳定，补偿后的残差满足高斯噪声分布等。由于帆板扰性振动、光压等不可避免导入到最后姿态输出环节，所以还需要额外识别扰性和光压等造成的低频分量并将其剔除(Xiong K. et al. 2012; 吕振铎等 2013)，否则这个低频分量会与星敏感器的低频分量耦合在一起。而这些假设，由前述误差的分类来看，是较难满足的，表现在：(1)视场空间误差和热变形误差无强周期概念，温度对器件的影响还有类似磁铁励磁和去磁表现出来的磁滞效应，导致所谓的低频频率实际上一直是变化的，即便在同一轨道位置、同一姿态、外界干扰全部一样，星敏的误差输出也不会重复(参见图 1.2)；(2)扰性振动等因素导致的低频误差本身难以识别出来，预先将其剔除很困难(参见图 3.4)；(3)运算量比较大。实际上低频误差亦是时变的，用姿态滤波或控制方法难以彻底解决星敏感器测量问题。相反，如果星敏感器测量精度达到了甚高水平，此时若姿态输出误差出现较大波动，则可以断定该误差主要来自扰性振动等外界因素，此时再做滤波或低频分解则非常有效，报道的成功案例是 DayStar。DayStar 用在气浮平台上，由于风速的影响，整个平台振动频率在 3.5Hz 以下，根据香农采样定理，需要两倍频的采样才能将这个振动频率识别出来，而 DayStar 的姿态输出频率为 10Hz，完全满足这个条件(Truesdale N. A. et al. 2013)。如图 3.4 和 3.5 所示，在经过低频分解和滤波后，DayStar 姿态输出角呈现高频误差特性，随时间呈现一定变化。

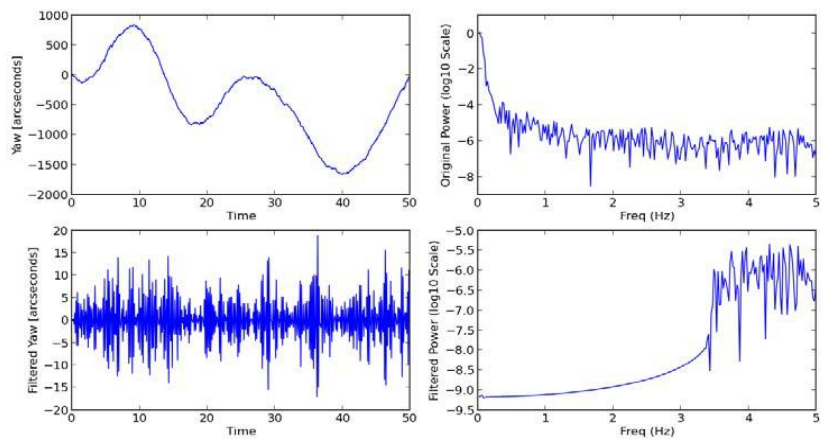

图 3.4 使用 FFT 和高通椭圆滤波器后的偏航角误差(Truesdale N. A. et al. 2013).

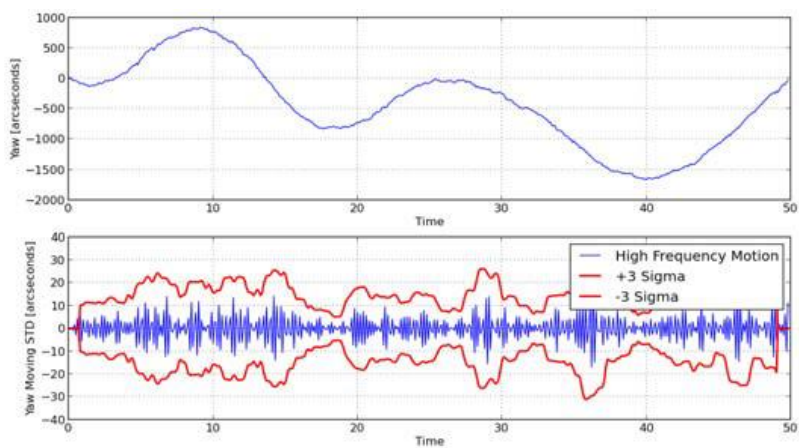

图 3.5 偏航角误差随时间的分布图(Truesdale N. A. et al. 2013)

以上分析可得出 2 个结论：(1)滤波或分解技术难以彻底解决星敏感器本身产生的低频误差，主要适合解决非星敏感器产生的低频误差；(2)解决由外界挠性导致的低频误差，星敏感器姿态输出频率必须高于振动频率 2 倍，否则会产生频谱混叠。由于星敏感器更新速率有限，外界干扰产生的高频非偏置误差可通过滤波抑制，但高频偏置误差则必须使用其他方法解决。

## 3.3 单星定位误差降低方法

设 $x_i, y_i$ 为视场内实际星点位置，$x_0, y_0$ 是星场中心位置，星敏感器测量得到的对应观测矢量为 $\tilde{\mathbf{b}}_i$。根据 Samaan's 定义，恒星在焦平面的投影模型为：

$$\tilde{\mathbf{b}}_i = \frac{1}{\sqrt{(x_i - x_0)^2 + (y_i - y_0)^2 + f^2}} \begin{bmatrix} x_i - x_0 \\ y_i - y_0 \\ -f \end{bmatrix} \tag{3.6}$$

目前星敏感器主要使用 QUEST 或类似算法(后文均以 QUEST 为例)给出姿态信息,用的是星图匹配成功的跟踪星点(Markley F. L. et al. 2000)。已知星点天球参考矢量为 $\mathbf{r}_i$,卫星姿态方向余弦阵为 $A$,$e_i$ 为测量误差,满足:$e_i^T \mathbf{b}_i = 0$,$E(e_i) = 0, \mathrm{var}(e_i) = \sigma_i^2$,则目前的姿态确定算法可表述为:

$$\tilde{\mathbf{b}}_i = A\mathbf{r}_i + e_i = \mathbf{b}_i + e_i \tag{3.7}$$

该问题可等价为 Wahba 问题,求解正交矩阵 $A$,使得指标函数:

$$J(A) = \sum_i \frac{1}{2} \beta_i (A\mathbf{r}_i - \mathbf{b}_i)(A\mathbf{r}_i - \mathbf{b}_i)^T \beta_i \tag{3.8}$$

达到最小,其中 $\beta_i$ 为加权因子,满足:$\beta_i^2 = \sigma_i^{-2} / \sum_i \sigma_i^{-2}$。

有大量的算法用于解决 Wahba 问题,在此不再详细介绍(Crassidis J. L. 2002)。由姿态确定算法如 QUEST 可知,单星精度决定了最终的姿态精度。因此,研究造成单星定位误差的各种因素和误差降低技术非常有必要。前述星敏感器工程流程(图 1.1)实际上也是一个误差传递过程,形象的表示可参见图 3.6。

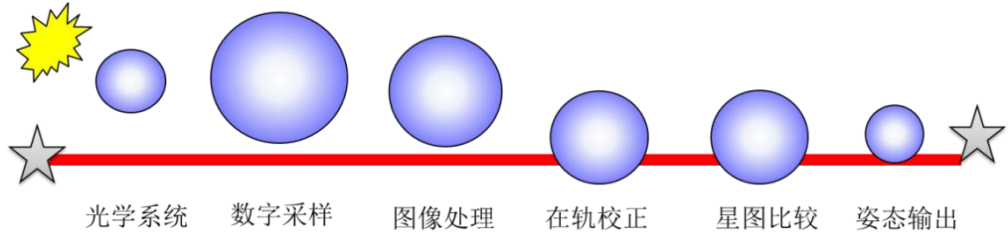

图 3.6 单星测量误差流程示意图

恒星经过这些过程,与理想位置会出现偏差,用公式可表述为:

$$\hat{x} = x_{true} + \sum_p e_p, \sum_p e_p = e_{algo} + e_n + e_l + e_m + e_{prnu} + e_{conf} + e_{catalog} \cdots + e_f \tag{3.9}$$

其中,$x_{true}$ 为星点真实位置,$e_p$ 表示单次测量中某种误差源 $p$ 带来的误差量,包括算法($e_{algo}$)、器件噪声($e_n$)、光学镜头畸变($e_l$)、CCD 器件的光响应不一致($e_{prnu}$)、温致焦距漂移($e_f$)、像移($e_m$)、速度畸变和抖动,系统装调($e_{conf}$)、星表($e_{catalog}$)等带来的位置误差(Zhang J. et al. 2016)。

假设这些误差源 $e_p$ 在时空域是独立分布的随机变量,服从均值 $\mu_{x_p}$、方差为 $\sigma_{x_p}^2$ 的某种分布,则所带来的误差可表达为 $\delta x_p = \sqrt{\mu_{x_p}^2 + \sigma_{x_p}^2}$ ($\left|\mu_{x_p} - \sigma_{x_p}\right| \le \delta x_p \le \left|\mu_{x_p} + \sigma_{x_p}\right|$ 或 $\left|\mu_{x_p} + \sigma_{x_p}\right| \le \delta x_p \le \left|\mu_{x_p} - \sigma_{x_p}\right|$)。对于小的偏差 $\mu_{x_p}$,要么忽略不计,

要么包含在 $\delta x_p$ 中。这样，要使单星测量误差要低于 0.02 像素，可以通过某种方式限制各个误差源的 $\mu_{x_p}$ 和 $\sigma^2_{x_p}$ 值。这样，有别于前述误差定义，上述误差源按其产生的偏差和方差大小可细分为常值误差、偏置误差和随机测量误差。

(1)常值误差即视轴偏差

与系统装调、引力释放和温度相关。温度不变时有偏差($\mu_{x_{conf}}$)，温度变化时有方差($\sigma^2_{x_{conf}}$)，需利用星敏感器与有效载荷的相对位置信息进行周期性标定。

(2)偏置误差包括空域畸变误差、空域校正残差，时域偏置误差等

空域畸变误差来自于镜头畸变、辐射损伤、器件老化等，可采用高阶多项式进行平场校正，这个是大的偏差。空域校正残差主要来自镜头畸变校正残差、CCD 器件的光响应不一致、时间抖动、模型误差(CCD 光谱响应和 PSF 分布)、模型限制、星点角速率不确定性(像移补偿)、图像截断误差(质心面积损失)、由角加速度导致的图像模糊、棱镜色差等，时域偏置误差主要来自温致焦距漂移(航天器工作环境变化)等。后两者不能校正的部分即为前文所谓的低频误差。

(3)随机测量误差包括 NEA 误差(方差)和质心误差(包括偏差和方差)

NEA 误差源于器件噪声、来自太阳、地球和亮星等的杂光干扰、时间抖动、模型误差(恒星光谱分布)、跟踪星等的不均匀性等，质心误差源于背景估计误差、亚像元质心算法、像移、速度畸变和抖动等。可通过采用更高精度的质心算法、可变积分时间设计解决部分质心误差问题。如 AST-301 采用新的质心算法，使用 TDI 技术和 IMA 技术改善像移影响。HAST 通过可调帧率技术减弱像移影响。

通过后端姿态输出结果反推，其研究加深了人们对于总体误差的理解，即误差不是个孤立的概念，而是多个误差源的综合。但这个思路试图将人们引向通过滤波方法和控制技术来解决精确定位问题，通过分析该解决方案更适合解决外界扰性导致的低频误差。鉴于此，论文放弃这个思路，转攻单星定位技术，在前端通过解决单星定位精度问题降低误差。由公式(3.9)可知，对每种误差源进行准确建模，再根据误差源特点对其产生的误差做限制可以实现精确定位。以下章节将分别对算法(第四章)、噪声(第四章)、信号模型(第四章)、运动(第五章)、速度抖动(第五章)、焦距漂移(第六章)、星点光谱中心变化(第六章)、星等分布(第七章)等因素进行分析，获得降低各类误差源定位误差的根本方法。

# 第四章 甚高精度星点定位技术研究

单星定位精度决定了星敏感器最终的指向精度。提高单星定位精度不仅要从硬件上考虑，软件上更需找到稳定的高精度星点提取算法。静态情况下，星点提取误差多受采样、噪声、模型畸变等影响，存在 S 误差、噪声误差、模型误差等问题，这些误差总体高于 0.1pixel，远大于甚高精度星点定位需求，优异的星点定位算法应能解决这一问题。本章分别展开了 IWCOG 算法优化、探测器 CRLB 特性、高采样 PSF 数据匹配算法理论精度、不同衍射形貌的高斯近似等研究，基本上覆盖并给出了对上述问题的解答，同时应用相关研究结果，给出了提高星敏感器静态精度的可能路径。

## 4.1 IWCOG 算法优化理论

目前国际上已有在轨使用的甚高精度星敏感器产品中，SIRTF 使用的 AST-301 星跟踪器、Pléiades 使用的 SED36 星跟踪器、Herschel 使用的 STR 星跟踪器、World-View 系列使用的 HAST 星跟踪器等姿态定位精度均达到亚角秒量级，其关键技术之一是单星提取精度的提高，国际上报道的单星提取精度可达 0.02pixel。目前研究的成熟的提取算法一类如质心法(Center of Gravity, COG)、带阈值的质心法、灰度平方加权法实时性较好，但均方误差(MSE)在 0.1pixel 左右；另一类如高斯曲面和抛物面拟合法均方误差可低于 0.1pixel，但不具备实时性，同时两类算法受噪声、运动等因素干扰较大，均不满足甚高精度成像敏感器的需求。

高斯形貌拟合法比常规质心法提取的精度要高，主要源于以下两点：(1)形貌拟合保留了信号"曲线"信息，而质心法则暗示信号分布图应符合阶梯形貌；(2)形貌拟合通过迭代方式修正与实际形貌偏差，当形貌与真实信号接近相同时迭代终止。而质心法只有足够的采样点情况下才能保证其为无偏估计，由于 CCD 探测器尺寸和光谱响应能力限制，离焦技术下的星点采样一般仅覆盖少数像元，不满足无偏估计要求。曲面拟合法计算量大，不满足实时要求。为实现更高精度

定位，发现国外自适应光学领域在研究 Shack-Hartmann 传感器波前定位中[1]，使用了迭代加权质心算法(Iteratively Weighted Center of Gravity, IWCOG)，并用仿真方法研究了其定位误差与信号参数、信噪比的关系、算法运行效率、参数误差随迭代次数的变化等(Baker K. L. et al. 2007; Vyas A. et al. 2009; Nightingale A. M. et al. 2013)，给出了若干定心技术其定位误差与信号参数关系表达式，这些研究表明相比前 2 类算法该算法有很大优势(Thomas S. et al. 2006)，目前尚未发现该算法在星敏感器产品上直接应用的报道(包括国内自适应光学领域)。设 $s(x, y)$ 为信号源，$\omega(x, y)$ 为加权函数，$\omega(x, y) = g(\|(x - x_c)^2 + (y - y_c)^2 / \sigma_\omega\|^2)$，$x_c, y_c$ 为像元中心坐标，$\sigma_\omega$ 为加权函数尺度，$K, L$ 为信号所占像元尺度。

由文献可知(Baker K. L. et al. 2007)，IWCOG 算法可以表述为：

$$x_c = \frac{M_{10}}{M_{00}}, y_c = \frac{M_{01}}{M_{00}}, \sigma_\omega = \frac{1}{M_{00}}\sqrt{M_{20}M_{00} + M_{02}M_{00} - M_{10}^2 - M_{01}^2} \tag{4.1}$$

当 $(x_c, y_c)$ 不再改进时停止，即上一次迭代结果与本次迭代结果基本相同。

该算法迭代方式能逼近真实点源位置，迭代次数可控。存在的问题是，Baker 给出的公式稍显复杂，由于峰值相位和噪声的不同，该算法仍然有些耗时，仿真角度未能说清算法定位机理和优越性，无法确定是否存在更好的算法问题，考虑到实时应用的需求，研究实用的优化算法也是必须解决的问题。

### 4.1.1 无噪声时的优化算法

为方便推导，公式(4.1)写成一维形式可表述为：

$$\mu = \frac{\sum_i x_i s_i \omega_i(x_i, \mu)}{\sum_i s_i \omega_i(x_i, \mu)} \tag{4.2}$$

其中，$s_i$ 为一维信号样本点。

通过化简，可知(4.1)式解最终满足下式：

$$\sum_{i=1}^{n} (x_i - \mu) s_i \omega(x_i, \mu) = 0, n \geq 3 \tag{4.3}$$

由文献可知(Anderson J. et al. 2000)，最小二乘高斯拟合算法满足下式：

---

[1] 准确地讲，先研究了类似的 Meanshift 算法，大约半年后才发现 Vyas A. 的论文及 IWCOG。

$$\mu = \arg\min\left\{\chi^2(\mu) = \sum_i \frac{[s_i - I_0 g_i(x_i, \mu)]^2}{\sigma_i^2}\right\} \tag{4.4}$$

其中，$\arg\min$ 指求使 $\chi^2$ 取最小值时的 $\mu$ 值，$\sigma_i^2$ 为像素噪声，假定满足：

$\sigma^2 = \sigma_i^2 = \sigma_j^2$，$I_0$ 为总信号量，$g_i(x_i, \mu)$ 满足高斯函数定义，尺度为 $\sigma_s$。

公式(4.4)的解需满足 $\partial\chi^2/\partial\mu = 0$，即，

$$\sum_i \frac{[s_i - I_0 g_i(x_i, \mu)]}{\sigma^2} I_0 g_i'(x_i, \mu) \approx \sum_i \frac{(\mu - x_i)[s_i - I_0 g_i(x_i, \mu)]}{\sigma^2} \frac{I_0 g_i(x_i, \mu)}{\sigma_s^2} = 0 \tag{4.5}$$

令 $\omega_i = g_i(x_i, \mu)$，$M_k(\mu) = \sum_i (x_i - \mu)^k I_0 g_i(x_i, \mu)\omega(x_i, \mu)$ 表示加权 $k$ 阶中心

矩，比较公式(4.3)和式(4.5)可知，最小二乘高斯拟合算法多出了一项：

$$M_1(\mu) = \sum_i (x_i - \mu) I_0 g_i(x_i, \mu)\omega_i(x_i, \mu) \approx 0 \tag{4.6}$$

公式(4.6)为奇函数，当 $s_i$ 与 $g_i(x_i, \mu)$ 重合时值为 0，由此可得(4.2)式。

由此可知，IWCOG 算法即为最小二乘高斯拟合算法在像素噪声一致情况下的解。在均匀高斯噪声下，最小二乘高斯拟合算法与迭代加权质心算法定位误差近似等价。为求得 IWCOG 算法优化版本，对最小二乘高斯拟合进行二次微分（K.J. Mighell 2002），可得，

$$\chi^2(\mu + \delta) = \sum_{k=0}^{\infty} \frac{1}{k!}(\delta \bullet \nabla)^k \chi^2(\mu) \approx \chi^2(\mu) + \delta \bullet \nabla \chi^2(\mu) + \frac{1}{2}\delta \bullet H \bullet \delta \tag{4.7}$$

其中，$H_{jj'} = \dfrac{\partial^2}{\partial\mu^{(2)}\partial\mu^{(1)}}\chi^2(\mu)$，$\delta$ 为步长。

公式(4.4)等价于求 $\delta^* = \min\arg\left\{\chi^2(\mu + \delta)\right\}$。由 $\partial\chi^2/\partial\delta = 0$ 可得新的迭代公式为：

$$\mu_{t+1} = \mu_t + \delta^*, H \bullet \delta^* = -\nabla\chi^2(\mu) \tag{4.8}$$

$\delta^*$ 可化简为：

$$\delta^* = 2 \cdot \sum_i (x_i - \mu)g_i\omega(x_i, \mu_t) \Big/ \sum_i g_i\omega(x_i, \mu_t) = \frac{2M_1(\mu)}{M_0(\mu)} \tag{4.9}$$

连续型的公式亦如此，将离散和变为积分形式即可。

由计算过程可知，上述仅计算出 $\omega$ 的中心点，对于 $\omega$ 的尺度 $\sigma_\omega^2$ 仍然需要算法得出。由此 $\sigma_\omega^2$ 可定义为：

$$\sigma_\omega^2 = \alpha \cdot \frac{\sum_i (x_i - \mu)^2 g_i(x_i, \mu)\omega(x_i, \mu)}{\sum_i g_i(x_i, \mu)\omega(x_i, \mu)} = \alpha \cdot \frac{M_2(\mu)}{M_0(\mu)} \tag{4.10}$$

$\alpha$ 系数是为保证 $\sigma_\omega^2$ 收敛于期望值。为了证明 $\sigma_\omega^2$ 收敛性需要引理 1(数学归纳法)。

**引理 1:** (1)如果 $0 < B_0 \le P \le D, \dfrac{P-B_0}{D-B_0} > \dfrac{Q}{B_0} > 0, PB_k >> DQ$ ，则 $B_{k+1} = \dfrac{PB_k - DQ}{B_k - Q}$ 序列单调递增必收敛至 $P+Q$ 。

(2)若 $P \ge D \ge B_0 > 0$ ， $PB_k >> DQ > 0$ ，则 $B_{k+1} = \dfrac{PB_k + DQ}{B_k + Q}$ 序列单调递增必收敛至 $P-Q$ 。

**性质 1:** 若 $\sigma_\omega^2$ 收敛于 $\sigma_s^2$ ，则 $\alpha \equiv 2$ 。

**证明：**

由公式(4.10)可得 $\dfrac{1}{\alpha}\left(\dfrac{1}{\sigma_s^2} + \dfrac{1}{\sigma_\omega^2}\right) = \dfrac{1}{\sigma_\omega^2}$ ，根据引理 1，当 $\alpha > 1$ 时， $\sigma_\omega^2$ 单调有界一定收敛。

令 $\sigma_{\omega 0}^2 = b\sigma_s^2$ ，则 $\sigma_{\omega n}^2 = \left[(\alpha-1) + \dfrac{1+(1-\alpha)/b}{(\alpha^n-1)/(\alpha-1)+1/b}\right]\sigma_s^2$ ，由此可得 $\alpha = 2$ 时， $\sigma_\omega^2$ 收敛至 $\sigma_s^2$ 。更具体的来说，当 b>1 时， $\sigma_\omega^2$ 单调递减有下界收敛至 $\sigma_s^2$ ，当 b<1 时， $\sigma_\omega^2$ 单调递增有上界收敛至 $\sigma_s^2$ ，当 b=1 时， $\sigma_\omega^2$ 恒收敛至 $\sigma_s^2$ ，所以无论加权函数 $\sigma_{\omega 0}^2$ 初始尺度如何选择， $\sigma_\omega^2$ 最终可收敛于 $\sigma_s^2$ 。

由每次迭代 $\sigma_{\omega k}^2$ 之间的关系也可得 $\sigma_\omega^2$ 一阶校正公式为: $\sigma_\omega^2 = \dfrac{\sigma_{\omega k}^2 \sigma_{\omega k+1}^2}{2\sigma_{\omega k}^2 - \sigma_{\omega k+1}^2}$ 。

从仿真过程看，由于 IWCOG 算法迭代次数可控，所以一般原有的 $\sigma_s^2$ 迭代都会趋于真实值。由于 $\omega$ 的指数衰减特性，计算窗口只需 $|x_i - \mu| \le 3\sigma_s$ 即可，这说明每次迭代参与计算的窗口数有限，从而加快单次迭代速度。

## 4.1.2 含噪声时的优化算法

此时 IWCOG 算法可表述为：

$$M_1'(\mu) = \sum_{i=1}^{n}(x_i - \mu)G_i\omega(x_i, \mu) = 0, G_i = s_i + e_i, n \geq 3 \tag{4.11}$$

设噪声源于背景，以下将其分两种情况。

(1)若 $e_i$ 为高斯噪声，则满足：

$$E(e_i) = 0, E(e_i e_j) = \sigma_n^2 \delta_{ij}, E(G_i) = s_i \tag{4.12}$$

由此可知，$\sigma_\omega^2$ 可表述为：

$$\sigma_\omega^2 = 2\frac{I_0\,\sigma_q^3\big/\sigma_s + \sigma_p^3}{I_0\,\sigma_q\big/\sigma_s + \sigma_p}, \frac{1}{\sigma_q^2} = \frac{1}{\sigma_\omega^2} + \frac{1}{\sigma_s^2}, \frac{1}{\sigma_p^2} = \frac{1}{\sigma_\omega^2} + \frac{1}{\sigma_n^2} \tag{4.13}$$

根据引理 1 可知 $\sigma_\omega^2$ 数列递增有上界收敛于 $\sigma_s^2 - \sigma_n^2$。

又 $M_0'(\mu) = \sum_i G_i\omega(x_i, \mu) = M_0(\mu) + \sigma_p$，$H \approx -\frac{1}{2}(M_0(\mu) + \sigma_p)$，由此得到迭代公式

为：

$$\mu_{t+1} = \mu_t + 2M_1'/M_0' \tag{4.14}$$

(2)若 $e_i$ 为泊松噪声，均值和方差为 $\lambda_n$，则满足：

$$E(e_i) = \lambda_n, E(e_i e_j) = \lambda_n \delta_{ij}, E(G_i) = s_i + \lambda_n \tag{4.15}$$

由此可知，$\sigma_\omega^2$ 可表述为：

$$\sigma_\omega^2 = 2\frac{I_0\,\sigma_q^3\big/\sigma_s + \sqrt{\lambda_n}\sigma_p^3}{I_0\,\sigma_q\big/\sigma_s + \sqrt{\lambda_n}\sigma_p}, \frac{1}{\sigma_q^2} = \frac{1}{\sigma_\omega^2} + \frac{1}{\sigma_s^2}, \frac{1}{\sigma_p^2} = \frac{1}{\sigma_\omega^2} + \frac{1}{\lambda_n} \tag{4.16}$$

根据引理 1 可以证明 $\sigma_\omega^2$ 数列递增有上界收敛于 $\sigma_s^2 - \sqrt{\lambda_n}\sigma_p \approx \sigma_s^2 - \lambda_n$。

又 $M_0'(\mu) = \sum_i G_i\omega(x_i, \mu) = M_0(\mu) + \lambda_n$，$H \approx -\frac{1}{2}(M_0(\mu) + \lambda_n)$，由此得到迭代公式

为：

$$\mu_{t+1} = \mu_t + 2M_1'/M_0' \tag{4.17}$$

### 4.1.3 特殊形貌畸变时的优化情况

#### 4.1.3.1 一阶偶畸变形貌

考虑 $G_i = s_i + q_i(\mu)$，$q_i(\mu) = \dfrac{I_0}{\sqrt{2\pi}\sigma_s} a_q (x_i - \mu)^2$，设 $\omega_{up}$、$\omega_{down}$ 分别为计

算

$M_2(\mu)$、$M_0(\mu)$ 是使用的加权函数，此时若下式：

$$a_q << \frac{1}{\sqrt{233 I_0}\sigma_s}, a_q^2(x_i-\mu)^4 << \sqrt{4\pi}\sigma_s^3 \tag{4.18}$$

成立，可证明 $\sigma_\omega^2$ 在 $\omega_{up}$、$\omega_{down}$ 满足某种条件时调递增有上界收敛于 $\sigma_s^2$。分以下

两种情况，论证有无噪声情况下均可使用原迭代公式(较小形貌畸变)。

(1)设 $\omega_{up} = \omega + 3a_q\sigma_s^2, \omega_{down} = \omega - a_q\sigma_s^2$，则，

$$M_0'(\mu) = \sum_i g_i \omega_{down} = \frac{I_0}{\sqrt{2}} - \frac{18 a_q^2 \sigma_s^4 I_0}{\sqrt{2\pi}}, H = -\frac{I_0}{2\sqrt{2}} - \frac{162 a_q^2 \sigma_s^4 I_0}{\sqrt{2\pi}} \tag{4.19}$$

于是，

$$H = -\frac{1}{2}M_0'(\mu) - \frac{153 a_q^2 \sigma_s^4 I_0}{\sqrt{2\pi}} \approx -\frac{1}{2}M_0'(\mu) \text{ 且 } M_1'(\mu) \approx M_1(\mu) \tag{4.20}$$

(2)设 $\omega_{up} = \omega_{down} = \omega - 7a_q\sigma_s^2$，则，

$$M_0'(\mu) = \frac{I_0}{\sqrt{2}} + 8 a_q\sigma_s^2 I_0 + \frac{126 a_q^2 \sigma_s^4 I_0}{\sqrt{2\pi}}, H = -\frac{I_0}{2\sqrt{2}} - 4 a_q\sigma_s^2 I_0 - \frac{378 a_q^2 \sigma_s^4 I_0}{\sqrt{2\pi}} \tag{4.21}$$

于是，

$$H = -\frac{1}{2}M_0'(\mu) - \frac{315 a_q^2 \sigma_s^4 I_0}{\sqrt{2\pi}} \approx -\frac{1}{2}M_0'(\mu) \text{ 且 } M_1'(\mu) \approx M_1(\mu) \tag{4.22}$$

#### 4.1.3.2 一阶奇畸变形貌

考虑 $G_i = s_i + q_i(\mu)$ ，$q_i(\mu) = \dfrac{I_0}{\sqrt{2\pi}\sigma_s} a_q (x_i - \mu)$，由 $q_i(\mu)$ 为奇函数，可得，

$$M_2'(\mu) \approx M_2(\mu), M_0'(\mu) \approx M_0(\mu), H \approx -\frac{1}{2}M_0'(\mu) \tag{4.23}$$

可知 $\sigma_\omega^2$ 等价于理想情形，$\sigma_\omega^2$ 单调递增有上界一定收敛于 $\sigma_s^2$。并且迭代方程仍然为：

$$\mu_{t+1} = \mu_t + 2M_1'/M_0' \tag{4.24}$$

### 4.1.4 IWCOG 仿真分析

令 $\sigma_s = 0.7$ 或 $\sigma_s = 1.2$ pixel，M=7pixel，采用蒙特卡罗法在不同位置(相位)生成星点灰度图，加入(或不加入)一定读出噪声，然后分别采用 COG 算法、高斯拟合法(Gaussian Curve Fitting，GCF)、IWCOG 算法计算质心，最后减去真实的质心获得绝对误差，分析星点尺度、噪声、相位对定位误差的影响，仿真结果参见图 4.1-4.2。

固定相位和尺度，随机加入噪声，改变 M 数为 3~17，分别运行 COG 法、IWCOG 算法、GCF 算法$10^6$ 次，计算三种算法的运行时间(单位为 1，比例尺=100：100：1，即图 4.3 中高斯拟合算法实际运行时间为标称值的 100 倍)，仿真结果参见图 4.3。

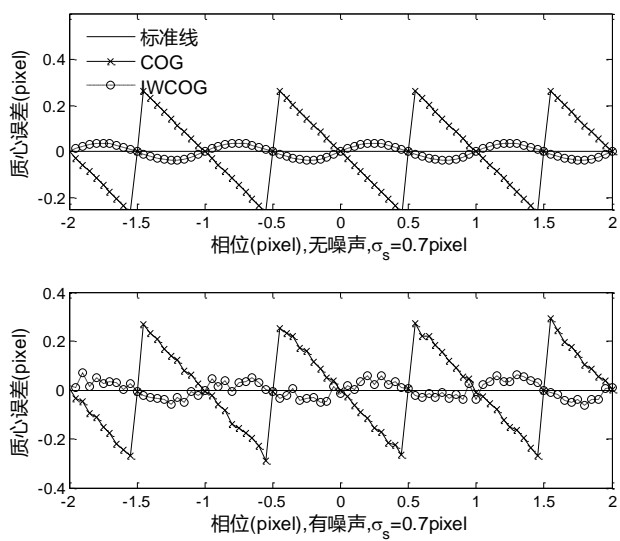

图 4.1 有无噪声时 COG 和 IWCOG 算法提取误差

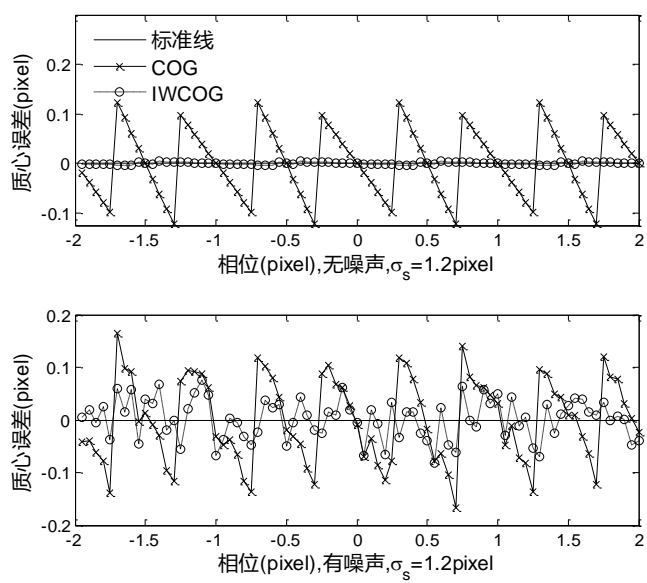

图 4.2 有无噪声时 COG 和 IWCOG 算法提取误差

由图 4.1 和 4.2 可以看出，IWCOG 法在有无噪声时均表现优异，无噪声时定位误差在 0-0.002pixel，由采样造成的 S 误差几乎可以忽略，在有噪声时误差仍然比 COG 要低。由图 4.3 可以看出，无论参与运算的窗口数量多少，高斯拟合算法最为耗时，是其他算法的 10 倍以上，随着窗口数增加，质心法和高斯拟合算法计算时间缓慢增加，IWCOG 算法增加较快，IWCOG 算法在 M 为 3-11 时，耗时为质心法的 2-5 倍。

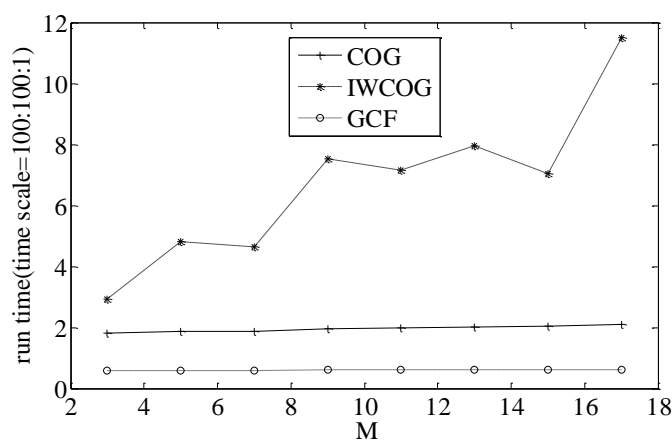

图 4.3 算法单次运行时间与 M 的关系

### 4.1.5 算法失效案例

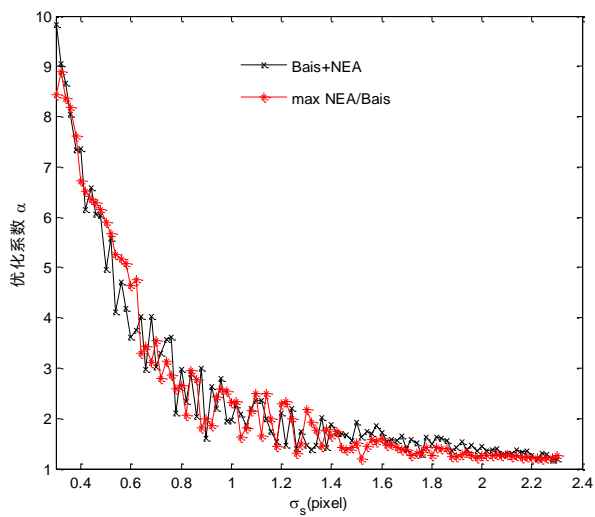

图 4.4 以 NEA 指标优化后的 $\alpha$ 系数

当运行仿真程序时发现，当星点尺度小于 0.7pixel 时，由于噪声分布的不均匀性，像点在迭代时计算的中心值若偏离实际位置太远，则程序有可能不会收敛而出现较大误差，发现解决这个问题的关键是尺度迭代中 $\alpha$ 系数，当 $\alpha$ 较大时很少出现上述情况。于是以绝对误差为比较指标，通过搜索的方式找到能够适应所有情况的系数，其结果参见图 4.4。

由图可知，当 $\sigma_s \geq 0.7$ pixel 时，$\alpha$ 系数基本为 2，这和理论推导基本一致，当 $\sigma_s \leq 0.7$ pixel 时，星点完全处于亚采样状态，此状态下易受外界干扰，如宇宙辐射、热像素点等，此时 $\alpha \geq 2$ 才能保证星点快速收敛。

## 4.2 IWCOG 与 Meanshift 算法

对比图像视觉领域经典的 Meanshift 算法(Fukunage K. et al. 1975; Cheng Y. 1995; Comaniciu D. et al. 2003)，IWCOG 与其有巨大的相似性：均采用迭代加权技术、均用于定位等。本节将展示二者的关联，并由图像相关性原理同样导出了 IWCOG 算法，并证明高斯加权函数的最优性(张俊等，2015，光学学报第 2 期)。

### 4.2.1 经典的 Meanshift 理论

基于Meanshift理论的技术是解决图形聚类、分割、模态识别和非刚性物体跟踪等图像视觉问题的有力工具。Meanshift框架最早由FuKunaga和Hostetler发明的一种无参估计算法(Fukunage K. et al. 1975)，主要思想是沿着密度梯度的上升方向寻找概率分布的峰值。最初的Meanshift表达式为：

$$\hat{m}(x) = \sum_i x_i g(\cdot) \Big/ \sum_i g(\cdot) \tag{4.25}$$

其中，$g(\cdot) = g\left(\left\|(x-x_i)/\sigma_s\right\|^2\right)$ 是核函数，$x_i$ 为样本点坐标，$n$ 为样本数，$\hat{m}(x)$ 为估计的样本中心。

样本点随距离样本中心的不同，其对均值 $\hat{m}(x)$ 的贡献是不相同的。Cheng最先考虑了这个因素(Cheng Y. 1995)，提出的更为一般的Meanshift表达式为：

$$\hat{m}(x) = \sum_{i=1}^n x_i g(\cdot)\omega(x_i) \Big/ \sum_{i=1}^n g(\cdot)\omega(x_i) \tag{4.26}$$

其中，$\omega(x_i)$ 为权函数，其余参数定义同公式(1)，以下类同。

按Cheng的算法原理，公式(4.26)中的 $\omega(x_i)$ 含义为离样本中心不同距离的样本点具有不同的权重，一般越近则权值越大，表示对均值 $\hat{m}(x)$ 的贡献越多。Cheng的算法扩大了Meanshift算法的应用范围，解决了聚类、模态识别、图像分割等问题。但是Cheng并未指出如何选取 $\omega(x_i)$。

Commanniciu 为解决非刚性物体的跟踪问题，使用 Bhattacharyya 系数定义了候选目标和跟踪目标的相似性，重新解释并导出了 $\omega(x_i)$ 公式，成功解决了视觉跟踪问题(Comaniciu D. et al. 2003)。但是，目前上述算法研究主要集中在样本点比较大的情况。实际星点覆盖像元数有限，属于小样本，上述算法不能算出星点的真实位置，且原算法主要解决聚类与跟踪的问题，与星点定位并不相同。IWCOG 算法在点源定位上取得了成功，主要改变是 IWCOG 算法改进了公式(4.26)加权方法，即将每步迭代的均值作为权函数新的中心点再次迭代计算(原算法是将窗口移动到每次迭代的中心点)，最终得到星点真实位置。可见，加权函数 $\omega$ 如何选取和解释是 Meanshift 算法能否扩展于其他应用的关键，以下首次结合 Meanshift 原理，通过形貌的互相关性研究，给出算法公式，重新阐释加权函数含义，给出算法的收敛性。

### 4.2.2 IWCOG 算法与互相关

星点在探测器焦面成像后，覆盖若干像元形成固定形貌的灰度直方图，在星点形貌较为理想情况下，可假定其形貌符合 $\omega(x_i, x)$，然后不断移动 $\omega(x_i, x)$ 中心位置逼近星点的灰度直方图，可知当二者互相关系数达到最大时，此中心即为所求。

设欲跟踪的实际星点形貌为 $g(x_i, x)$，离散化采样为 $g_i$，候选目标形貌为 $\omega(x_i, x)$，中心点为 $x$，离散化为 $\omega_i$，则跟踪形貌 $g(x_i, x)$ 与目标形貌 $\omega(x_i, x)$ 的互相关矩阵表示为 $\rho(s(g), s(\omega)) = \mathrm{cov}(s(g), s(\omega))$，互相关的意义是使得跟踪形貌和目标形貌中心点一致，可通过迭代并令每步迭代后的星点与目标形貌满足：$\mathrm{E}_g(x_i) = \mathrm{E}_\omega(x_i) = m(x)$，则可得，

$$\mathrm{cov}(s(g), s(\omega)) = \mathrm{E}\left\{\left[x_i - \mathrm{E}_g(x_i)\right]\left[x_i - \mathrm{E}_\omega(x_i)\right]\right\} = \sum_{i=1}^{n}(x_i - m(x))(x_i - m(x))g_i\omega_i \quad (4.27)$$

为使二者相似度最大，只需 $\partial\mathrm{cov}(s(g), s(\omega))/\partial m(x) = 0$，从而得出，

$$\hat{m}(x) = \sum_i x_i g_i \omega_i \Big/ \sum_i g_i \omega_i \quad (4.28)$$

比较公式(4.26)和(4.28) 可知，公式(4.28)中的 $\omega(x_i, x)$ 物理含义为目标形貌，脱离了 Meanshift 算法原本的定义，质心法相当于跟踪 $\omega(x_i, x) \equiv 1$ 的目标形貌，这与实际星点形貌不符。迭代含义即令 $\hat{m}(x) \to x$ 进行下一次计算，也即跟踪形貌 $g(x_i, x)$ 通过目标形貌 $\omega(x_i, x)$ 不断追踪逼近。

按 Cheng 的推导，影子核 $\hat{q}_K(x)$ 取得最大值即得到灰度直方图最大密度点，此时 $\hat{m}(x) - x \approx 0$，$x$ 即为所求。这要求 $\omega(x_i, x)$ 满足：

$$\omega^* = \arg\max\left\{\hat{q}_K(x) = \frac{1}{n}\sum_{i=1}^{n}K(x_i, x)\omega(x_i, x)\right\} \quad (4.29)$$

利用施瓦兹不等式可得，

$$\sum_i K(x_i, x)\omega(x_i, x) \le \left[\sum_i K^2(x_i, x)\right]^{1/2} \cdot \left[\sum_i \omega^2(x_i, x)\right]^{1/2} \quad (4.30)$$

等式成立当且仅当 $K(x_i, x) = C\omega(x_i, x)$ 。所以，对于高斯星点形貌 $g(x_i, x)$，其影子核函数 $\kappa$ (或 $\omega$) 仍然为高斯函数。由于星点灰度直方图是 $\omega(x_i, x)$ 积分近似，IWCOG 算法以星点灰度直方图灰度值代替它。

对于理想情况 $K(x_i, x) = C\omega(x_i, x)$ 时算法的收敛性证明参见性质 2。

**性质 2：** 若 $K(x_i, t)$ 为凸函数，$t_{j+1} \neq t_j$，则 $\hat{q}_K(t_{j+1}) > \hat{q}_K(t_j)$，$\hat{q}_K(t)$ 收敛于一点。

**证明：** 当 $K(x_i, t) = C\omega(x_i, t)$ 时，$\hat{q}_K(t) = \dfrac{C}{n} \sum_{i=1}^{n} K^2(x_i, t)$，$K^2(x_i, t)$ 为凸函数，$d$ 为维数，则，

$$\hat{q}_K(t_{j+1}) - \hat{q}_K(t_j) = \frac{Cc_{k,d}^2}{n\sigma^{2d}} \cdot \left[ \sum_{i=1}^{n} k^2 \left( \left\| \frac{t_{j+1} - x_i}{\sigma} \right\|^2 \right) - \sum_{i=1}^{n} k^2 \left( \left\| \frac{t_j - x_i}{\sigma} \right\|^2 \right) \right] \tag{4.31}$$

由凸函数性质 $\hat{q}_K(t_{j+1}) - \hat{q}_K(t_j) \geq \hat{q}'_K(t_j)(t_{j+1} - t_j)$，可令，

$$\Lambda \left( \left\| \frac{t_j - x_i}{\sigma} \right\|^2 \right) = k \left( \left\| \frac{t_j - x_i}{\sigma} \right\|^2 \right) \bullet k' \left( \left\| \frac{t_j - x_i}{\sigma} \right\|^2 \right) \tag{4.32}$$

则，

$$\hat{q}'_K(t_j)(t_{j+1} - t_j) = \frac{Cc_{k,d}^2}{n\sigma^{2d+2}} \sum_i \Lambda \left( \left\| \frac{t_j - x_i}{\sigma} \right\|^2 \right) \left[ \|t_j - x_i\|^2 - \|t_{j+1} - x_i\|^2 \right] \tag{4.33}$$

$$= \frac{Cc_{k,d}^2}{n\sigma^{2d+2}} \sum_i \Lambda \left( \left\| \frac{t_j - x_i}{\sigma} \right\|^2 \right) \left[ |t_j|^2 - 2t_j x_i - |t_{j+1}|^2 + 2t_{j+1} x_i \right] \tag{4.34}$$

$$= \frac{Cc_{k,d}^2}{n\sigma^{2d+2}} \left( \sum_i \Lambda \left( \left\| \frac{t_j - x_i}{\sigma} \right\|^2 \right) |t_j|^2 - 2t_j t_{j+1} \sum_i \Lambda \left( \left\| \frac{t_j - x_i}{\sigma} \right\|^2 \right) + |t_{j+1}|^2 \sum_i \Lambda \left( \left\| \frac{t_j - x_i}{\sigma} \right\|^2 \right) \right) \tag{4.35}$$

$$= \frac{Cc_{k,d}^2}{n\sigma^{2d+2}} \sum_i \Lambda \left( \left\| \frac{t_j - x_i}{\sigma} \right\|^2 \right) \left[ |t_j| - |t_{j+1}| \right]^2 > 0 \tag{4.36}$$

由此可得，

$$\hat{q}_K(t_{j+1}) - \hat{q}_K(t_j) \geq \hat{q}'_K(t_j)(t_{j+1} - t_j) > 0, \hat{q}_K(t_{j+1}) > \hat{q}_K(t_j) \tag{4.37}$$

同比可证

$$\hat{q}_K(t_{j+m}) > \hat{q}_K(t_j), \hat{q}_K(t) = \frac{C}{n} \sum_{i=1}^{n} K^2(x_i, t) < \frac{C}{n} \sum_{i=1}^{n} \max(K^2(x_i, t)) \tag{4.38}$$

可知，$\hat{q}_K(t)$ 单调递增有上限必收敛，从而证明了 IWCOG 算法的收敛性。

本节从 Meanshift 理论角度出发研究了高斯加权迭代算法，提出了基于形貌"追踪"的权函数定位机制，证明了算法的收敛性。

## 4.3 探测器件 CRLB 特性和 IWCOG 算法最优性

由于成像器件存在大量噪声，由第 2.1 节可知，CRLB 可用于确定信号参数受噪声的影响程度，估计光电成像器件的定位误差，研究定位方法的无偏性、有

效性。当一个估计方法达到器件的 CRLB 时，可认为其是最小方差无偏估计器 (MVUE)。成像器件的定位误差方面，Winick 最先利用 CRLB 理论推出了 CCD 类成像器件一维和二维高斯点源在泊松噪声情况下的 CRLB 指标，但未给出高斯噪声下的 CRLB(Winick K. A. 1986)。Vyas、Thomas 等仿真研究了 IWCOG 算法，但未研究该算法是否具备 CRLB 特性。由于数值仿真手段难以确定算法是否具备 CRLB 特性，可知如下问题仍需要研究：(1)在理想成像模型下，含有噪声的光电器件用于定位能做到多高精度；(2)不同的噪声模型，其光电器件的定位误差下限是否一致；(3)是否存在定位方法能够实现光电器件的定位极限；(4)常规定位技术如质心法运算效率高、最小二乘高斯拟合算法定位精度最好，二者能否用于高精度定位场合，满足器件定位极限实时性算法应该是什么。

本节将通过理论研究和仿真验证回答上述问题。首先通过星点信号建模研究光电成像器件在泊松和高斯噪声下的定位精度即 CRLB 指标，其次，重点研究迭代加权质心法、最小二乘高斯拟合法、质心法的定位误差，将其与器件的 CRLB 指标进行对比，从理论层次上揭示了三种定位方法的优缺点，并通过仿真验证这一结论(张俊等 2015，光学学报第 8 期)。

## 4.3.1 光电器件的 CRLB 特性

### 4.3.1.1 CRLB 和成像敏感器的点源信号模型

由第 2.1 节可知，若样本 PDF $p(s;\theta)$ 满足下式：

$$\mathrm{E}\left[\frac{\partial}{\partial\theta}\ln p(s;\theta)\right]=0 \tag{4.39}$$

则任何一个无偏估计器满足：

$$\mathrm{var}(\hat{\theta}) \geq -1/I(\theta), \tag{4.40}$$

其中，$I(\theta)=\mathrm{E}\left[\dfrac{\partial^2}{\partial\theta^2}\ln p(s;\theta)\right]=-\mathrm{E}\left[\left(\dfrac{\partial}{\partial\theta}\ln p(s;\theta)\right)^2\right].$

公式(4.39)-(4.40)中的 $s$ 为信号采样点，$\theta$ 为样本待估计参数，$p(s;\theta)$ 为样本估计的似然函数，$I(\theta)$ 为 Fisher 信息量，表示样本的集中程度，$\mathrm{var}(\hat{\theta})$ 为 $\hat{\theta}$ 方差 (variance)。

公式(4.40)表明不论采用何种参数估计方法，由于噪声因素的影响，样本某参数的估计方差不小于 $1/I(\theta)$，$1/I(\theta) \triangleq CRLB(\theta)$。当一种估计法所得参数 $\theta$ 的最小方差满足 $\text{var}(\hat{\theta}) = 1/I(\theta)$ 时，该估计方法对于参数 $\theta$ 即为最小方法无偏估计(MVUE)方法。这是本文对 IWCOG 算法和 GLSF 算法最优性的判定依据，也是 COG 算法不具备高精度的根本原因。

星敏感器采用 CCD、APS CMOS 等光电成像器件作为图像传感器，存在的噪声类型和来源是多种多样的，器件的定位误差下限与器件参数的关系只能通过对星点信号建模得到。

文献中将光电成像敏感器的点源信号描述为理想二维高斯模型：

$$S(x, y, x_c, y_c) = (2\pi\sigma_s^2)^{-1} \exp\left[\frac{-(x - x_c)^2}{2\sigma_s^2}\right] \exp\left[\frac{-(y - y_c)^2}{2\sigma_s^2}\right] \tag{4.41}$$

由此可得相应的一维模型为：

$$S(x, x_c) = (2\pi\sigma_s^2)^{-1/2} \exp\left[\frac{-(x - x_c)^2}{2\sigma_s^2}\right], S(y, y_c) = (2\pi\sigma_s^2)^{-1/2} \exp\left[\frac{-(y - y_c)^2}{2\sigma_s^2}\right] \tag{4.42}$$

为简化分析，采用一维模型，则第 $i$ 个像元产生的理想形貌为：

$$g_i(x_c) = I_0 \int_{x_i - \Delta x/2}^{x_i + \Delta x/2} S(x, x_c) dx \tag{4.43}$$

其中，$x_c, y_c$ 为点源真实坐标；$\sigma_s$ 为点源的尺度，$\Delta x$ 为像元尺度，$I_0$ 为信号能量的归一化系数。光电成像器件的噪声主要分泊松和高斯噪声两类，分别对像元采样进行噪声建模，分析成像器件的 CRLB 特性。

### 4.3.1.2 泊松噪声下的 CRLB 指标

像元信号模型满足：

$$s_i = \lambda_i + e_i, \lambda_i = I_0 g_i(x_c) \tag{4.44}$$

公式(4.44)中 $e_i$ 为泊松噪声，满足：

$$\text{E}(e_i) = \lambda_n, \text{E}(e_i e_j) = \lambda_n \delta_{ij}, \text{E}(s_i) = \lambda_i + \lambda_n, \text{var}(s_i) = \lambda_i + \lambda_n \tag{4.45}$$

成像单元能量信号满足泊松分布，其似然函数为：

$$\ln p(s; x_c) = \ln \prod_i p(s_i; x_c) = \sum_i [-\lambda_i - \lambda_n] + \sum_i s_i \ln[\lambda_i + \lambda_n] - \sum_i \ln s_i! \tag{4.46}$$

由此可得，

$$\Gamma = \frac{\partial \ln p(s; x_c)}{\partial x_c} = \sum_i \frac{I_0 s_i g'(x_c)}{I_0 g(x_c) + \lambda_n} \tag{4.47}$$

由公式(4.39)和(4.47)可得，

$$E[\Gamma^2] = \sum_i [I_0 g_i'(x_c)]^2 / (I_0 g_i(x_c) + \lambda_n) \tag{4.48}$$

由公式(2.17)-(2.18)可得，像元信噪比 $S/N$ 为：

$$R^2 = \sum_i I_0^2 g_i^2(x_c) / \mathrm{var}(s_i) \approx \begin{cases} I_0^2 / (2\sqrt{\pi} \lambda_n \sigma_s) & \lambda_i \ll \lambda_n \\ I_0 & \lambda_i \gg \lambda_n \end{cases} \tag{4.49}$$

由公式(4.40)和(4.48)化简可得 Fisher 量 $I(x_c)$ 为：

$$I(x_c) \approx \begin{cases} I_0^2 / 4\sqrt{\pi} \lambda_n \sigma_s^3 & \lambda_i \ll \lambda_n \\ I_0 / \sigma_s^2 & \lambda_i \gg \lambda_n \end{cases}, \tag{4.50}$$

由公式(4.50)，可得定位误差为：

$$\sigma_x = \begin{cases} \sqrt{2} \sigma_s / R & \lambda_i \ll \lambda_n \\ \sigma_s / R & \lambda_i \gg \lambda_n \end{cases} \tag{4.51}$$

由公式(4.50)-(4.51)可知，存在泊松噪声时，定位误差下限与信噪比成反比，与点源尺度成正比；由公式(4.51)可知，定位误差大小与泊松噪声背景项是否占主导地位有关，相差一个系数 1.414。

### 4.3.1.3 高斯噪声下的 CRLB 指标

像元信号模型同公式(4.44)，$e_i$ 为高斯噪声，满足：

$$E(e_i) = 0, E(e_i e_j) = \sigma_n^2 \delta_{ij}, E(s_i) = \lambda_i, \mathrm{var}(s_i) = \lambda_i + \sigma_n^2 \tag{4.52}$$

成像单元能量信号满足高斯分布，其似然函数为：

$$\ln p(s; x_c) = \ln \prod_i p(s_i; x_c) = -\sum_i \ln \sqrt{2\pi} \sigma_n - \sum_i [s_i - \lambda_i]^2 / 2\sigma_n^2 \tag{4.53}$$

由此可得，

$$\Gamma = \frac{\partial \ln p(s; x_c)}{\partial x_c} = \sum_i \frac{I_0}{\sigma_n^2} (s_i - I_0 g_i(x_c)) g_i'(x_c) \tag{4.54}$$

由公式(4.39) 和(4.54) 可得 Fisher 量 $I(x_c)$ 为：

$$I(x_c) = \sum_i I_0^2 (g_i'(x_c))^2 / \sigma_n^2 \approx I_0^2 / (4\sqrt{\pi} \sigma_n^2 \sigma_s^3) \tag{4.55}$$

又信噪比 $S/N$ 定义同公式(4.49)，可得，

$$R^2 = I_0^2 / 2\sqrt{\pi}\sigma_n^2\sigma_s \qquad (4.56)$$

可以验证式(4.54)满足式(4.39)的定义，结合公式(4.56)可得，

$$\sigma_x = \sqrt{2}\sigma_s / R \qquad (4.57)$$

由(4.56)-(4.57)可知，高斯噪声下定位误差下限与信噪比成反比，与点源尺度成正比。这与泊松噪声背景占比较大时的 CRLB 性能类似。

由上可知，要实现点源的定位精度 $\sigma_x$ 优于 0.02pixel，在 $\sigma_s = 1\mathrm{pixel}$ 情况下，只需 $R \geq 70.71$，考虑到离散情况，从后续仿真图 4.5 可以看出，只需 $R \geq 50$。CRLB 指标反映了影响定位误差的核心参数，这一指标能够用于器件选型。

## 4.3.2 几种定位技术的精度分析

质心法、最小二乘高斯拟合法是仿真最多研究最深入的两种常规定位方法，迭代加权质心算法是近几年在国外在 Shack-Hartmann 传感器波前定位正在研究的算法。三种方法的定位精度虽然可以通过大量仿真近似得到，理论上确定三种算法的定位误差并相互比较更有必要。

### 4.3.2.1 迭代加权质心算法精度

由前面可知，迭代加权质心算法满足：

$$\mu = \frac{\sum_i x_i s_i \omega_i(x_i, \mu)}{\sum_i s_i \omega_i(x_i, \mu)} \qquad (4.58)$$

其中，$\mu$ 为算法迭代的终值，$s_i$ 为像素灰度值，$\omega$ 为加权函数，化简后可得公式：

$$\sum_i (x_i - \mu) s_i \omega_i(x_i, \mu) = 0 \qquad (4.59)$$

可知公式(4.59)满足公式(4.39)的定义。

令

$$T(x_c) = \sum_i (x_i - x_c) I_0 g_i \omega_i(x_i, x_c) \qquad (4.60)$$

可得(van Trees H. L. 2004)，

$$T(x_c) = T(\mu) + \frac{\partial T}{\partial \mu}(x_c - \mu) + o(\Delta^2) \qquad (4.61)$$

以下分别就泊松噪声和高斯噪声计算公式(4.61)值，从而得到迭代加权质心算法的 $\text{var}(x_c)$。

(a) 泊松噪声

当 $\lambda_t \ll \lambda_n$ 时，化简公式(4.61)可得，

$$\text{var}(x_c) = \text{var}(T)/(\partial T/\partial \mu)^2 = \frac{4\sqrt{\pi}\sigma_s^3 \lambda_n(\lambda_n+1)}{I_0^2} = \frac{\lambda_n+1}{I(x_c)} > \frac{1}{I(x_c)} \tag{4.62}$$

当 $\lambda_t \gg \lambda_n$ 时，同理可得，

$$\text{var}(x_c) < \frac{4\sqrt{\pi}\sigma_s \lambda_n(\lambda_n+1)g_i(x_c)}{I(x_c)\lambda_n} \leq \frac{2\sqrt{2}(\lambda_n+1)}{I(x_c)} \tag{4.63}$$

(b) 高斯噪声

高斯噪声满足公式(4.52)定义，化简公式(4.61)可得，

$$\text{var}(x_c) = \text{var}(T)/(\partial T/\partial \mu)^2 = \frac{4\sqrt{\pi}\sigma_n^2 \sigma_s^3}{I_0^2} \equiv 1/I(x_c) \tag{4.64}$$

由公式(4.62)-(4.64)可知，当存在泊松噪声时，迭代加权质心算法定位误差方差接近误差极限，泊松噪声背景占主导时，该算法接近 MUVE 算法，其他情况下，定位误差有上限。公式(4.64)表明高斯噪声下该算法的定位方差等于 CRLB，是一种 MVUE 方法。以上分析可认为迭代加权质心算法是一种最优估计方法，这也从理论上证实了 Thomas 等的论断。

## 4.3.2.2 最小二乘高斯拟合算法精度

由文献可知，最小二乘高斯拟合算法满足下式：

$$x_c = \arg\min\left[\chi^2 = \sum_i \frac{[s_i - I_0 g_i(x_i,\mu)]^2}{\sigma_i^2}\right] \tag{4.65}$$

其中，$\arg\min$ 指求使 $\chi^2$ 取最小值时的 $\mu$ 值，$\sigma^2 = \sigma_i^2 = \sigma_j^2$ 为像素噪声。

公式(4.65)的解满足 $\partial\chi^2/\partial\mu = 0$，即，

$$\sum_i \frac{[s_i - I_0 g_i(x_i,\mu)]}{\sigma^2} I_0 g_i(x_i,\mu) \approx \sum_i \frac{(\mu-x_i)[s_i - I_0 g_i(x_i,\mu)]}{\sigma^2} \frac{I_0 g_i(x_i,\mu)}{\sigma_s^2} = 0 \tag{4.66}$$

令 $\omega_i(x_i, \mu) = g_i(x_i, \mu)$ ，化简可得，

$$x_c = \frac{\sum_i x_i[s_i - I_0 g_i(x_i, \mu)]\omega_i(x_i, \mu)}{\sum_i s_i\omega_i(x_i, \mu)} \tag{4.67}$$

上式即为差分迭代加权定位算法，化简公式(4.67)，当 $x_c$ 接近真值时，公式即为(4.58)。可见，最小二乘高斯拟合算法与迭代加权质心算法定位误差近似等价。更深入的研究表明由于离散采样，迭代加权质心算法定位误差稍大于最小二乘高斯拟合算法。

### 4.3.2.3 质心法定位精度

同比与式(4.59)，质心算法满足下式：

$$\sum_i (x_i - \mu)s_i = 0, \tag{4.68}$$

可知公式(4.68)满足公式(4.39)的定义。

令

$$T(x_c) = \sum_i (x_i - x_c)I_0 g_i \tag{4.69}$$

分别按 4.3.2.1 节计算方法，重新计算质心算法的定位方差。

(a) 泊松噪声

当 $\lambda_t \ll \lambda_n$ 时，化简公式(4.69)可得，

$$\text{var}(x_c) = \frac{M\lambda_n(\lambda_n + 1)}{I_0^2} \gg \frac{\lambda_n + 1}{8 \cdot I(x_c)} \gg \frac{1}{8 \cdot I(x_c)} \tag{4.70}$$

当 $\lambda_t \gg \lambda_n$ 时，同理可得，

$$\text{var}(x_c) = \frac{M\lambda_n(\lambda_n + 1)}{I_0^2} \gg \frac{\sqrt{\pi}\sigma_s^3\lambda_n(\lambda_n + 1)}{2I_0^2} = \frac{\sqrt{\pi}\sigma_s\lambda_n(\lambda_n + 1)}{2I(x_c)I_0} \tag{4.71}$$

(b) 高斯噪声

$$\text{var}(x_c) = M\sigma_n^2/I_0^2 \gg 1/8 \cdot I(x_c) \tag{4.72}$$

其中 $M = \sum_i (x_i - \mu)^2$.

从(4.70)-(4.72)可知，COG 法随噪声和像元数呈大幅变化，不具备 CRLB 性能，虽然计算简单，但并不是一种 MVUE 算法，这是其在成像器件点源定位中存在较大定位误差的主要原因。

需要指出的是，更精确的对比需采用离散形式公式(4.48)和(4.55)计算器件极限误差，然后研究上述三种定位算法误差能否达到 CRLB 性能，通常离散形式的 CRLB 要稍小于本节推导的公式。

由以上理论分析可知，在考虑噪声误差情况下，最小二乘高斯拟合算法定位误差同于迭代加权质心算法，低于质心算法，几乎不存在 S 误差，但在窗口效应和采样作用下，最小二乘高斯拟合算法模型可等效为无穷小像元采样，因而其定位精度应稍高于迭代加权质心算法，在星敏感器等应用场合该误差可忽略不计。因而在器件噪声、采样窗口、光学参数等综合条件下，三种算法的定位精度：GLSF 与 IWCOG 相同，远高于 COG。不难分析，反映器件硬件性能的 NEA 误差应与此类似。而 IWCOG 法作为高斯噪声下的 MUVE 方法，由于具有实时性，因而适用于更高精度的点源快速定位。

### 4.3.2.4 算法仿真比较与讨论

为保证仿真结果的可靠性、一般性，若不加说明，本节采用的仿真流程、方法、参数和参数取值参见表 4.1，除采用绝对误差验证三种算法定位误差外，同时采用 NEA 指标间接验证算法性能。

表 4.1 蒙特卡洛模拟仿真参数

| 参数 | 取值范围 | 参数 | 取值范围 |
|---|---|---|---|
| $I_0$ | 100 | $\Delta$ (pixel) | $10^{-3}$ |
| $\sigma_s$ (pixel) | [0.6,1.3] | $M$ (pixel) | 3、5、7 |
| $\mu$ (pixel) | $[-0.5, 0.5]$ | $R$ | [2,100]或无噪声 |

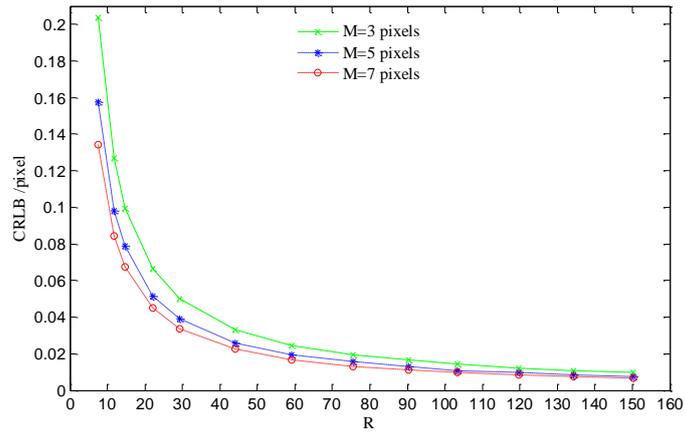

图 4.5 高斯噪声下的定位误差

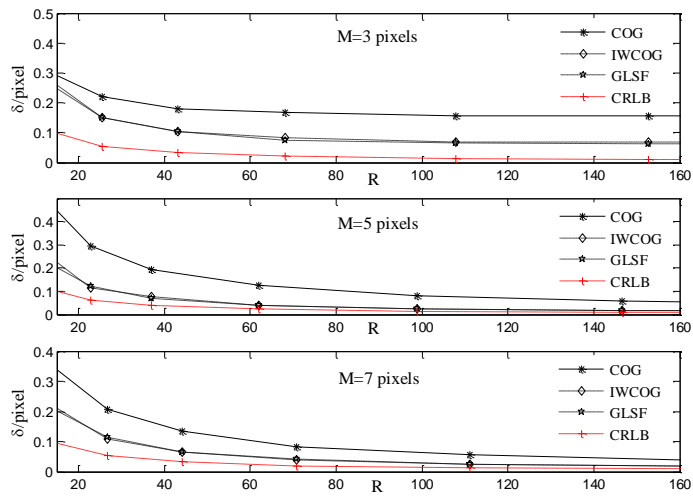

图 4.6 不同算法下，M 和 R 对点源定位误差的影响(随机相位、$\sigma = 1pixel$ )

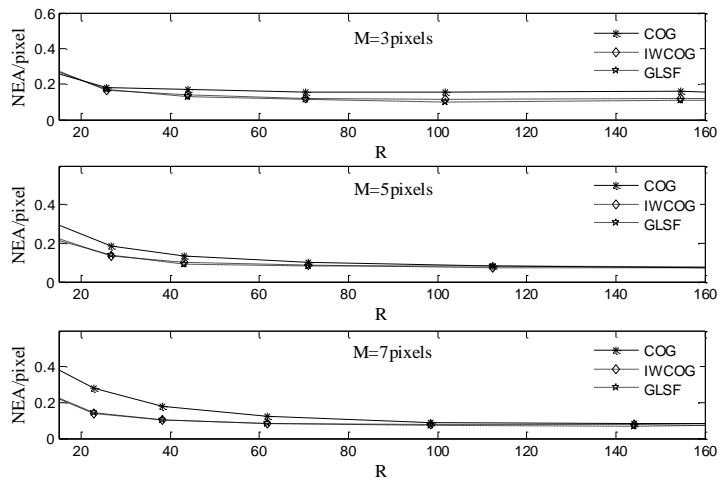

图 4.7 不同算法下，M 和 R 对点源 NEA 误差的影响(随机相位、$\sigma = 1pixel$ )

令 $\sigma = 1pixel$ 不变，用蒙特卡洛法随机给出 $\mu$ 模拟生成 10000 个星点，改变 $M$ 和

$R$ 值，按公式(4.55)计算离散情况下不同 $M$ 值的 CRLB(图中 CRLB 指

$\sqrt{CRLB(x_c)}$，下同)，所得结果参见图 4.5，同时计算质心法、迭代加权质心法和最小二乘高斯拟合算法计算绝对误差，并与图 4.5 $M = 7\,pixel$ 的 CRLB 进行对比，所得结果参见图 4.6。再次计算质心法、迭代加权质心法和最小二乘高斯拟合算法的 NEA 误差，所得结果参见图 4.7。

图 4.5 说明对于同一星点，窗口大小对 CRLB 指标影响较小，第 2 节推导的公式虽为理想情况，也是实际值的高度近似，便于物理测试中的误差分析。由图 4.6 可知，IWCOG 算法的定位误差等同于最小二乘高斯拟合算法，远低于质心法，接近器件的 CRLB 指标；图 4.7 说明 IWCOG 算法在 NEA 指标上也有优势。由于最小二乘高斯拟合算法是目前研究精度最高的算法，这侧面印证了 IWCOG 的 MVU 特性，由于 IWCOG 耗时仅比质心法高 3-5 倍，因而考虑误差和时间因素，IWCOG 算法是最好的算法。

### 4.3.2.5 NEA 试验特性对比

通过在地面采集多幅单个星点图像，由于平台的不稳定和大气视宁度影响，无论采用何种算法都无法确定每次星点图像真正的质心位置，可以用 NEA 指标判断多幅图像在不同噪声或其他影响因素下质心定位稳定度，对于高精度的定位算法，其 NEA 指标也会比低精度的定位算法稳定度高，这样可以间接说明算法的定位精度。本实验采集不同位置多幅单星点图像，然后分别用质心法、IWCOG 法和最小二乘高斯迭代拟合算法计算 NEA 指标，如图 4.8 所示。可以预知，质心法的 $\sigma_x$ 会远大于 IWCOG 算法、最小二乘高斯拟合算法。

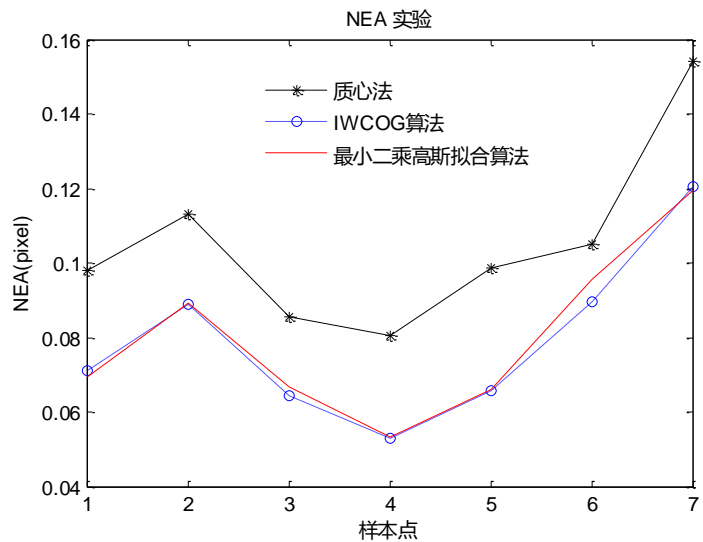

图 4.8 NEA 试验对比(数据源自武延鹏)

从图 4.8 可以看出，IWCOG 算法的 NEA 特性等同于最小二乘高斯迭代拟合算法，高于质心法，由于最小二乘高斯迭代拟合算法是精度最高的算法，试验同样印证了 IWCOG 的 MVU 特性。

### 4.3.3 S 误差校正解析

星敏感器为实现亚像素定位，采用了离焦技术，通常尺度 $\sigma_s \in [0.8, 2]$，所购 CCD 器件读出噪声每个像元多在 5-10e，武延鹏试验数据显示，信噪比 $R \in [10, 80]$，对两个参数取平均即 $\sigma_s = 1, R = 50$，根据公式 $\delta x_n = \dfrac{\sqrt{2}\sigma_s}{R}$ 可知，由噪声导致的器件理论误差为 $\delta x_n = 0.028 \, \text{pixel}$。由图 4.9 可看出，当 $M = 7 \, \text{pixels}$ 时质心法的定位误差 $\delta x = 0.14 \, \text{pixel}$，而一般 S 误差范围是 $\delta x_{algo} \in [0, 0.16]$。由公式 (3.9) 知，理想情况下质心法定位误差可表述为 $\delta x = \bar{\delta} x_{algo} + \bar{\delta} x_n$，令 $\bar{\delta} x_{algo} = 0.08$，可推出 $\bar{\delta} x_n = \sqrt{0.14^2 - 0.08^2} = 0.115 \, \text{pixel}$。由此可看出 $\bar{\delta} x_n = 4.11 \delta x_n$，可知质心法的抗噪声能力远不及 IWCOG。根据 Liebe 的公式 (3.5)，在使用质心法作星点定位情况下，为提高静态精度，最直接的方式是校正 S 误差(杨君等 2010)。有研究人员推出了 S 误差解析式，有些论文报道通过一些特殊方式，S 误差降低了 60%以上，这实质上校正了 S 误差。但是由上面分析可以看出，绝大多数情况下，$\bar{\delta} x_n > \delta x_{algo}$，即噪声产生的误差通常大于 S

误差，可以推知，由质心法计算出来的相位误差绝大部份来自噪声，绝非算法本身，而噪声是随机的，这种情况下将 S 误差降低 60%必定代价很高！但是同时看到 Anderson J. 在校正天文图像 S 误差中却取得了成功(Anderson J. et al. 2000)，这似乎与前述分析矛盾。实际上，哈勃获得的天文图像信噪比远大于 50，由噪声导致的理论误差低于 0.02pixel，而质心法 S 误差通常在 0.05pixel 以上，作者也没有用质心法做定位，而是通过多幅图像插值再迭代方式找到真正的质心，这个已经跟最小二乘高斯拟合法和 IWCOG 法非常相近，最后精度达到了 0.02pixel，接近理论精度。

S 误差本质上源于采样，所有算法都有(图 4.1-4.2)，但幅度和频率不一样，最小二乘高斯拟合法和 IWCOG 法的 S 误差幅值在 0.002pixel，而质心法幅值在 0.1pixel 左右。所以，与要求星敏感器定位精度相比，IWCOG 不需补偿 S 误差。

本节通过研究 CRLB 理论，并对算法比较，获知以下结论：IWCOG 算法在高斯噪声下是一种最优算法，在泊松噪声下是一种近优算法，能够广泛用于点源成像的定位工作。CRLB 理论和 NEA 试验论证了这一点，该算法 S 误差可忽略，而质心法不满足 CRLB 性能，S 误差较大。根据前述研究，证明 IWCOG 算法完全满足实时性、鲁棒性、精确度的星载要求。不仅如此，导出的 CRLB 定位指标显示了定位误差与信噪比的关系，可用于器件选型和试验分析处理。

## 4.4 IWCOG 多星定位误差特性

第 4.3 节 CRLB 理论表明，星点的定位精度与 PSF 尺度($\sigma$)成正比，与信噪比($R$)成反比。实际的星点尺度是用窗口像元数($M$)来表征的，由图 4.5 可知，不同的窗口其 CRLB 值略微不同。与此类同，不同的定位算法所得精度与窗口像元数、相位($\mu$)也有关连。因此，针对不同的定位算法，研究定位精度与信号尺度、信噪比、相位及窗口像元数关系很有必要。星敏感器采用多星定位机制，研究各定位算法在多星提取的统计规律很有意义。后面采用蒙特卡罗法随机生成多个星点，分析大量随机采样下的算法运行特点，分析 IWCOG 算法在多星点提取中的统计优势。

同 4.1.4 节仿真流程，分别用 COG 算法、高斯拟合法(Gaussian Curve Fitting，GCF)、IWCOG 算法计算质心，最后减去真实的质心计算样本的误差，系统分析

各参数对定位误差的影响。表 4.1 是全文仿真中使用的参数和取值范围。本节以 $\sigma$、$\mu$、M 及 $R$ 等为自变量，以 $\delta$ 或 $\varepsilon$ 为因变量进行蒙特卡罗仿真，若不加说明，模拟参数和取值均按表 4.1 执行。

### 4.4.1 R 和 M 对定位误差的影响

用蒙特卡罗法随机给出 $(\mu, \sigma)$ 模拟生成 10000 个星点灰度直方图，改变 R 和 M 值，仿真结果参见图 4.9。由图 4.9 可知，质心法(COG)提取误差呈现 3pixel>>5 pixel >7 pixel，IWCOG 算法提取误差呈现 3 pixel >>5 pixel≈7 pixel，高斯拟合算法(GCF)提取误差呈现 3 pixel <5 pixel <7 pixel(R≥20)。同等窗口下，IWCOG 算法提取误差最小，R<6 时 IWCOG 算法提取误差低于质心法 1.5~2 倍，在 R>6 时 IWCOG 算法提取误差低于质心法 2.5~3.5 倍；当 R=40、M=3 时，IWCOG 算法提取误差为 0.1pixel，质心法为 0.22pixel，M=5 时分别为 0.026、0.089pixel，M=7 时分别为 0.02、0.068pixel，由此可知增加计算窗口大小能明显降低这两种算法的提取误差，但当 M≥5 时这种趋势变缓，IWCOG 算法表现尤其明显，说明当达到 M=5 时，IWCOG 算法提取精度对窗口大小依赖程度不大。而高斯拟合法只在 R>60 且 M=3 时优于质心法，在信噪比较低和窗口太大时算法受到的干扰很大。

为了解样本分布特点，固定参数 M=5，分别对 R=10、40、100 再进行仿真，将图 4.9 中的样本分布展开，可得图 4.10-4.12。其图 4.10-4.12 中(a)~(c)显示样本点质心误差情况(只显示前 500 个样本点)，(d)~(f)显示对应的质心误差直方图，以下类同。

由于是蒙特卡罗采样，三种算法的样本误差分布较为均匀，图 4.10-4.12 中(a)~(c)充分显示了这一特点。同时，直方图分布(d)~(f)可以看出，质心法误差呈泊松形貌，IWCOG 算法呈 L 形貌，高斯拟合算法介于二者之间，可见 IWCOG 算法更利于实现更高精度。具体分析可知，随着信噪比的提高，三种算法提取的低误差样本数增加明显，尤以 IWCOG 算法增加最为显著，在 R=100 时，IWCOG 算法提取误差低于 0.02pixel 的样本数占比达 73%，而质心法不足 14%；R=40 时低于 0.02pixel 误差的样本占比分别为 44%、0.8%，低于 0.04pixel 误差的样本占比分别为 72%、1.9%，从概率上讲，IWCOG 算法在多星定位中具有很大优势，如 R=40 时使用 IWCOG 法提取 5 颗星，误差同时高于 0.04pixel 的概率为 $(1\text{-}72\%)^5=0.17\%$，而质心法为$(1\text{-}1.9\%)^5=90.9\%$，误差同时高于 0.02pixel 的概率为 $(1\text{-}44\%)^5=5.5\%$，质心法为$(1\text{-}0.8\%)^5=96.1\%$。

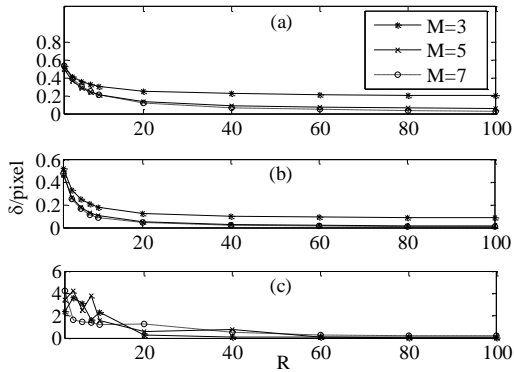

图 4.9 M 和 R 对星点定位精度的影响(随机 $\sigma$ 和 $\mu$)：(a)COG；(b)IWCOG；(c)GCF

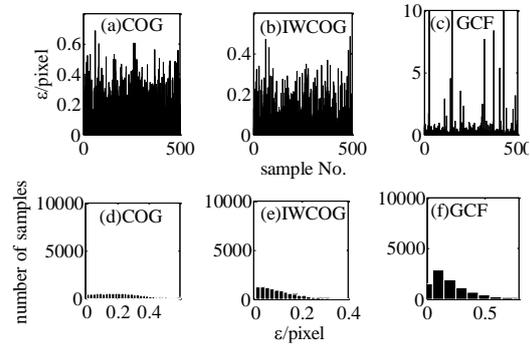

图 4.10 R 等于 10 对星点定位精度的影响 (随机 $\sigma$ 和 $\mu$)：(a)~(c)为样本点质心提取 误差，(d)~(f)为样本点质心误差直方图

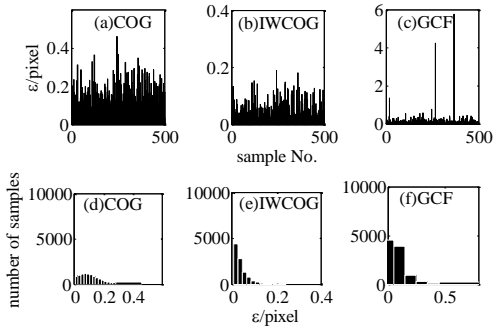

图 4.11 R 等于 40 对星点定位精度的影响(随 机 $\sigma$ 和 $\mu$)：(a)~(c)为样本点质心提取误差， (d)~(f)为样本点质心误差直方图

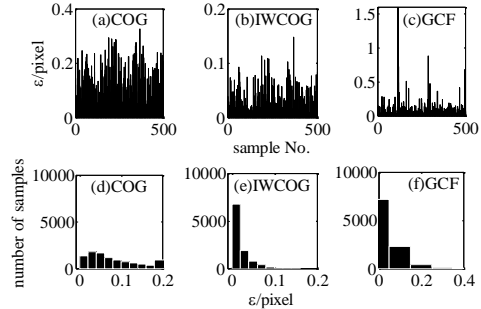

图 4.12 R 等于 100 对星点定位精度的影响 (随机 $\sigma$ 和 $\mu$)：(a)~(c)为样本点质心提取 误差，(d)~(f)为样本点质心误差直方图

## 4.4.2 $\sigma$ 和 M 对定位误差的影响

用蒙特卡罗法随机给出$(\mu, R)$模拟生成 10000 个星点灰度直方图，改变星 点 PSF 尺度 $\sigma$ 和 M，仿真结果参见图 4.13。

由图 4.13 可知，同等窗口下，IWCOG 算法提取误差最小，M=3、5、7时IWCOG 算法提取误差低于质心法2、3、2.5倍。由于星点信噪比一般高于10， M=5、7 时定位误差在0.031、0.028pixel，同比质心法误差高达0.104、0.087pixel。

为了解样本分布特点，分别就 $\sigma = 0.8$，M=3、5、7和 $\sigma = 1.2$，M=5进行仿 真所得参见图4.14-4.17。

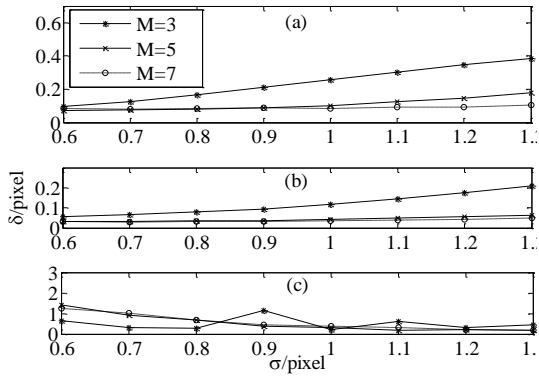

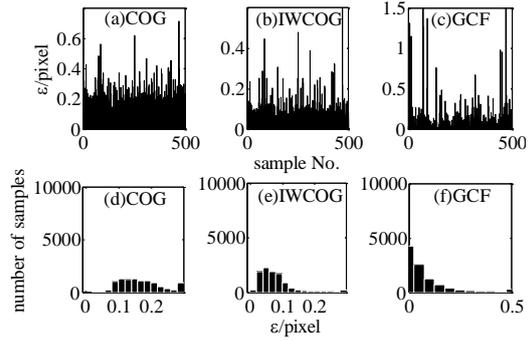

图 4.13 M 和 $\sigma$ 对星点定位精度的影响(随机 R 和 $\mu$)：(a)COG；(b)IWCOG；(c)GCF

图 4.14 M=3 和 $\sigma=0.8$ 对星点定位精度的影响(随机 R 和 $\mu$)：(a)~(c)为样本点质心提取误差，(d)~(f)为样本点质心误差直方图

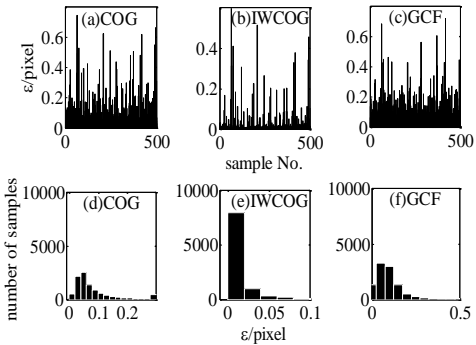

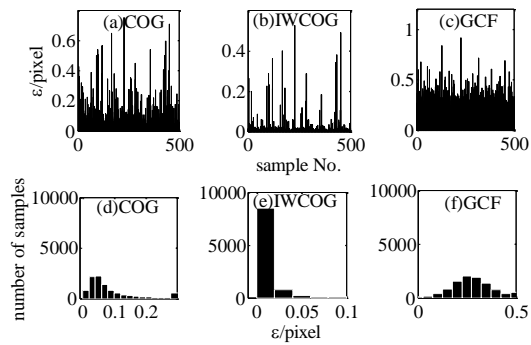

图 4.15 M=5 和 $\sigma=0.8$ 对星点定位精度的影响(随机 R 和 $\mu$)：(a)~(c)为样本点质心提取误差，(d)~(f)为样本点质心误差直方图

图 4.16 M=7 和 $\sigma=0.8$ 对星点定位精度的影响(随机 R 和 $\mu$)：(a)~(c)为样本点质心提取误差，(d)~(f)为样本点质心误差直方图

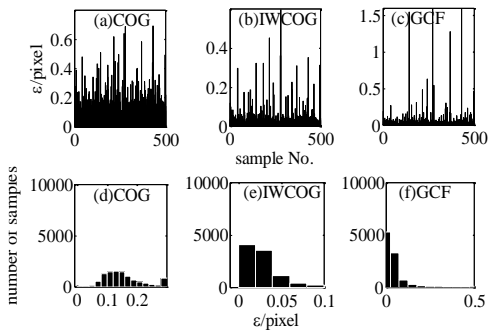

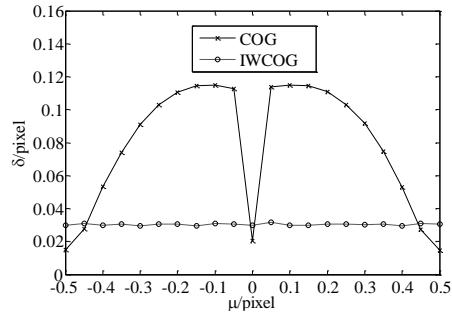

图 4.17 M=5 和 $\sigma=1.2$ 对星点定位精度的影响(随机 R 和 $\mu$)：(a)~(c)为样本点质心提取误差，(d)~(f)为样本点质心误差直方图

图 4.18 不同相位 $\mu$ 的相位误差(随机 R 和 $\sigma$)

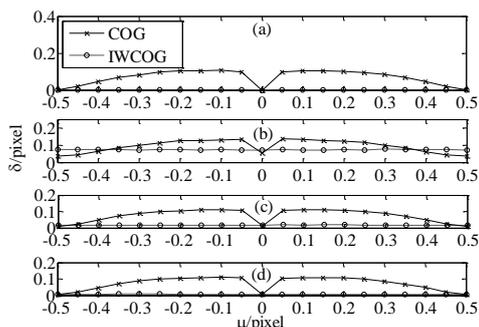

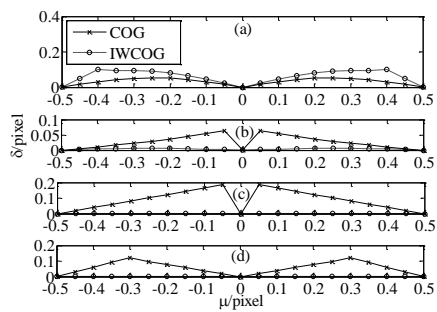

图 4.19 R 对相位误差的影响(随机 $\sigma$)：
(a)无噪声 (b) $R$=10;(c) $R$=40; (d)
$R$=100

图 4.20 $\sigma$ 对相位误差的影响(不含噪声)：
(a) $\sigma$=0.3; (b) $\sigma$=0.6; (c) $\sigma$=0.9; (d) $\sigma$=1.2

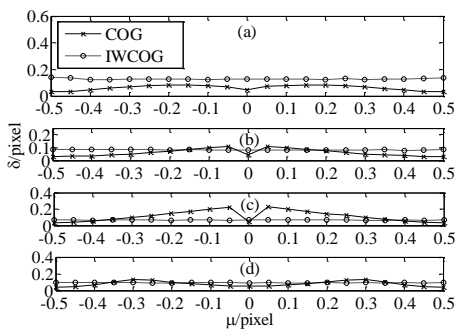

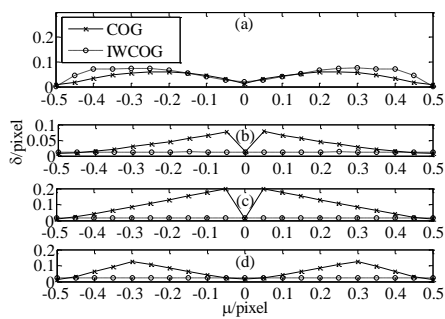

图 4.21 R=10 时 $\sigma$ 对相位误差的影
响：(a) $\sigma$=0.3; (b) $\sigma$=0.6; (c) $\sigma$=0.9;
(d) $\sigma$=1.2

图 4.22 R=40 时 $\sigma$ 对相位误差的影响：(a)
$\sigma$=0.3; (b) $\sigma$=0.6; (c) $\sigma$=0.9; (d) $\sigma$=1.2

从图4.14-4.17中(a)~(c)可以看出，蒙特卡罗采样下三种算法样本点提取误差分布较为均匀；误差直方图分布(d)~(f)显示，不论星点PSF大小和M数如何，质心法误差灰度直方图均呈泊松形貌，而IWCOG算法提取误差在M=3时近似泊松形貌，M=5、7时为典型的L形貌，高斯拟合算法提取误差在M=3像元时为L形貌，M=5、7时逐渐转为泊松形貌。由此可见，随着计算窗口数M的增加，IWCOG算法用于星点提取所得较大定位误差概率极小。M=5、7时质心法和IWCOG算法的低误差样本数增加明显，尤其以IWCOG算法增加最为显著，M=5时IWCOG算法提取误差低于0.02pixel的样本数占比达79%，而质心法不足0.5%；M=7时两种算法提取误差低于0.02pixel的样本占比分别为85%、0.85%。

另外，当 M 大于 $3\sigma$ 时，质心法与 IWCOG 算法提取误差总体对 PSF 尺度不太敏感，而当 M 与 $3\sigma$ 相近时(如 $\sigma$=0.9、M=3)，高斯拟合算法提取误差低于 0.05pixel 的样本数最为密集，这显示其对 PSF 大小不同的依赖特性。由于本节模拟的平均 R≈50，高斯拟合算法总体误差依然最高，从图 4.13 可以看出。

### 4.4.3  $\sigma$ 和 R 对定位误差的影响

用蒙特卡罗法随机给出 $(\sigma, R)$ 模拟生成10000个星点灰度直方图(包括无噪声情况)，改变相位 $\mu$，步长0.05pixel，模拟的结果参见图4.18。

用蒙特卡罗法随机给出 $\sigma$ 模拟生成10000个星点灰度直方图，分别对无噪声(运算一次即可)、R=10、40、100再进行仿真，质心法和IWCOG算法仿真结果参见图4.19。最后固定信噪比参数，分别就无噪声和 $R$=10 进行仿真(增加 $\sigma = 0.3$ 作对比)研究PSF尺度 $\sigma$ 对相位误差的影响，仿真结果参见图4.20-4.22。

由图 4.18 可知，在随机 R 和 PSF 尺度下，质心法提取的相位误差呈近似正弦规律，波动幅度为 0.1pixel，周期为 1/2pixel，IWCOG 算法提取的相位误差为 0.03pixel。从图 4.19 可知，高斯噪声大小对两种算法提取的相位误差形貌影响较小，对相位误差波动幅度影响较大，随着信噪比的降低，质心法提取的相位误差波动幅度明显拉窄。而从 4.20-4.22 可知，在相同 R 和算法下，不同的 PSF 尺度其算法提取的相位误差形貌有很大差异，而在相同的 PSF 尺度、不同的 R 下其算法提取相位误差形貌基本相同，综合图 4.18-4.22 可知，相位误差波动幅度主要受噪声影响，而相位误差形貌主要受 PSF 尺度 $\sigma$ 的影响。图 4.18 中质心法提取的相位误差波动幅度相当于图 4.19 中三种 R 曲线的线性叠加，而质心法的形貌相当于图 4.20-4.22 中同样 R 下形貌的叠加效果，这个结论同样适用于 IWCOG 算法。因而，如何降低噪声是降低相位误差的根本路径，研究 PSF 形貌则有助于做精确有效的补偿算法。由于 IWCOG 算法相位误差波动幅度与 PSF 相关性较小，因而更易补偿(由 4.3.3 知不必补偿)。

从图 4.18-4.22 同时可以看出，IWCOG 算法的相位误差同比最小；随着信噪比提高，两类算法误差呈下降趋势，在 R=40 时，IWCOG 算法提取的相位误差为 0.018pixel，质心法提取的相位误差波动幅度为 0.1pixel；无噪声时最大相位误差分别为 0.001、0.1pixel，这与 Nightingale 仿真结果基本相同，印证了 IWCOG 算法优于质心法的特点。

由图 4.20-4.22 可知， $\sigma \geq 0.6$ 时 IWCOG 算法提取的相位误差同比最低，呈常值分布， $\sigma = 0.3$ 时 IWCOG 算法提取误差稍高于质心法，主要因为 PSF 尺度较小时，星点所占像元数不到 2 个像元，灰度值基本由中心点占据，由于 IWCOG

算法加权函数的衰减性导致计算外围灰度值更低，只有相比质心法向外移才能达到收敛条件，模拟同时表明，PSF 尺度大于 5 时，星点采样数足够，IWCOG 算法同样不再有优势，这表明该算法的误差降低与 PSF 尺度关系很大，$\sigma$=0.6~1.2 时 IWCOG 算法与质心法相比，相位误差下降更快，因而更适合高精度星点提取。

本节通过对质心法、IWCOG 法、高斯拟合法蒙特卡罗仿真，研究发现由 IWCOG 算法提取的误差直方图呈现 L 分布，多星同时出现较大定位误差概率极小，质心法的误差直方图呈现泊松形貌分布，多星同时出现较高误差的概率极大；在 R=40、M=5 时，IWCOG 法提取误差低于 0.02pixel 的样本占比分别为 44%，低于 0.04pixel 的样本占 72%，远高于质心法的 0.8%和 1.9%，因而 IWCOG 算法适于星敏感器中的多星定位。通过研究 R、PSF 大小对精度的影响及与相位误差的关系，得出降低噪声、适当增加计算窗口数、估计 PSF 大小并做适当补偿算法是提高星点提取误差的有效方式，分析表明 IWCOG 算法更易做到这一点。本节对关键参数均采用蒙特卡罗采样达 10000 次，所得出的结论不失一般性、可靠性和实用性。

## 4.5 高采样 PSF 数据匹配算法理论及特性

前述定位算法中，无论是 IWCOG 类算法、质心法、最小二乘高斯拟合算法等均以星点形貌近似符合高斯对称分布为假设前提，实际星点受光学像差等因素影响，星点 PSF 多偏离高斯对称分布，由此导致的质心偏差已不可忽略，为实现更高精度定位，提出一种细分提取算法—高采样 PSF 数据匹配(郝云彩等 2016)，通过实验得到高采样星点 PSF 图像(像元尺度小)并与低采样星点 PSF 图像(像元尺度大)做匹配，当匹配达到最佳时，以高采样星点 PSF 图像质心作为低采样星点的真实质心(郝云彩等 2012)，这样做的好处是高采样星点 PSF 图像星点质心受噪声、采样窗口、算法等影响较小且包括更为精确的畸变信息，有利于后续使用最小二乘法等做更高精度的区域校正，但该算法形式如何，与 IWCOG 算法有何不同需要仔细分析。

本节从匹配角度出发，利用最小二乘导出具体的算法公式，同时将其与迭代加权算法、最小二乘高斯拟合算法做比较，研究算法收敛速度和收敛精度，揭示其本质内涵，并讨论其与其他两类算法的优劣特性。

### 4.5.1 算法公式和优化

依然采用欧拉 2 范数距离定义为高采样星点 PSF 图像与低采样星点 PSF 图像匹配程度：

$$d(s,g) = \sum_{i=1}^{n} [s_i - \sum_j g_{ij}(r_0, r)]^2 \tag{4.73}$$

其中 $i$ 为低采样单元格数，$j$ 为每个低采样单元格包括的高采样数，$s$、$g$ 分别为低采样、高采样灰度分布图。

当两采样图像匹配达到最佳时即为所求，用数学公式可描述为：

$$r_0^* = \arg\min d(s,g) = \arg\min \sum_{i=1}^{n} [s_i - \sum_j g_{ij}(r_0, r)]^2 \tag{4.74}$$

此时满足 $\partial d(s,g)/\partial r_0 = 0$，即：$\sum_{i=1}^{n} [s_i - \sum_j g_{ij}(r_0, r)]g'_{ij}(r_0, r) = 0$

将上式展开得，

$$\sum_i^n \sum_j [e_{ij} + g'_{ij}(r_0, r)(r - r_0)]g'_{ij}(r_0, r) = 0 \tag{4.75}$$

其中 $e_{ij}$ 包括高采样像元噪声和泰勒展开误差 $e_{ij,high}$、$\varepsilon_{ij,high}$，低采样像元噪声和泰勒展开误差 $e_{ij,low}$、$\varepsilon_{ij,low}$，高低采样像元灰度差分误差 $e_{ij,diff}$。

当匹配达到最佳时，$e_{ij}$ 只包括前四种，均为高斯白噪声，于是可得，

$$\sum_i^n \sum_j e_{ij} = \sum_i^n \sum_j (e_{ij,low} + e_{ij,high} + \varepsilon_{ij,high} + \varepsilon_{ij,low}) \approx 0 \tag{4.76}$$

即 $\sum_i^n \sum_j e_{ij}(r - r_0) = 0$，带入(4.75)，可得，

$$\sum_i^n \sum_j \{[e_{ij} + g'_{ij}(r_0, r)](r - r_0)\}g'_{ij}(r_0, r) = 0 \tag{4.79}$$

由 $g_{ij}(r_0, r) = \Delta_{ij} + g'_{ij}(r_0, r)$，$\Delta_{ij}$ 为截断误差，可得，

$$\sum_i^n \sum_j [g_{ij}(r_0, r) + e_{ij} - \Delta_{ij}](r - r_0)[g_{ij}(r_0, r) - \Delta_{ij}] = 0 \tag{4.78}$$

若 $e_{ij,diff} \neq 0$，则，

$$\sum_i^n \sum_j [g_{ij}(r_0, r) + e_{ij} - \Delta_{ij}](r - r_0)[g_{ij}(r_0, r) - \Delta_{ij}] = -\sum_i^n \sum_j e_{ij,diff}(r - r_0)[g_{ij}(r_0, r) - \Delta_{ij}] \tag{4.79}$$

对比公式(4.3)可知，当 $g(\cdot) = g_{ij}(r_0, r) + e_{ij} - \Delta_{ij}$，$\omega(\cdot) = g_{ij}(r_0, r) - \Delta_{ij} = g'_{ij}(r_0, r)$ 时，上式算法可以等效为：

$$f(r_0) = \sum_i^n \sum_j (r - r_0)g(\cdot)\omega(\cdot) = 0, i \geq 3, j \geq 1 \tag{4.80}$$

由前文可知，可以以差分快速算法计算中心点：

$$r_{k+1} = r_k + 2 \cdot \sum_{i=1}^{n} [s_i - \sum_j g_{ij}(r_0, r)] g'_{ij}(r_0, r) \Big/ \sum_i^n \sum_j^n g_{ij}(r_0, r) g'_{ij}(r_0, r) \qquad (4.81)$$

必须指出的是，这里所获的差分 IWCOG 方法最初并非按公式(4.67)推出，而是在做高采样 PSF 数据匹配理论时获得的，保留了郝云彩提出的差分思想。

## 4.5.2 算法精度和收敛速度

以下分高采样和低采样器件有无噪声，$g'_{ij}(r_0, r)$ 是否符合高斯分布分为以下几种情况讨论。设 $\Delta_g$ 为 $g'_{ij}(r_0, r)$ 不为高斯形貌时的模型误差，假设各噪声不相关，满足高斯分布，即，

$$\mathrm{E}(e_{ij}\Delta_{ij}) = 0, \mathrm{E}(\Delta_{ij}) = 0, \mathbf{D}(\Delta_{ij}) = \sigma^2_{\Delta_{ij}}, \mathrm{E}(e_{ij}\Delta_g) = 0, \mathrm{E}(\Delta_{ij}\Delta_g) = 0, \mathbf{D}(\Delta_g) = \sigma^2_{\Delta_g} \geq 0$$
$$\mathrm{E}(e_{ij}\Delta_{ij}) = 0, \mathrm{E}(\Delta_{ij}) = 0, \mathbf{D}(\Delta_{ij}) = \sigma^2_{\Delta_{ij}}, \mathrm{E}(e_{ij}\Delta_g) = 0, \mathrm{E}(\Delta_{ij}\Delta_g) = 0, \mathbf{D}(\Delta_g) = \sigma^2_{\Delta_g} \geq 0 \quad (4.82)$$

### 4.5.2.1 无噪声、高斯形貌

由题可知，$\sigma^2_{low} = \sigma^2_{high} = 0$，$\sigma^2_{\Delta_{ij}} \sim 0$，$g'_{ij}(r_0, r)$ 符合高斯分布。由公式(4.78)可得，本节算法：$g(\cdot) = g_{ij}(r_0, r) + e_{ij} - \Delta_{ij} = g_{ij}(r_0, r) - \Delta_{ij}$，$\omega(\cdot) = g'_{ij}(r_0, r)$，IWCOG 算法：$g(\cdot) = g_{ij}(r_0, r) + \Delta_{ij} + e_{ij,low} = g_{ij}(r_0, r) + \Delta_{ij}$，$\omega(\cdot) = g'_{ij}(r_0, r)$。可知本节算法达到最佳匹配，当 $e_{ij,diff} = 0$ 时，本节算法与 IWCOG 算法精度一致。由公式(4.79)可知，当 $e_{ij,diff} \neq 0$ 时，存在像元泰勒展开误差，精度下降。

### 4.5.2.2 有噪声、高斯形貌

由题可知，$\sigma^2_{low} > 0, \sigma^2_{high} > 0, \sigma^2_{\Delta_{ij}} \sim 0, \sigma^2_{\Delta_{ij}} << \sigma^2_{low}, \sigma^2_{high}$，$g'_{ij}(r_0, r)$ 符合高斯分布。由公式(4.78)可得，本节算法：$g(\cdot) = g_{ij}(r_0, r) + e_{ij} - \Delta_{ij}$，$\omega(\cdot) = g'_{ij}(r_0, r)$，IWCOG 算法：$g(\cdot) = g_{ij}(r_0, r) + \Delta_{ij} + e_{ij,low}$，$\omega(\cdot) = g'_{ij}(r_0, r)$。可知当本节算法达到最佳匹配，即 $e_{ij,diff} = 0$，本节算法还受 $e_{ij,high}$ 影响。由公式(4.79)可知，$e_{ij,diff} \neq 0$，主要由高、低采样像元噪声误差导致，由此可知定位存偏差，精度较 IWCOG 算法低。

### 4.5.2.3 无噪声、非高斯形貌

由题可知，$\sigma_{low}^2=\sigma_{high}^2=0$，$\sigma_{\Delta_{ij}}^2\sim0$，$g_{ij}'(r_0,r)$ 非高斯分布，$G(r_0,r)$ 高斯分布。由公式(4.78)可得，本节算法：$g(\cdot)=g_{ij}(r_0,r)+e_{ij}-\Delta_{ij}=g_{ij}(r_0,r)-\Delta_{ij}$，$\omega(\cdot)=g_{ij}'(r_0,r)$ $=G(r_0,r)+\Delta_g$，IWCOG 算法：$g(\cdot)=g_{ij}(r_0,r)+\Delta_{ij}+e_{ij,low}=g_{ij}(r_0,r)+\Delta_{ij}$，$\omega(\cdot)=$ $G(r_0,r)$。可知当本节算法达到最佳匹配，即 $e_{ij,diff}=0$，本节算法除收像元泰勒展开误差影响外，还受非高斯模型误差影响。由公式(4.79)可知，$e_{ij,diff}\neq0$，主要由像元泰勒展开误差和两项非高斯模型误差共同导致，由此可知定位存偏差，精度低于 IWCOG 算法。

### 4.5.2.4 有噪声、非高斯形貌

由题可知，$\sigma_{low}^2>0,\sigma_{high}^2>0,\sigma_{\Delta_{ij}}^2\sim0,\sigma_{\Delta_{ij}}^2<<\sigma_{low}^2,\sigma_{high}^2$，$g_{ij}'(r_0,r)$ 非高斯分布，$G(r_0,r)$ 高斯分布。由公式(4.78)可得，本节算法：$g(\cdot)=g_{ij}(r_0,r)+e_{ij}-\Delta_{ij}$，$\omega(\cdot)=$ $g_{ij}'(r_0,r)=G(r_0,r)+\Delta_g$，IWCOG 算法：$g(\cdot)=g_{ij}(r_0,r)+\Delta_{ij}+e_{ij,low}$，$\omega(\cdot)=G(r_0,r)$。可知当本节算法达到最佳匹配，即 $e_{ij,diff}=0$，本节算法除受高、低采样像元噪声误差影响外，还受非高斯模型误差影响。由公式(4.79)可知，$e_{ij,diff}\neq0$，主要由一项高低采样像元噪声和两项非高斯模型误差共同导致，由此可知定位存偏差，精度低于 IWCOG 算法。

以上分析可知，由于高采样噪声难以忽略，即使 $e_{ij,diff}=0$，本节算法最终定位精度低于 IWCOG 算法，考虑到实际情况，$e_{ij,diff}\neq0$，可知本节算法导致的定位偏差较 IWCOG 算法大。最小二乘高斯拟合算法不同于本节算法，其优势在于：(1)仅考虑噪声误差情况下，高斯拟合模型不包括高采样噪声误差，因而其精度高于本节算法，与 IWCOG 法一致；(2)仅考虑窗口效应下，高斯拟合模型可以等效为无穷小像元采样，因而其精度高于 IWCOG 算法和本节算法，由 4.1.4 节知，其定位误差在 0.002pixel 量级；(3)仅考虑信号模型下，最小二乘采用高斯函数拟合与 IWCOG 采用高斯加权函数一致，因为精度一样，而本节算法则引入由非高

斯信号模型导致的算法偏差。最终的星点定位误差包括畸变误差和算法偏差两项，从前述定义看，由于算法偏差具有稳定性，因而 3 种算法在表述信号畸变误差上属于同一层次，区别仅在于不同的算法计算所得的畸变误差并不相同，但并不影响后续使用最小二乘法区域畸变校正。其劣势在于：最小二乘高斯拟合采用拟合加迭代法，单步浮点运算次数最多；本节算法由于需动态将高 PSF 数据转换为低采样数据，单步浮点运算次数较多；而 IWCOG 算法运算窗口为±3$\sigma$，单步浮点运算次数有限。由上述分析可知，最小二乘高斯拟合算法速度最慢，本节算法次之，IWCOG 最快。在低采样噪声、光学参数等同等条件下，考虑到高采样噪声，三种算法的精度为：最小二乘高斯拟合算法≈IWCOG>高采样 PSF 数据匹配法，不难分析，其 NEA 指标亦是如此。

### 4.5.3 高采样 PSF 数据匹配法 CRLB 特性

同样可以用 CRLB 理论估计这个方法的定位精度。对于背景和读出噪声占主导的信号模型，高斯信号定位的 CRLB 界为：

$$\sigma_x = \frac{\sqrt{2}\sigma_s}{R}, R = \sqrt{\sum_{i,j} \frac{(I_i g_i)^2}{\mathrm{var}(s_i)}}, s_i = I_i g_i + e_{ij,low} \tag{4.83}$$

由公式(4.83)可知，不计 $\Delta_{ij} \neq 0, \Delta_g \neq 0$ 引入的偏差，本节算法精度为：

$$\sigma_{x'} = \frac{\sqrt{2}\sigma_s}{R'}, R' = \sqrt{\sum_{i,j} \frac{(I_i g_i)^2}{\mathrm{var}(I_i g_i + e_{ij} + \Delta_{ij})}} \tag{4.84}$$

由上可知，高采样 PSF 数据匹配法在高斯噪声下不具备 MVU 性质，主要原因是高采样和泰勒展开余项引入的误差所致。但是该技术在获取 PSF 畸变导致的质心偏移上有很大优势(郝云彩等 2015)，仍然是很好的方法。

### 4.6 星敏感器静态精度改进方法研究

4.3 节证明了在星点形貌近似符合高斯分布情况下，IWCOG 相关算法能够获得或接近探测器的理论精度，基本上解决了高斯星点的甚高精度定位问题。由公式(4.51)和(4.57)可知，探测器的理论精度与星点尺度成正比，与信噪比成反比，这个特性为提高星敏感器静态精度提供了解决方向。

### 4.6.1 基本策略

静态理想情况下，由公式(3.9)知定位误差公式可表述为：

$$\hat{x} = x_{true} + e_{algo} + e_n \tag{4.85}$$

其中，$x_{true}$ 为星点真实位置，包括器件噪声($e_n$)、算法和采样误差($e_{algo}$)等误差源带来的位置误差。

由 4.1 节可知，在采用无偏算法情况下，$e_{algo} \sim 0.002\,\text{pixel}$，可以忽略。在存在噪声情况并采用无偏算法下，所产生的定位误差最小为 CRLB 界(以背景占主导为例)，可表述为：

$$\varepsilon = \frac{\sqrt{2}\sigma_s}{R}, R^2 \approx I_0^2 \big/ (2\sqrt{\pi}\sigma_n'^2\sigma_s) \tag{4.86}$$

其中 $\sigma_s$ 为信号的尺度，$R$ 为信噪比，$\sigma_n'$ 为本底噪声，$\varepsilon$ 为 $e_n$ 产生的误差量。

由文献可知，对于 CCD 类性探测器，某星等 $m$ 在 $T$ 时间内产生的光强度可表述为：

$$I_0 = F_0 T e^{-0.4(m-m_0)} \tag{4.87}$$

由此可知，要提高点源的定位精度 $\varepsilon$，应设法提高信噪比，降低星点尺度。提高信噪比的主要方式是增加积分时间、降低噪声。降低星点尺度的技术参见第 4.6.2 节。为进一步提高姿态输出精度，星敏感器采用多星定位技术，由于星等不一致，由公式(4.86)和(4.87)可知，每个星点在相同积分时间内定位精度不同，应考虑多星定位对平均星点误差的影响。

采用离焦技术的星点大小可表述为：

$$\sigma_s = \beta l, l = \lambda f / \pi D = c_0 d, d = \frac{A_{FOV}}{N_{pixel}} \tag{4.88}$$

其中，$l$ 为艾里斑大小，$D$ 为透镜的通光孔径，$f$ 为焦距，$\lambda$ 为星点光谱中心波长，$\beta$ 为实际星点尺寸与艾里斑比值。$A_{FOV}$ 为全视场角，$N_{pixel}$ 为探测器阵列长度(或者宽度)像元个数。

由公式(4.86)和(4.88)，Liebe 给出单星定位精度可重新表述为：

$$\varepsilon = d\delta \tag{4.89}$$

其中，$\delta$ 为星点质心提取误差。

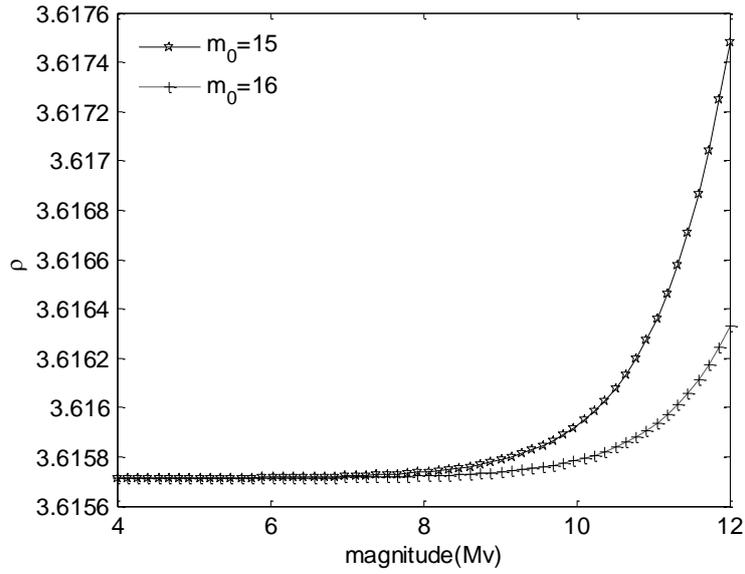

图 4.23 多星对质心提取误差的影响

在积分时间较短情况下，星敏感器背景($B$)和读出噪声贡献最大，假设不考虑读出噪声，则 $\sigma_n'^2 \approx BT$。由公式(4.86)-(4.89)可得，无星场畸变星敏感器的静态质心提取精度为：

$$\delta = \frac{2\pi^{1/4}\sigma_s^{1/2}c_0 B^{1/2}e^{0.4(m-m_0)}}{F_0 T^{1/2}} \triangleq \kappa e^{0.4(m-m_0)}, \kappa = \frac{2\pi^{1/4}\sigma_s^{1/2}c_0 B^{1/2}}{F_0 T^{1/2}} \tag{4.90}$$

由文献可知，任意天区星等 $m$ 星数近似服从指数分布，可表述为：

$$N_m = 6.58e^{1.08m}\frac{1-\cos(A_{FOV}/2)}{2} \tag{4.91}$$

由此可得，在多星参与姿态确定下，平均质心提取精度满足：

$$\bar{\delta} = \frac{\int_{m_1}^{m_2}\delta N_m dm}{\int_{m_1}^{m_2}N_m dm} = \kappa\rho \tag{4.92}$$

可知，

$$\rho = 0.73 \bullet \frac{e^{1.48(m_2-m_0)}-e^{1.48(m_1-m_0)}}{e^{1.08(m_2-m_0)}-e^{1.08(m_1-m_0)}} \tag{4.93}$$

为评估多星参与对 $\rho$ 的影响，选取 $m_1=4$，$m_2=12$，$m_0=15$或16，所得曲线参见图 4.23，由图 4.23 可见 $\rho$ 变化不大。由系数 $\kappa$ 可知，增加暗星数量并不能降低平均星点提取误差，改进 $\delta$ 主要方式是增加有效积分时间，降低星点所占像元数。增加有效积分时间的方法参见第五章 5.2 节。而降低星点所占像元数的则

存在弊端，即前文所述，在像元过少或亚采样状态下，定位误差会有轻微增加，这是设计应当考虑的，所以，最合适的像元数是适当即可，如5×5窗口尺度。

在多星参与姿态确定下，最终的多星定位精度为：

$$\varsigma = \frac{\overline{\delta}d}{\sqrt{N_{star}}} \propto \frac{\overline{c}_0 d}{\sqrt{N_{star}T}} \propto \frac{\beta \overline{\lambda} f}{\sqrt{N_{star}TD}} \tag{4.94}$$

所以，减小像元等效角 $\sigma_s$，增加参与姿态确定的恒星数量能够提高姿态确定精度。如何减小 $\sigma_s$，参见下节。

## 4.6.2 多缝干涉技术

首先研究衍射极限光学器件的定位原理，此时 $\sigma'_s = \beta' \frac{\lambda f}{D}$，$\beta' < \infty$。圆孔光学衍射信号为艾里斑斑函数：$\beta' = \frac{1.22}{3}$，$\sigma'_s = 0.41 \frac{\lambda f}{D}$，方孔光学衍射信号为 $\sin c^2$ 函数：$\beta' = \frac{1}{3}$，$\sigma'_s = \frac{\lambda f}{3D}$。由 4.7 节可知，若将方孔能量等效为圆形艾里斑，则 $\beta' = \frac{\sqrt{\pi}}{2} \frac{1}{3} \approx 0.3$，即 $\sigma'_s = 0.3 \frac{\lambda f}{D}$，则在星敏感器系统其他参数不变的情况下，可以提高点源图像的定位精度 1-0.3/0.41=27%，这个结论与自然基金中推想结果基本相同(郝云彩等 2012)。同样，对于采用离焦技术的星敏感器，$\beta = \xi \beta', \xi > 1$，$\xi$ 不变情况下，上述结论仍然成立，所以 $\sigma_s$ 难以降低。以上分析表明改变孔径形貌并不能大幅提高恒星敏感器定位精度，而减小 $\xi$ 是最有效率的方式，这也是为什么望远镜等衍射极限类器件能够达到较高定位精度的根本原因。

目前，国外已有小型干涉式星敏感器产品(IST)，其产品 X/Y 轴 $1\sigma$ 精度已到 $1''$，重量低于 200 克，最大工作速度 $5''/s$，这是传统光学镜头较难做到的，其蕴含新的理论和技术支撑。基于前述研究，本节提出另外一种基于衍射式光学的恒星敏感器，从成像系统定位角度提出多缝干涉式恒星敏感器突破亚角秒精度的可能性，探索其实现的技术途径。

根据光学衍射理论，点源矩缝干涉成像形成的夫琅和费衍射斑可表述为：

$$I(P) = I_0 \left( \frac{\sin \varphi}{\varphi} \right)^2 \left( \frac{\sin \gamma}{\gamma} \right)^2, \varphi = \frac{\pi a r_a}{\lambda f}, \gamma = \frac{\pi b r_b}{\lambda f} \qquad (4.95)$$

点源多缝干涉成像形成的多缝夫琅和费衍射斑可表述为：

$$I(P) = I_0 \left( \frac{\sin \varphi'}{\varphi'} \right)^2 \left( \frac{\sin N\gamma'}{\gamma'} \right)^2, \varphi' = \frac{\pi a r}{\lambda f}, \gamma' = \frac{\pi d_a r}{\lambda f}, d_a > a \qquad (4.96)$$

由上可知，点源矩缝阵列干涉成像形成的矩缝夫琅和费衍射斑可表述为：

$$I(P) = I_0 \left( \frac{\sin \varphi}{\varphi} \right)^2 \left( \frac{\sin \gamma}{\gamma} \right)^2 \left( \frac{\sin N_a \varphi'}{\varphi'} \right)^2 \left( \frac{\sin N_b \gamma'}{\gamma'} \right)^2, \varphi = \frac{\pi a r_a}{\lambda f}, \gamma = \frac{\pi b r_b}{\lambda f},$$
$$\varphi' = \frac{\pi d_a r_a}{\lambda f}, \gamma' = \frac{\pi d_b r_b}{\lambda f}, d_a > a, d_b > b \qquad (4.97)$$

N 个缝衍射结果是能量由单缝的主极转移到多缝的主极强上，半宽是单缝的 $\frac{1}{N}$，能量为单缝的 $N^2$，可知点源矩缝阵列衍射主极强半宽是单矩缝的 $\frac{1}{N_a N_b}$，能量为 $N_a^2 N_b^2$，即，

$$\sigma'_s = \frac{1}{N_a N_b} \sigma_s, I' = N_a^2 N_b^2 I_{sl} \approx N_a^2 N_b^2 \frac{1}{N_a N_b} \frac{a}{d_a} \frac{b}{d_b} \frac{1}{N_a N_b} I = \kappa_a \kappa_b I \qquad (4.98)$$

其中 $\kappa_a, \kappa_b$ 为横向和纵向的遮光比，$\kappa_t = \frac{t}{d_t}, t = a, b$。

可见，综合效果是，来自矩缝阵列的所有能量最终集中在主极强上，而尺度缩放为原来的 $\frac{1}{N_a N_b}$，N 缝夫琅和费衍射模拟效果参见图 4.24 和 4.25。由图所示，N=20 尺度为 N=10 的 0.5 倍，为 N=5 的 0.25 倍。由 4.7 节可知，多缝干涉式星敏感器星象分布可等效为高斯信号，这样可以使用原有的定位算法和精度估计公式。根据公式 $\varepsilon = \frac{\sqrt{2}\sigma_s}{R}$，可得 $\varepsilon = N_a \kappa_a \cdot N_b \kappa_b \varepsilon'$。在信噪比难以提升的情况下，可以通过衍射方式达到更高精度的定位。

在系统设计其他参数不变情况下，对于理想光学系统，多缝干涉式与传统光学信号尺度满足：$\sigma'_s = \frac{1}{N_a N_b} \sigma_s$，如果原有系统不变，由于尺度大幅缩小，星点

在焦平面上成像在单位像素内，无法形成一定尺度高斯斑点。所以，需要在原有离焦基础上，将该信号尺度再度放大 $N_a N_b$ 倍，这可再利用离焦方法得到，离焦距离大约为原来离焦量的 $N_a N_b$ 倍。采用离焦方法有缺陷，衍射式星敏感器光学系统应需要新的设计。如衍射掩模的形貌，安装的位置是在镜头前、镜头后校正单元前、还是 CCD 前，怎么分布等均需详细研究。另外，本节尚未对不同光谱做相应讨论，这也是遗留问题。

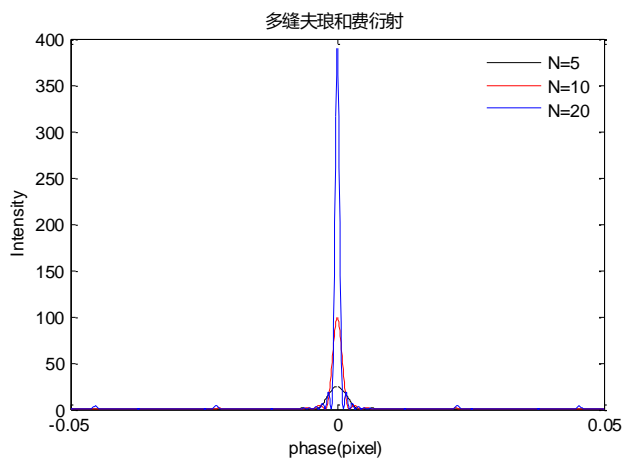

图 4.24 N 缝夫琅和费衍射比较(曲线)

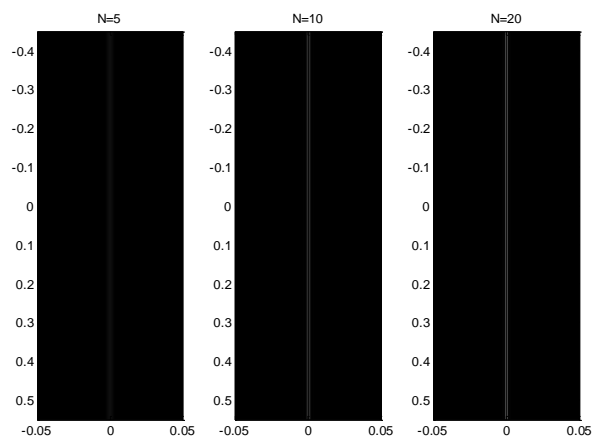

图 4.25 N 缝夫琅和费衍射比较(二维图)

由多缝干涉理论可知，干涉掩模阵列有一定的遮光比，在同样积分时间内，星点能量传送效率有所下降，但是，从精度提升效率正比与掩模阵列数目，所以，若实现同等精度的星敏感器，星敏感器视场可按阵列数目比例缩放，可见这种星敏感器易实现小视场。由前文可知，小视场下提升精度的主要途径是增加暗星数量。由于视场小，几何畸变也更容易校正。多缝干涉模式有助于实现星敏感器的小型化、一体化、高精度。

## 4.7 不同衍射形貌的高斯近似

高斯函数三个参数 $(I, x, \sigma_s)$ 表述星点的能量、位置和形貌信息，比起夫琅和费方孔和圆孔衍射函数，描述的斑点能量分布更简单，更易做算法研究。将高斯函数 $I_g = I_0 e^{-\frac{x^2}{2h^2}}$，夫琅和费方孔衍射斑 $I_s = I_0 \left( \dfrac{\sin(x/h)}{x/h} \right)^2$，圆孔衍射斑：

$I_c = I_0 \left( \dfrac{2J_1(x/h)}{x/h} \right)^2$，展开可得，

$$I_g = I_0 \left( 1 - \frac{1}{2}\left(\frac{x}{h}\right)^2 + \frac{1}{8}\left(\frac{x}{h}\right)^4 - \frac{1}{48}\left(\frac{x}{h}\right)^6 \cdots \right) \tag{4.99}$$

$$I_s = I_0 \left( 1 - \frac{1}{3}\left(\frac{x}{h}\right)^2 + \frac{2}{45}\left(\frac{x}{h}\right)^4 - \frac{1}{315}\left(\frac{x}{h}\right)^6 \cdots \right) \tag{4.100}$$

$$I_c = I_0 \left( 1 - \frac{1}{2^2 2!}\left(\frac{x}{h}\right)^2 + \frac{1}{2^4 3! 2!}\left(\frac{x}{h}\right)^4 - \frac{1}{2^6 4! 3!}\left(\frac{x}{h}\right)^6 \cdots \right) \tag{4.101}$$

从泰勒展式可知，夫琅和费衍射斑可以用高斯函数来模拟，有两种方法做高斯等效:(1) $I = 0$，求得高斯半径为 $\sigma_s$ (张新宇等 2013)；(2) $I = \dfrac{1}{2\pi\sigma_s^2}$，求得高斯半径为 $\sigma_s$。

对于圆孔衍射斑，$\sigma_s = \dfrac{0.41\lambda f}{ND}$；若采用第二种等效办法，即满足

$\displaystyle\int_0^{2\pi}\int_0^{\infty} x I_c(x) dx d\theta = 1$，$I_0 = \dfrac{\pi N^2 D^2}{4\lambda^2 f^2}$，则由 $\dfrac{1}{2\pi\sigma_s^2} = I_0$，可得 $\sigma_s = \dfrac{\sqrt{2}\lambda f}{N\pi D}$。对于方孔衍射斑，$\sigma_s = \dfrac{\lambda f}{3Nd}$，$\sigma_s = \dfrac{\lambda f}{\sqrt{2\pi}Nd}$。对方孔多缝衍射斑采用两种方法做模拟所得结果参见图 4.26。

由图 4.26 可知，两种方法均能较好得到高斯近似形式，第二种方法近似效果更佳，主要因为能量模拟能够将方孔衍射斑次斑的能量吸收进来，因而更准确。

然后用第二种方法分别对方孔和圆孔多缝衍射做模拟，所得结果参见图 4.27-4.30，图 4.28 与图 4.27，图 4.30 和图 4.29 不同之处在于，图 4.27 和图 4.29 仅对衍射因子做高斯近似，图 4.28 和图 4.30 又对干涉因子采用高斯近似。

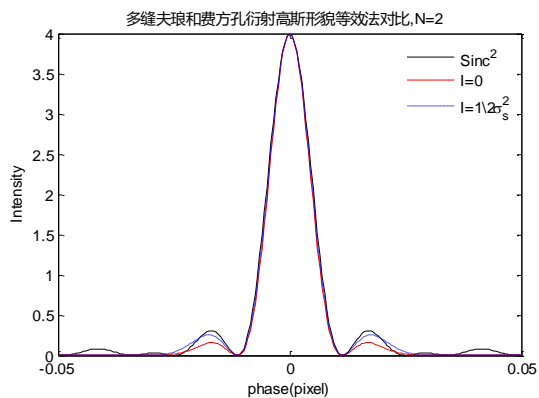

图 4.26 N=2 夫琅和费方孔衍射高斯等效图：(1)红色，$I = 0$；(2)蓝色，$I = \dfrac{1}{2\pi\sigma_s^2}$

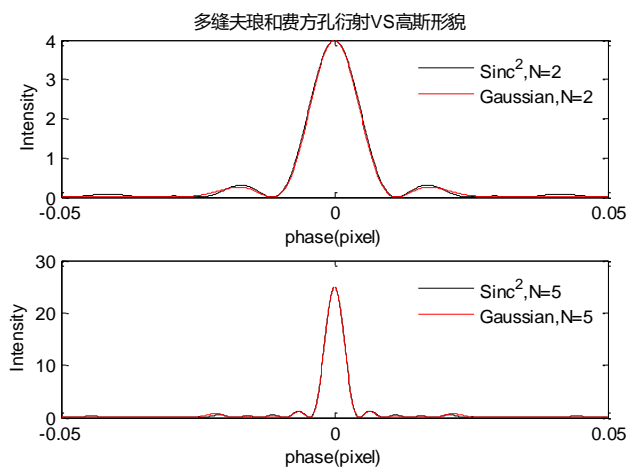

图 4.27 N 缝夫琅和费方孔衍射高斯等效图，干涉因子为 sinc 函数：(1)黑色曲线，sinc 函数形貌；(2)红色曲线，高斯形貌

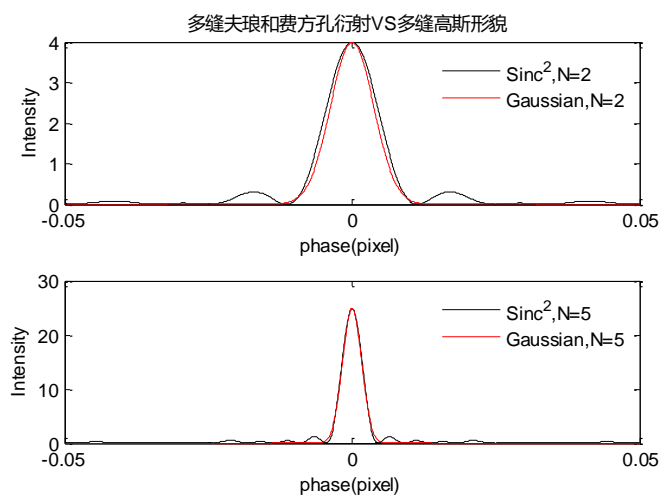

图 4.28 N 缝夫琅和费方孔衍射高斯等效图，干涉因子为高斯函数：(1)黑色曲线，sinc 函数形貌；(2)红色曲线，高斯形貌

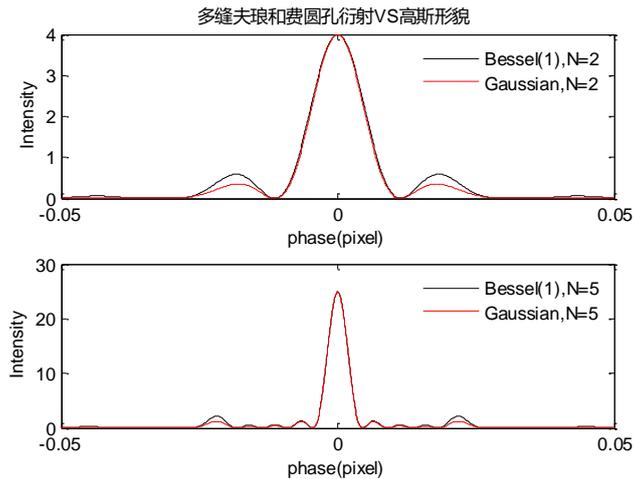

图 4.29 N 缝夫琅和费圆孔衍射高斯等效图，干涉因子为 sinc 函数：(1)黑色曲线，sinc 函数形貌；(2)红色曲线，高斯形貌

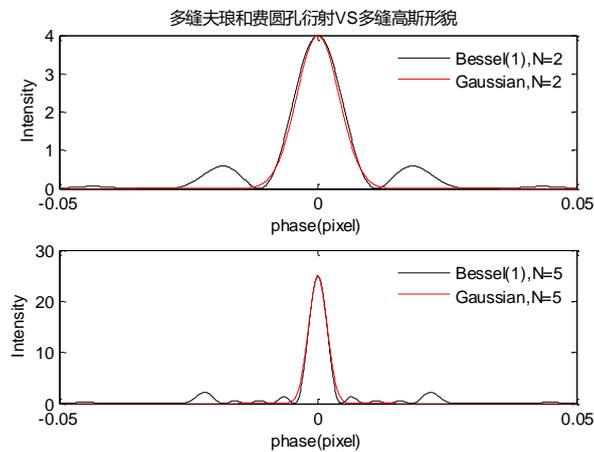

图 4.30 N 缝夫琅和费圆孔衍射高斯等效图，干涉因子为高斯函数：(1)黑色曲线，sinc 函数形貌；(2)红色曲线，高斯形貌

比较图 4.28 与图 4.27、图 4.30 和图 4.29 可以看出，干涉因子采用高斯形式后，会抹平次级斑的信息，但从仿真结果看，由于高斯函数适宜做快速算法，这种近似做法仍然值得肯定，当然，采用符合形貌的定位算法能进一步提高星点提取精度；比较图 4.28 与图 4.30、图 4.27 和图 4.29 可进一步看出，在系统设计其他参数都不变的情况下，方孔比圆孔更适于高精度星点提取，这与 4.6 节所得结论相同。由夫琅和费衍射信号与高斯信号的等效性，可知本论文研究的单星提取算法同样适用于多缝干涉式星敏感器。

## 4.8 本章小结

通过研究 IWCOG 优化算法、CRLB 理论等内容，由此得到的主要结论如下：

(1)IWCOG 优化算法和差分算法适合甚高精度星点提取

理论和部分试验均证实了 IWCOG 算法和最小二乘高斯拟合法在高斯噪声下满足成像器件的 CRLB 特性，是 MVUE 方法。该算法 S 误差在 0.002pixel，无需校正，在多星定位上有优势，精度可达 CRLB 指标。而质心法不满足 CRLB 性能，S 误差较大，在没有先验信息下不能校正。IWCOG 耗时仅比质心算法高 2-5 倍，比最小二乘高斯拟合算法低上千倍。针对不同衍射星点形貌给出两种高斯近似理论，表明 IWCOG 算法能够实现各类星敏感器甚高精度星点定位。

(2)CRLB 精度指标有助于星敏感器设计和技术研发

CRLB 理论分析表明，单星点定位精度与信号尺度成正比、与信噪比成反比，这个关系指出了定位精度与设计参数之间的本质联系。为此本文探索了可能提高静态定位精度的若干途径，得出增加有效积分时间、减小星点尺度是提高单星静态精度的主要方式。增加星数不能提升平均单星定位精度，对指向精度的贡献仍然服从 Liebe 公式。CRLB 精度指标可用于器件选型和不同型号设计之间理论对比，为未来星敏感器精度提升提供思路。

# 第五章 全精度高动态技术研究

运动误差是制约星敏感器动态精度提升的根本环节(Chen Y. 1987)。遥感卫星，科学探索和国防任务需要频繁的机动操作，要求高机动下如 1º/s 保持较高甚至静态精度。但是，一旦星敏感器速率超过 0.1º/s，在未采取任何措施下，精度会很快下降，动态性能已经是限制先进空间任务能否完成的核心指标之一，提高动态性能甚至实现全精度动态技术已经是星敏感器的研究热点。动态情况下，星点会发生像移，随角速率、视场位置不同而不同，如何实现实时的运动畸变识别和补偿是解决动态星点精确提取和提升动态性能的关键。本节提出了基于单帧解析补偿的动态技术和基于多帧解析补偿的全精度动态技术，试图解决单帧或多帧情况下的运动畸变补偿问题。

## 5.1 单帧解析补偿动态技术研究

研究了单帧运动状态下星点的动态模型，提出一种新的方法得到匀速运动和加速运动下的动态补偿公式和定位精度。首先为运动中的星点图像建模，分析畸变、抖动、加速度产生的条件，其次基于 CRLB 理论给出匀速运动下补偿公式和定位精度，然后针对其他速度情况，提出一个新的策略获得一般的补偿公式和定位精度，估计由运动抖动带来的误差，通过仿真来分析验证本节技术(Zhang J. et al. 2015, AO)。

### 5.1.1 运动中的星点图像模型

一旦星敏感器角速率超过一定值，如 0.1º/s，星图质量下降很快。为量化运动影响，应首先为运动星点建立模型。设 $(x_c(t), y_c(t))$ 为 CCD 相平面坐标系星点位置，是角速率和时间的函数。$f$ 为光学系统离焦后的有效焦距，星点的方向可以定义为：$\bar{O}(t) = [x_c(t) \quad y_c(t) \quad -f]'$。

为获得精确的 $x_c(t)$，需先推出姿态矩阵 $\mathbf{C}(t)$。根据文献(Pasetti A. et al. 1999; Samaan M. A. et al. 2002)，如果姿态矩阵 $\mathbf{C}(t)$ 在 $t$ 时刻可做泰勒二阶展开，则可表述为：

$$\mathbf{C}(t) = [I - \omega^\times t + \frac{1}{2}(\omega\omega' - \omega' \omega I - \dot{\omega}^\times)(t)^2]\mathbf{C}(0) \tag{5.1}$$

其中，$\omega' = \begin{bmatrix} \omega_{xt} & \omega_{yt} & \omega_{zt} \end{bmatrix}$，$\omega_t^\times$ 表示 $\omega_t$ 的叉乘矩阵，$t$ 表示积分时间。

假设 $(x_0, y_0)$ 是零时刻运动星点的初始位置，$\Delta_1 = \omega_y x_0 t, \Delta_2 = \omega_x y_0 t, \Delta_1, \Delta_2 \ll f$ (Pasetti A. et al. 1999)。由于焦距和角速度一直在漂移或抖动，从公式(5.1)可得，运动星点的实时位置和速度可建模为：

$$x_c(t) = x_0 + v_1 t + 0.5 a_1 t^2 \tag{5.2a}$$

$$a_1 = f(\dot{\omega}_y - \omega_x \omega_z) + (\omega_x \omega_y - \dot{\omega}_z) y_0 - (\omega_y^2 + \omega_z^2) x_0 \tag{5.2b}$$

$$v_1 = v_0 + \varsigma_{\omega v} + \varsigma_{fv}, v_1 = f\omega_y + \omega_z y_0 + x_0(\omega_y x_0 - \omega_x y_0)/f \tag{5.2c}$$

$$v_0 = \bar{f}\,\bar{\omega}_y, \varsigma_{\omega v} = \bar{f}\,\delta\omega_y, \varsigma_{fv} = \delta f \,\bar{\omega}_y \tag{5.2d}$$

$$y_c(t) = y_0 + v_2 t + 0.5 a_2 t^2 \tag{5.2e}$$

$$v_2 = -f\omega_x - \omega_z x_0 + y_0(\omega_y x_0 - \omega_x y_0)/f \tag{5.2f}$$

$$a_2 = -f(\dot{\omega}_x + \omega_y \omega_z) + (\omega_x \omega_y - \dot{\omega}_z) x_0 - (\omega_x^2 + \omega_z^2) y_0 \tag{5.2g}$$

其中，$a_1, a_2, v_1, v_2$ 是星点在 CCD 焦平面的加速度和速度，单位分别为 pixels/s$^2$ 和 pixels/s。$v_0$ 为平均速度，$\varsigma_{\omega v}$ 表示速度抖动项，满足 $N(0, \sigma_{\omega v}^2)$ 分布。$\varsigma_{fv}$ 表示速度畸变项(van Bezooijen R. W. H. 2003)，$\bar{f}$ 是 $f$ 的平均值，$\delta f$ 是 $f$ 的漂移量。$\bar{\omega}_{yt}$ 是 $\omega_{yt}$ 平均值，$\delta\omega_{yt}$ 是 $\omega_{yt}$ 的抖动。

公式(5.2a)-(5.2g)可知，速度畸变与初始位置有关，这在 Pasetti 和刘朝山的论文中也可以看到，这种畸变，又叫空域畸变。假设 $\omega_z = \dot{\omega}_z = 0$，$y_0(\omega_y x_0 - \omega_x y_0)$ $\ll f$ 或 $x_0(\omega_y x_0 - \omega_x y_0) \ll f$ 成立，当 $y_0\omega_y t - x_0\omega_x t > 2f$ 或 $\Delta\omega_y > 2\omega_y$ 时，由加速度导致的畸变又称为时域畸变，该畸变量可能超过由速度导致的空域运动畸变误差。即航天器不处于制动状态，远离视场中心的区域也会存在加速度分布，而在

制动初始时刻，加速度的影响同样不可忽略。由公式(5.2b)可知，速度畸变主要来自于焦距估计误差，由于焦距受温度、色差效应影响较为显著，而在轨校正过程不能对这个变量实时校正，造成焦平面速度估计存在偏离，该项是温度的函数。速度抖动主要源于航天器的高频振动，表现在航天的实时角速度总是在预定值附近震荡，造成焦平面速度估计在均值附近振动，该项是震荡幅度的函数，需要陀螺提供平稳的姿态信息。

根据冈萨雷斯和伍兹的著书《数字图像处理》(Gonzalez R. C. et al. 2009)，运动中的星点可建模为：

$$S(x,T) = \int_0^T \frac{I_0}{2T\sqrt{\pi}\sigma_s} \exp\left[\frac{-(x-x_c(t))^2}{2\sigma_s^2}\right] dt \tag{5.3}$$

其中 $T$ 是曝光时间，$x_c(t)$ 是 $t$ 时刻在 CCD 焦平面的星点位置，$I_0$ 时间 $T$ 内光通量，$\sigma_s$ 是高斯信号尺度。

## 5.1.2 匀速状态下的运动补偿量和定位精度

在运动情况下，由公式(3.9)知定位误差公式可表述为：

$$\hat{x} = x_{true} + e_v, e_v = e_n + e_m \tag{5.4}$$

其中，$x_{true}$ 为星点真实位置，$e_v$ 包括器件噪声($e_n$)、像移、速度畸变和抖动($e_m$)等误差源带来的位置误差。

由此可知，在其他部件处于理想状态下，运动下的定位误差源包括器件噪声和运动。器件噪声是造成定位误差的最根本因素(Auer L. H. et al. 1978; Winick K. A. 1986; Chen Y. 1987; Rao C. R. 1992; Gai M. et al. 2001; Lindegren L. 2013)，在航天器处于静止状态时，静态星点的定位精度主要与器件噪声有关。由第 4 章 CRLB 理论可知，静态星点的定位精度为：

$$\delta x_n = \sqrt{CRLB(x)} = \frac{\sqrt{2}\sigma_s}{R}, R^2 \approx \frac{I_0^2}{2\sqrt{\pi}\sigma_n'\sigma_s} \tag{5.5}$$

其中，$R$ 是信噪比(SNR)，$\sigma_n'^2 = \sigma_n^2 + \lambda_n$。

为求得运动下的补偿量和定位精度，可将运动星点建模为静态星点。由公式(5.2a)可知，运动状态下像移量与 $x_0, a$ and $v_0$ 有关。当 $a=0$，星斑的中心 $x_c(\tau)$ 即是轨迹的中心点，

$$x_c(\tau) = x_0 + \Delta x, \Delta x = L/2, L = v_1 T \tag{5.6}$$

其中 $L/2$ 即为许多文献提及的运动补偿量(CMC)。

反观公式(5.1)，航天器运动会导致像移图像的尺度扩大。但是，矩形的运动轨迹 $x_c(t)$ 并不适合预测运动对星点尺度造成的影响，很现实的方案是将矩形轨迹等效为高斯函数 $N(x_c(\tau), \sigma_r^2)$。由此可得，最终的尺度即使二者的和，即，

$$\sigma_s'^2 = \sigma_s^2 + \sigma_L^2, \sigma_L^2 = \frac{1}{T}\int_0^T [x(t) - x_c(\tau)]^2 dt = \frac{L^2}{12} \tag{5.7}$$

公式(5.6)和公式(5.7)表明运动图像可等效为质心为 $x_c(\tau)$ 和尺度为 $\sigma_s'^2$ 的星点函数。由前述推导可得，运动状态下的定位误差为：

$$\delta x_v = \sqrt{2}\sigma_s'/R', R'^2 \approx \frac{I_0^2}{2\sqrt{\pi}\sigma_n'\sigma_s'} \tag{5.8}$$

公式(5.8)显示当 $L \ll 2\sqrt{3}\sigma_s$ 时，运动导致的定位误差可忽略不计；当 $\sigma_r^2 \geq \sigma_s^2$ 即 $L \geq 2\sqrt{3}\sigma_s$，星点定位最小误差快速增长，运动导致的定位误差成主要误差项，该误差在积分时间不变情况下正比于 $L^{3/2}$。这种情况下，捕获的星图数据只能用于输出姿态，其精度为一定速度下的 CRLB 值。如果星图数据用于在轨校正，就会引入额外的误差。一旦角速率超过一定值，如 $5°/s$，星点会被大幅拉长，只有最亮的星用于定位，而误差远高于静态情况下的CRLB指标。为获得更低的CRLB定位误差，由公式(5.7)可知，可采用 TDI 和可变帧率技术. 可以看出对于 TDI 技术， $L=1, \sigma_r^2 = 1/12$。因为 $\sigma_s^2 \gg 1/12$，这就是 TDI 技术能够获得静态精度的原因(van Bezooijen R. W. H. et al. 2002; van Bezooijen R. W. H. 2003)。可变帧率技术是另外一个有效的实现方法，该法通过调节积分时间限制 $\sigma_r^2$ 增长达到这一点 (Michaels D. L. et al. 2004)。DBA 算法，与可变帧率类似，通过将整个有效积分时间分为若干段，单段时间像移因此减少，降低了整体像移对最后定位精度的影响。

由公式(5.2a)-(5.2g)可知，时域畸变是影响更高精度定位的因素之一，研究加速度下的运动补偿量和定位精度有助于获得更高的动态性能(Zhang J. 2015, AO)。

### 5.1.3 加速度状态下的运动补偿量和定位精度

获得可变速率下的运动补偿量和定位误差并不容易，需要用一个思想实验来实现。假设积分时间内存在个理想的闪光灯，它的开启停止成像过程是完美的，则曝光时间可以被切为若干小段。所以整个策略就是 $T$ 可以用无数个 $T_p$ 组成，其中 $\sum_p T_p = T$，$T_p \propto 1/v_{T_p}$。对于每个间隔 $T_p$，速度可以视作常数，星点在 $T_p$ 内遵循高斯形貌分布，与静态的一致，尺度为 $\sigma_s$。当 $T_p$ 跳入 $T_{p+1}$ 时，星点位置 $x_p$ 会立即进入 $x_{p+1}$，因此公式(5.3)的信号模型可表述为：

$$S(x,T) = \sum_p S_p(x,T_p) \tag{5.9a}$$

$$x_c(T_m) = x_0 + v_1 T_p + 0.5 a_1 T_p^2 \tag{5.9b}$$

于是一般运动补偿量(GMC) 可表述为：

$$\Delta x = (\int_0^T \frac{1}{v_1 + a_1 t} dt)^{-1} \int_0^T \frac{x_0 + v_1 t + 0.5 a_1 t^2}{v_1 + a_1 t} dt - x_0 = \frac{1}{4a_1} \frac{v_T^2 - v_1^2}{\ln v_T - \ln v_1} - \frac{v_1^2}{2a_1} \tag{5.10}$$

注意当 $v_1 = 0, a_1 \neq 0$ 时，公式(5.10)定义无效，这种情况下，$\Delta x \approx 0.125 a_1 T^2$。

运动下的星点尺度也可用同样策略得到，

$$\sigma_{s,a}'^2 = \sigma_s^2 + (\int_0^T \frac{1}{v_1 + a_1 t} dt)^{-1} \int_0^T [x_0 + v_1 t + 0.5 a_1 t^2 - (x_0 + \Delta x)]^2 \frac{1}{v_1 + a_1 t} dt = \sigma_s^2 + \sigma_L^2 \tag{5.11}$$

其中，

$$\sigma_L^2 = (\Delta U)^2 - \frac{\Delta U(v_T^2 - v_1^2)}{2a_1(\ln v_T - \ln v_1)} + \frac{v_T^4 - v_1^4}{16 a_1^2 (\ln v_T - \ln v_1)}, \Delta U = \Delta x + v_1^2/2a_1, v_T = v_1 + a_1 T$$

注意当 $v_1 = 0, a_1 \neq 0$ 时，公式(5.11) 无效，这种情况下，$\sigma_L^2 \approx 0.015625 a_1^2 T^4$。

当 $a = 0$ 时，$\Delta x = v_0 \tau$，$\sigma_{s,a}'^2 = \sigma_s'^2$。公式(5.11) 推导与公式(5.7)，公式(5.10) 推导和(5.6)的结果相同，CMC 是 GMC 的特例，表明这个新的策略假设是成立的，与事实相符。在新的运动补偿下，加速度下的像移星点定位精度能够接近光电器件的 CRLB 指标。

根据公式(5.8)，就得到了加速度下近似的定位误差：

$$\delta x_{v,a} = \sqrt{2}\sigma_{s,a}'/R', R'^2 \approx \frac{I_0^2}{2\sqrt{\pi}\sigma_n'^2 \sigma_{s,a}'} \tag{5.12}$$

这个策略同样可以评估运动抖动效应。此时的运动轨迹可定义为：

$$x_c(t) = x_0 + \Delta x = x_0 + (v_0 + \varsigma_v)t \qquad (5.13)$$

其中，$\varsigma_v$ 表示抖动，满足 $E(\varsigma_v) = 0$，$var(\varsigma_v) = \sigma_v^2$。

因此，可得 $\sigma_{s,j}'^2$，

$$\sigma_{s,j}'^2 = \sigma_s^2 + \sigma_L^2, \sigma_L^2 = \frac{(\Delta x)^2}{3} + \frac{(\sigma_v T)^2}{12} \qquad (5.14)$$

这样，抖动下的定位误差为：

$$\delta x_{v,j} = \sqrt{2}\sigma_{s,j}'\big/R', R'^2 \approx \frac{I_0^2}{2\sqrt{\pi}\sigma_n'^2\sigma_{s,j}'} \qquad (5.15)$$

## 5.1.4 仿真分析

为评估新的运动补偿公式性能，做了一系列仿真试验。每个测试用蒙特卡洛法随机生成 1000 星点，以研究多星定位误差随信噪比 R 和速度的变化情况。信噪比 R 定义为：$R = I_0/(\sqrt{K}\sigma_n)$。其中，$K = \lceil 2\sqrt{\pi}\sigma_s' \rceil$ 是星点的有效尺度。

设 $\mu_{real,i}$ 表示第 $i^{th}$ 个星点真实位置，$\mu_{calc,i}$ 表示计算位置。定位的绝对误差定义为 $\varepsilon_i = |\mu_{real,i} - \mu_{calc,i}|$。因而，平均定位误差可定义为 N 个星点的均方误差。$N = 1000$ 保证获得稳定的 $\delta$。蒙特卡洛模拟使用的参数和值参见表 5.1。

第 1 个仿真论证公式(5.3)，显示不同速度和加速度情况下的星点形貌。使用不同的 $v_1$ and $a_1$ 生成星点形貌参见图 5.1. 图中显示当 $a_1 = 0$ 时，星点呈对称形貌，当 $a_1 \neq 0$ 时，出现不同程度的扭曲。第 2 个仿真验证公式(5.2a)，显示由加速度

表 5.1 蒙特卡洛模拟参数和值

| 参数 | 值 | 参数 | 取值范围 |
|---|---|---|---|
| $I_0$ | 100 | $v_1$ | $[-200, 400]$ |
| $\sigma_s$ (pixel) | 1 | $a_1$ | $-1000$ or $[0,1000]$ |
| $T$ (s) | 0.1 | 视场 | $5° \times 5° (1024 \times 1024)$ |
| $N$ | 200 | $R$ | $[5,120]$ 或无噪声 |

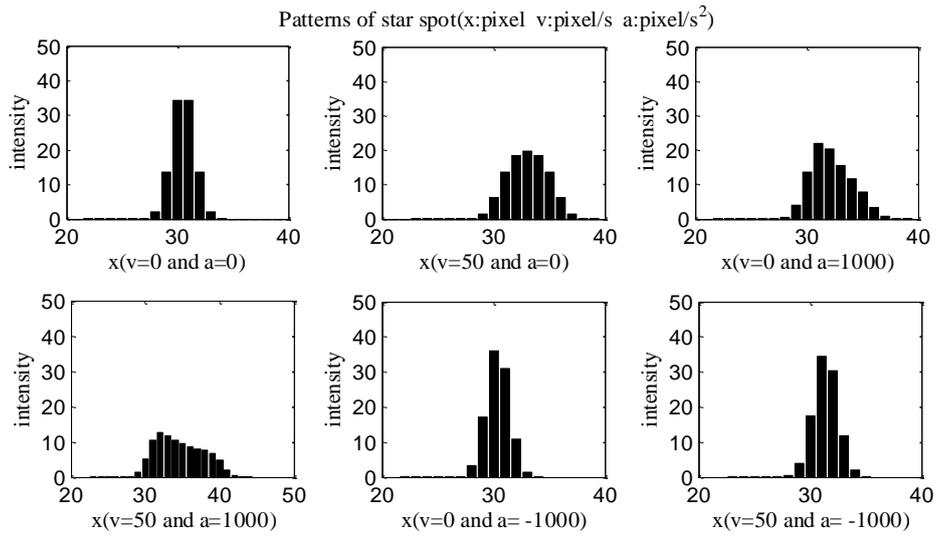

图 5.1 不同速度和加速度下星点形貌

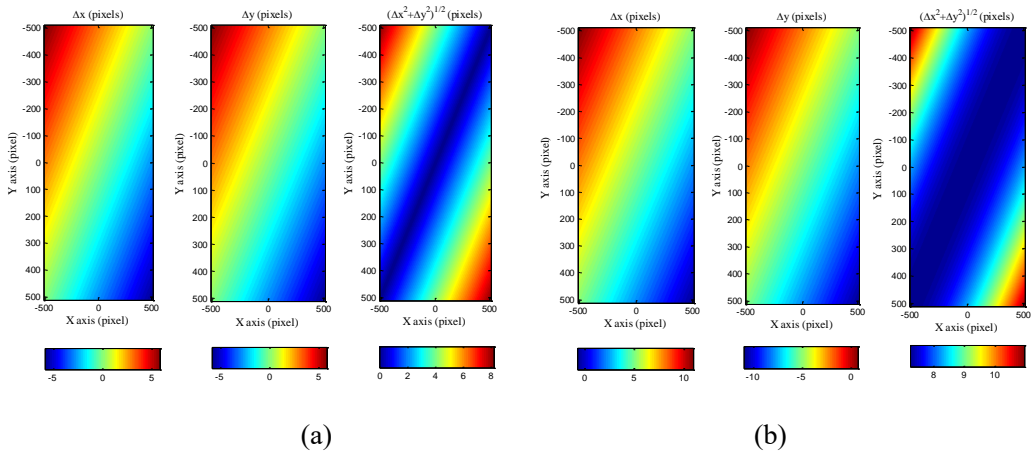

图 5.2 不同速度和加速下的焦平面像移分布：(a) $v_1 = v_2 = 200, a_1 = a_2 = 0$；(b)

$$v_1 = v_2 = 200, a_1 = a_2 = 1000$$

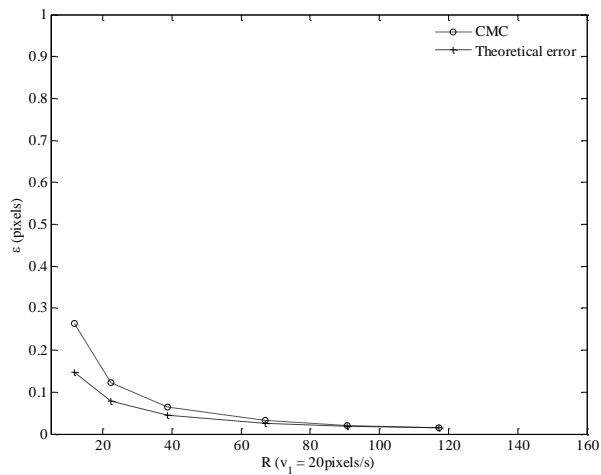

图5.3 速度 $v_1 = 20\,\text{pixels/s}$ 下，运动补偿后 $R$ 对像移星点定位精度的影响

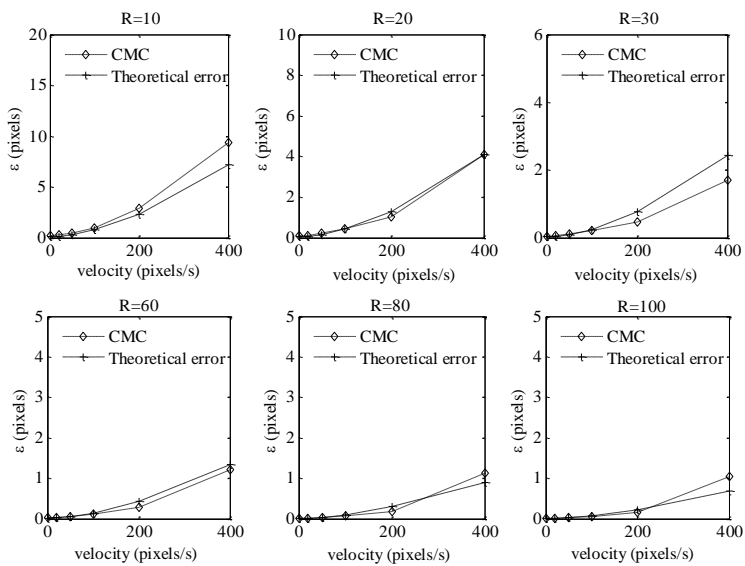

图 5.4 速度和 $R$ 对星点定位精度的影响

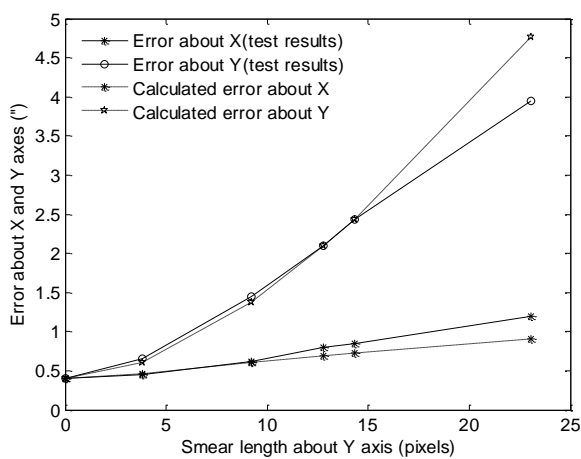

图 5.5 动态定位测试数据和拟合数据对比(实线为测试数据，虚线为拟合数据)

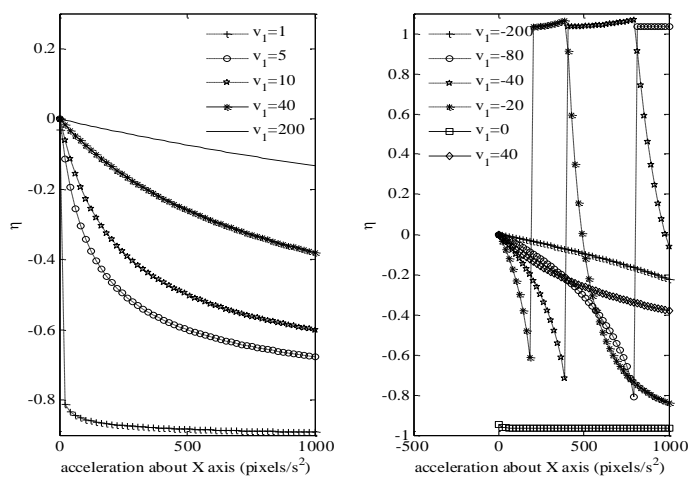

图 5.6 不同加速度和速度下的质心偏率(左图速度与加速度同向，右图不同向)

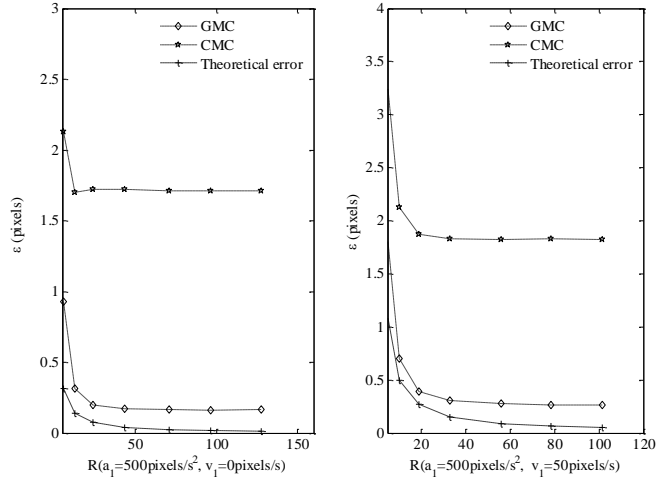

图5.7 不同 $R$ 和速度下，两种运动补偿策略(GMC and CMC)对定位精度的影响：左图，$v_1 = 0$

pixels/s ，$a_1 = 500$ pixels/s²；右图，$v_1 = 50$ pixels/s ，$a_1 = 500$ pixels/s²

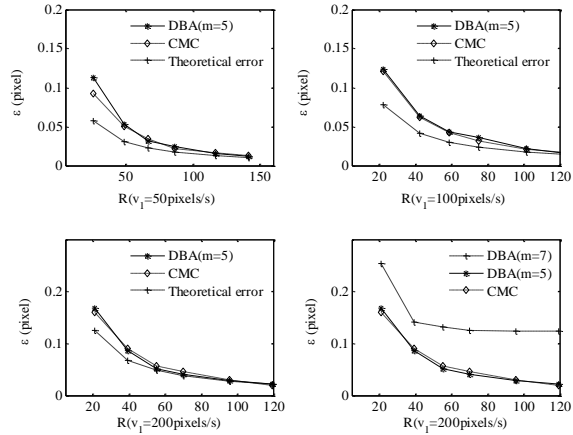

图5.8 DBA算法与本文的方法精度对比，精度是速度和 $R$ 的函数，积分时间0.1s，m定义参见文献
(Pasetti A. et al. 1999).

引起的运动畸变，参见图 5.2。由图可知，当 $a_1 = a_2 = 0$ 时，相比于 $L = v_1 T = 10$

pixels，视场边缘的畸变量与速度运动的位移相当。为验证匀速运动定位误差公式，

固定 $v_1 = 20$ pixels/s，不同信噪比下的 RMS 误差参见图 5.3。很明显由 IWCOG

确定像移星点质心所得误差与一定速度下的CRLB值相同。第4章证明了 IWCOG

算法的 MVU 特性。

　　重复以上仿真流程，使用不同的信噪比 R 和速度。其定位误差 $\delta$ 由图 5.4 所

示。当速度低于一定值时，图 5.4 显示定位误差是常值，与 $x_0$ 和 $v_1$ 不相关。当速

度超过一定值，运动误差占首要地位，与 $v_1^{3/2}$ 成正比。图 5.4 显示与理论预测符合很好。

为进一步证明运动下 CRLB 指标的正确性，用本节公式拟合试验数据。这是动态测试的常规试验。星敏感器位于转台上，对准夜空某一块天区。Y 轴不动，以不同速率转动 X，经过星图采样、提取、配备环节输出姿态，所得结果参见图 5.5 实线部分。采用公式(5.8) 为 X/Y 轴建模，$\xi_x = \sqrt{2}\sigma_s'/R'$，$\xi_y = \sqrt{2}\sigma_s/R'$，其中 $v_1T$ 为像移量，可得 $\sigma_s = 1.22\,pixel$。由模型计算出的定位误差参见图 5.5 虚线部分。在像移量较小时，图 5.5 显示计算的误差几乎与测试数据相同，说明 CRLB 指标可用于在线计算实际误差。当像移量较大时，误差有一定出入，这可能是像移畸变不均匀所致。

为评估新的策略对定位精度的性能，可将其与原有的补偿公式进行比较。原有的补偿公式为路径的一半，即 $\Delta x_{old} = L/2, L = v_0 T + 0.5aT^2$。这样可定义质心偏率为：$\eta = |(\Delta x_{old} - \Delta x)/\Delta x|$。质心偏率表示真实星点中心与路径一半的偏离程度。由公式(5.10)，$\eta$ 是 $v_1$ 和 $a_1$ 的函数。设定 $v_1$ 为 1、5、10、40、200、0、-20、-40、-80、-200 pixels/s，加速度范围为 0 到 1000 pixels/s²，计算的 $\eta$ 参见图 5.6。由图可知，对于匀速运动的星点，质心偏率为 0，而对于加速运动的星点，可分为两种情况，对于速度与加速度同方向的，则加速度越大，质心偏率越大；对于加速度与速度相反的，质心偏率随加速度有个峰值，然后两边下降。由此可知，为获得较好的定位精度，准确的加速度补偿是必须的。

为此，在 $v = 0\,pixel/s, a = 500\,pixel/s^2$ 和 $v = 50\,pixel/s, a = 500\,pixel/s^2$ 下，采用不同加速度补偿策略，所得定位精度参见图 5.7。GMC 由公式(5.10)定义，CMC 即为 $\Delta x_{old}$。图 5.7 显示新补偿公式下，加速度的运动误差低于旧的补偿方案。表明新的策略是获得扭曲星点图像质心偏离的很好方法。但是，图中也显示误差高于推导的 CRLB 值，采用的定位方法不能获得 $a \neq 0$ 下的理论精度，可能只是处理扭曲星点图像的亚优方法。这需要进一步的研究证实这个论断。

为进一步说明方法的实用性，将其与 DBA 方法进行比较，结果参见图 5.8。由图 5.8 可知，DBA 在像移量相对于 Binning 指数成整数倍时定位较好，在有取

整截断误差时较差(如 m=7)。当存在取整误差时，多帧不能有效地叠加在一起，定位偏差不可避免。本节方法没有这个问题，所有情况下都可达到 CRLB 精度。

## 5.2 多帧解析补偿动态技术研究

### 5.2.1 单帧解析补偿技术存在的问题

由 5.1 节可知，为降低像移影响，降低积分时间是有效方式，但是，为提高星敏感器静态性能，根据 CRLB 理论，星点静态下的定位精度与信噪比成反比，在既定的硬件工艺下，提高信噪比唯一方式是延长积分时间。所以，二者要求存在一定矛盾。而且，延长积分时间技术存在如下问题：(1)运动速率较小时，星点采样易出现饱和情况，星图失真；(2)运动速度较大时，像移十分严重，精度很快变差；(3)延长积分时间意味着姿态输出速率更低；(4)空域和时域畸变更严重。而解决这些问题最有效的方式仍然是降低积分时间。另外，为提供更佳的姿态控制，姿态输出率应尽可能高，最好能达 10Hz 以上，角速率达 1deg/s 时能保持较高甚至静态精度。但是，由于速度畸变、饱和情况、运算速度等的影响，这三种指标多互相牵制，提高一项指标，其他两项指标会有不同程度下降。正因如此，Liebe 曾经说过"由于摄入的光子数有限，未来的精度和姿态输出率难以有大幅提升"，但是基于并行技术的发展，本节提出一种可行方法可同时提高三项指标，从而突破 liebe 的观点(Liebe C. C. 2002)。本节在前文研究基础上，给出一种多帧解析补偿方法(MFAC)，能大幅提高动态性能和姿态输出率，同时保持甚至提高精度。

### 5.2.2 多帧解析补偿方法

本节研究的主要目标是研究多帧解析补偿方法获得高动态、高姿态更新率和高精度的方案。该技术以单帧解析补偿技术为基础，通过解析补偿高动态下的运动偏置误差，提高了动态性能，通过多帧解析叠加，避免了精度下降，通过星点位移矢量解析校正，增加了姿态输出率，这三种策略并行不悖。除上述优势外，该技术能有效避免采样象素饱和，解决由单帧星点数据缺失导致的精度下降或姿态不能输出问题。该技术与 TDI、DBA 等技术一样，除要求相平面速度和精确

的时钟信息外，无额外的需求，因而易于实施。仿真结果证实了所提出方法的正确性、有效性和鲁棒性，能实现上述目标。在像移量一定时，多帧解析补偿算法与时间的平方成反比，即在像移量尽量小情况下，倾向于较大的时间间隔而不是无限细分。

### 5.2.2.1 原理

根据第 5.1 节可知，假设星点在同一方向运动，积分时间内的速率是知道的，运动补偿后，采样初始时刻或末端时刻星点的位置即可确定。多帧解析补偿方法类似于 DBA 算法和可变帧率技术，将单帧再细分为多个小帧，用软件模拟 TDI 技术(Pasetti A. et al. 1999)。即整个积分时间划分为若干小段，每小段对应一帧，每帧内的运动星图可以等效为静止的。这样，整个积分时间内星图就可以等效为前述多帧的叠加。为方便讨论，如第 5.1 节所述，可以用单帧补偿技术得到每帧初始时刻位置，再根据速度信息将其补偿到整个积分时间内某个时刻，这样该技术即获得了星点在某时刻的真实位置。

如图 5.9 所示，该例将整个积分时间等分为 6 段 $\Delta T = \sum_{i=1}^{6} \Delta T_i$，通过将各时间段的星点位置反推到积分的开始时刻 $T_{S1}$，则 $\Delta T$ 内星点在 $T_{S1}$ 时刻位置可精确确定。

如图 5.10 所示，星敏感器一帧图像的处理时间包括积分时间($\Delta T_i$)、图像存储时间($\Delta T_r$)、观测星提取和匹配($\Delta T_p$)和姿态数据输出时间($\Delta T_c$)几个部分。并行工作模式导致不同的工作模式，姿态数据更新时间可表示为：
$\gamma = \max \left\{ \Delta T_i, \Delta T_r, \Delta T_p, \Delta T_c \right\}$。对于大多数星敏感器来说，决定并行工作模式的是积分时间和观测星提取与匹配时间，即 $\gamma = \Delta T_i$。为方便，本节仅讨论一种情况，即姿态数据更新时间由积分时间确定。

如图 5.11 所示，速度畸变包括空域畸变(P1-P3)和时域畸变(P4)，这也是一定要实施速度畸变估计和补偿的主要因素。以图 5.11 P4 为例，重新计算每一帧的相对于新的积分开始时刻 $T_{Si}$ 的位置，叠加起来即得到 $\Delta T_i$ 内星点在 $T_{Si}$ 时刻的星点位置。如此重复，即可提高姿态输出率。通过不断延长积分时间 $\Delta T$，可提高

定位精度。而减小 $\Delta T_i$ 或整个积分时间内的时间片数，在探测器读出噪声可忽略情况下，也可提高动态性能。有关补偿的问题请参见前面小节。

### 5.2.2.2 算法流程

由前述定义可知，获得这些动态技术所需的速度信息和时钟同步信息是精确定位的关键。速度信息可通过两种常规方式获得：跟踪模式下采用测量矢量差分法，或者使用陀螺信息在线计算。

获得这些速度分布信息后，即可根据单帧解析方法将不同星点的位置信息补偿到积分时间初始时刻。设某个星点在单帧的参数分布(匀速)$(\mathbf{v}_{\Delta T_i}, \mathbf{a}_{\Delta T_i}, \Delta T_i)$，$\mathbf{x}_i$

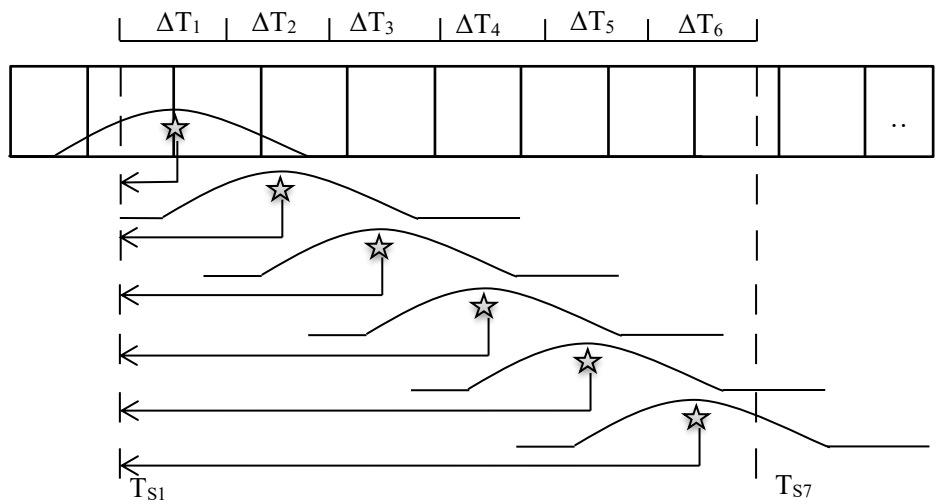

图 5.9 MFAC 模式下信号序列

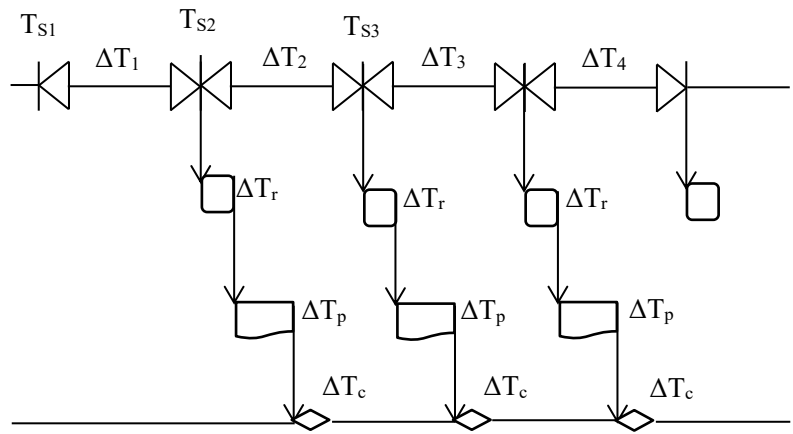

图5.10 星敏感器并行处理时间示意图

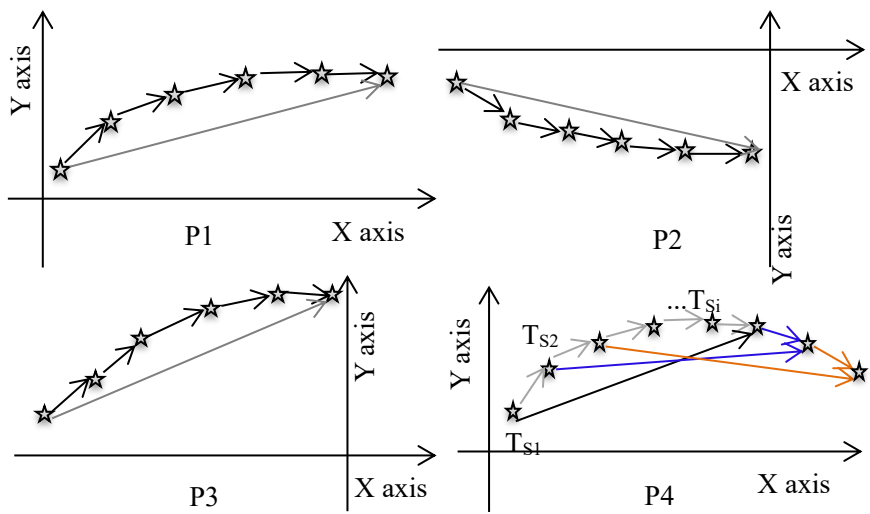

图5.11 不同星场位置的空域速度畸变(P1-P3)和时域速度畸变(P4)

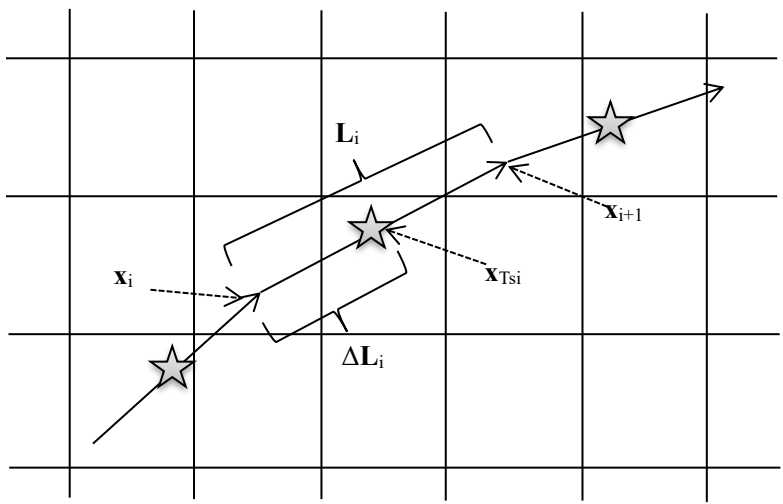

图5.12 MFAC例程变量定义示意图

为星点在 $T_{Si}$ 的位置，$\mathbf{x}_{i+1}$ 为 $T_{S(i+1)}$ 时刻的位置，$\mathbf{x}_{T_{Si}}$ 为第 $i^{th}$ 帧使用定位算法计算的中心位置，$\Delta\mathbf{L}_i$ 是运动补偿矢量，$\mathbf{L}_i$ 是星点在单个时间片内走过的位移。则可推知，

$$\mathbf{x}_1 = \frac{\Delta T_1}{\Delta T}(\mathbf{x}_{T_{S1}} - \Delta\mathbf{L}_1) + \frac{\Delta T_2}{\Delta T}(\mathbf{x}_{T_{S2}} - \mathbf{L}_1 - \Delta\mathbf{L}_2) + ... + \frac{\Delta T_6}{\Delta T}(\mathbf{x}_{T_{S6}} - \sum_{i=1}^{5}\mathbf{L}_i - \Delta\mathbf{L}_6) \quad (5.16a)$$

$$\mathbf{x}_2 = \frac{\Delta T_2}{\Delta T}(\mathbf{x}_{T_{S2}} - \Delta\mathbf{L}_2) + \frac{\Delta T_3}{\Delta T}(\mathbf{x}_{T_{S3}} - \mathbf{L}_2 - \Delta\mathbf{L}_3) + ... + \frac{\Delta T_7}{\Delta T}(\mathbf{x}_{T_{S7}} - \sum_{i=2}^{6}\mathbf{L}_i - \Delta\mathbf{L}_7) \quad (5.16b)$$

若 $\Delta T_i = \Delta T_j$，由此可得，

$$\mathbf{x}_2 = \mathbf{x}_1 + \mathbf{L}_1 + \frac{\Delta T_1}{\Delta T}(\Delta\mathbf{L}_1 - \Delta\mathbf{L}_7) \quad (5.17)$$

其中，$\Delta \mathbf{L}_i$ 按公式(5.10)计算。

若 $\Delta T_i \neq \Delta T_j$，则可以按公式(5.16a)-(5.16b)进行详细计算，本节提出的多帧解析补偿方法仍然适用。

由上可知，只需根据单帧补偿法计算并保存 $\Delta \mathbf{L}_i$ 数据即可较高的姿态输出率，其精度由 $\Delta T_i$ 数目来保证，动态性能由 $\Delta T_i$ 大小确定，均可通过前述公式计算出来。

仍然假定整个积分时间分为 6 段，则算法流程如下：

- 第 1 步：每帧积分时间内，估计整个相平面的各星点的参数分 $(\mathbf{v}_{\Delta T_i}, \mathbf{a}_{\Delta T_i}, \Delta T_i)$，$i = 1,2...,6$。

- 第 2 步：初始位置未知，则计算 $\mathbf{x}_{T_{Si}}$、$\mathbf{L}_i$、$\Delta \mathbf{L}_i$；初始位置已知，调到第 4 步；

- 第 3 步：按公式(5.16a)计算 $T_{S1}$ 星点的初始位置 $\mathbf{x}_1$；

- 第 4 步：估计整个相平面的速度分布 $(\mathbf{v}_{\Delta T_7}, \mathbf{a}_{\Delta T_7}, \Delta T_7)$，计算并保存 $\mathbf{L}_7$，$\Delta \mathbf{L}_7$，得到 $T_{S2}$ 星点的初始位置 $\mathbf{x}_2$；

- 第 5 步：令 $\mathbf{L}_i = \mathbf{L}_{i+1}$，$\Delta \mathbf{L}_i = \Delta \mathbf{L}_{i+1}$，$\mathbf{x}_{T_{Si}} = \mathbf{x}_{T_{S(i+1)}}$，调到第 4 步。

### 5.2.2.3 性能分析

(1)精度分析

整个时间段 $\Delta T$ 分为 N 个 $\Delta T_i$ 小段，$i = 1,2,\cdots N$。设 $(I_i, \mathbf{x}_{i0})$ 为每段星点信号参数，由于使用的是线性 CCD，$I_i \propto \Delta T_i$。则定位算法关于 $N$ 个信号整体亮度 $I$、真实位置估计满足：

$$\hat{I} = \sum_i I_i, \sum_i (\mathbf{x}_{i0} - \hat{\mathbf{x}}_0) I_i = 0 \tag{5.18}$$

设 $v$ 为 $\Delta T_i$ 内速度，$\sigma'_{n,i}$ 为第 $i^{th}$ 帧星点噪声。使用一维信号模型，该时间内的补偿量和定位精度为：

$$\sigma_{\mathbf{x}_{i0}} = \sqrt{\text{var}(\mathbf{x}_{i0})} = \frac{\sqrt{2}\sigma'_{s,i}}{R_i'^2}, R_i'^2 \approx \frac{I_i^2}{2\sqrt{\pi}\sigma_{n,i}'^2\sigma'_{s,i}}, \sigma_{s,i}'^2 = \sigma_s^2 + v^2\Delta T_i^2 / 12 \qquad (5.19)$$

仍然假设 $\Delta T_i = \Delta T_j$，$I_i = F\Delta T_i$，探测器噪声是时域不相关的，则公式(5.18)

可表示为：$\hat{\mathbf{x}}_0 = \sum_{i=1}^{N} \mathbf{x}_{i0} / N$。 此时定位误差为：$\text{var}(\hat{\mathbf{x}}_0) = \sum_{i=1}^{N} \text{var}(\mathbf{x}_{i0}) / N^2$，由此

得多帧解析补偿后星点定位误差为：

$$\sigma_{\hat{\mathbf{x}}_0} = \sqrt{\text{var}(\hat{\mathbf{x}}_0)} = \frac{1}{\sqrt{N}} \frac{\sqrt{2}\sigma'_{s,i}}{R_i'} = \frac{1}{\sqrt{N}} \frac{2\pi^{1/4}\sigma'_{n,i}\sigma_{s,i}'^{3/2}}{F\Delta T_i} = \frac{2\pi^{1/4}\sigma'_n\sigma_{s,i}'^{3/2}}{F\Delta T} = \frac{\sqrt{2}\sigma'_{s,i}}{R'},$$

$$R' = \frac{(F\Delta T)^2}{2\sqrt{\pi}\sigma_n'^2\sigma'_{s,i}}, \sigma'_n = \sqrt{N}\sigma'_{n,i} \qquad (5.20)$$

由此可见，理想情况下，多帧解析方法能够获得与 TDI 一样的精度。如某帧出现失效情况，则定位误差为：

$$\sigma_{\hat{\mathbf{x}}_0} = \frac{1}{\sqrt{N-1}} \frac{\sqrt{2}\sigma'_{s,i}}{R_i'} \approx \sqrt{\frac{N}{N-1}} \frac{2\pi^{1/4}\sigma'_n\sigma_{s,i}'^{3/2}}{F\Delta T} = \sqrt{\frac{N}{N-1}} \frac{\sqrt{2}\sigma'_{s,i}}{R'} \qquad (5.21)$$

如 N=6，可知精度下降 $\sqrt{\frac{N}{N-1}} - 1 = 9.5\%$，当 $N \gg 1$，可知精度基本保持。

可见与 TDI 相比，本节的方法有以下优势：

(1)鲁棒性增强。当某帧出现失效情况，在 N 值较大情况下，对定位精度影响不大，这极适合星敏感器工作于高动态模式。

(2)积分时间可调。根据视场内速度先验分布调整每帧积分时间或修改 N 值而不影响最终精度，可用于宽动态模式，即高速和低速下均可使用。

(3)速度畸变精确补偿。速度分布随不同位置和积分时间而变化，可利用前文知识进行精确补偿，而 TDI 则在硬件实施上困难，表明该方法可用于更复杂的运动场合和 CCD 器件上。

(4)提高了姿态输出速率。只要跟踪程序耗时能够有效控制，有效积分时间可分为更小的时间片而不影响定位精度。

可见，该方法对硬件的依赖降低，更灵活鲁棒，没有 CTI 效应，实现了软件化 TDI。与 DBA 相比，该算法另外优势在于无截断误差，同比精度更高，更适应复杂的运动场合。

以上分析假设 $\Delta T_i = \Delta T_j$，由于时间抖动，$\Delta T_i \neq \Delta T_j$ 情况更普遍，令 $\Delta T_i = \mu_{\Delta T_i}$

$\pm \sigma_{\Delta T_i}$，由前文公式，星点偏置误差和尺度为：$\sigma_{s,\Delta T_i}'^2 = \sigma_s^2 + v^2(\mu_{\Delta T_i} \pm \sigma_{\Delta T_i})^2 / 12 \approx$

$\sigma_s^2 + v^2 \mu_{\Delta T_i}^2 / 12$。由此可知，积分时间小幅偏离对尺度影响不大，与 $\sigma_s^2$ 或 $v^2 \mu_{\Delta T_i}^2 / 12$ 相比是小项。前述叠加原理仍然适用，但偏置误差不可忽略，必须精确补偿掉，叠加后的位置精度仍然接近 CRLB 值。

(2)效率分析

由于整个过程均为解析计算，与新提出的多帧复原算法、帧相关算法相比，无迭代或矩阵等复杂运算过程，效率更高。除要求精确的角速度分布和时钟信息后，该算法未引入额外的要求，算法有望在现有基础上改进直接投入使用。

### 5.2.2.4 帧差分法速度估计误差分析

前文指出陀螺信息可用来计算更相平面运动速度,但常规帧间差分法是求运动速度更简便的方法,跟踪模式下的测量矢量差分法通过识别同一颗星在两帧中的位置,得到星点的运行速率。由于星敏感器无陀螺累积误差,帧差法也是求取星场速度的常规方法。这个方法速度可定义为：$\mathbf{v} = \dfrac{d\mathbf{x}}{dt} = \dfrac{\mathbf{x}_2 - \mathbf{x}_1}{\Delta T_1}$。由公式(5.17)和(5.20)，速度矢量和精度可表述为：

$$\mathbf{v} = \mathbf{v}_{\Delta T_1} + 0.5\mathbf{a}_{\Delta T_1} \Delta T_1 + \frac{1}{\Delta T}(\Delta \mathbf{L}_1 - \Delta \mathbf{L}_7) \tag{5.22a}$$

$$\sigma_{\mathbf{v}} = \frac{\sqrt{2}\sigma_{\mathbf{x}1}}{\Delta T_1} = \frac{2^{3/2}\pi^{1/4}\sigma_n'\sigma_s'^{3/2}}{FN\Delta T_1^2}, \sigma_s'^2 \approx \sigma_s^2 + v^2 \Delta T_1^2 / 12 \tag{5.22b}$$

由此可见，若陀螺给出的角速率信息是精确的，由于速度空域畸变和时域畸变，常规帧间差分法引入 $\dfrac{1}{\Delta T}(\Delta \mathbf{L}_1 - \Delta \mathbf{L}_7)$ 偏置速度误差。若使用帧差法估计的速度分布用于后一帧星点初始位置估算，会引入一定的误差。该误差会在加速或减速阶段累积，而在其他情形有不同程度抵消，所以从总体上看，由帧差分法获得的速度信息不能长时间用于运动补偿。

帧差法估计的速度方差则由 CRLB 决定，在像移量和 N 值一定时，与帧间时间的平方成反比，即在像移量尽量小情况下，为获得更精确的速度估计，多帧解析补偿算法法倾向于较大的时间间隔而不是无限细分。大量文献研究表明，速度畸变在高速率和视场边缘处极为明显，表现为视场相关和时间相关性，因此使用陀螺信息实时估计相平面速度分布是十分必要的，由帧差法获得的速度估计不

适合速度补偿。换言之，仍然需要陀螺辅助速度补偿。较差的陀螺漂移一般小于0.1 度/h，由于积分时间多在 0.5s 以下，此时由漂移导致的误差为 0.05 角秒(本体坐标系)，与单个星点定位误差相比可忽略不计，无需补偿。如何不利用陀螺实现精确速度补偿仍然需要进一步研究。

### 5.2.3 仿真分析

为评估多帧解析补偿方法的有效性和正确性，做了一系列仿真试验。将同一积分时间分为若干段 N，根据公式(5.3)生成不同速度多个时间片的星点，并加入噪声，计算不同时间片初始点星点位置，以上重复 M 次以统计定位精度。这些测试中，为得到稳定的定位误差 $\varepsilon$，M 设为 200，平均定位误差即为 M 个星点的均方误差。仿真的全部参数和取值范围参见表 5.2。

为验证文中所提方法，改变速度和时间片 N，计算的定位精度参见图 5.13。图中显示对于较低的速度，时间片数量对定位误差几乎无影响，精度接近积分时间内的 CRLB 值。随着速度得增大，较高 N 值能保持定位精度不变，而较低

表 5.2  测试中的参数及取值

| 参数 | 取值范围 | 参数 | 取值范围 |
|---|---|---|---|
| $I_0$ | 1000 | $v$ (pixels/s) | $[10, 400]$ |
| $\sigma_i$ (pixel) | 1 | $a$ (pixels/s$^2$) | $0, \pm 1, 10, 50, 200$ |
| $\Delta T$ (s) | $[0.1, 0.5]$ | $N$ | $1, 2, 5, 10$ |
| $M$ | 200 | $R$ | $[5, 150]$ |

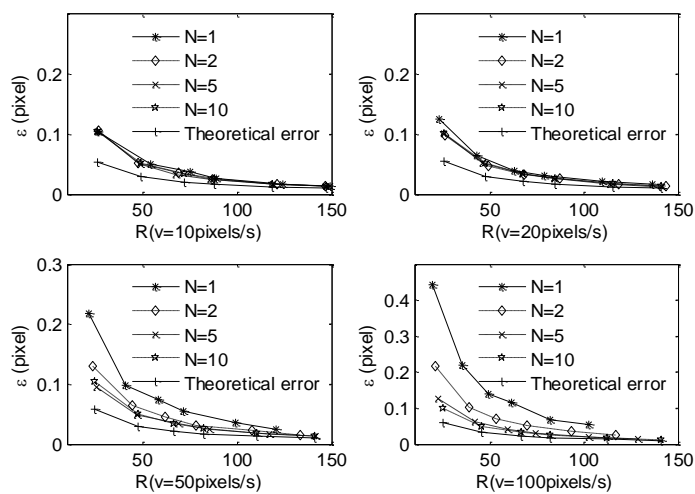

图 5.13  不同 N 和速度下多帧星图定位误差($\Delta T = 0.1s$)

的 N 值精度变差，说明高速运动下，较低 N 值的单帧像移量也加大，最终影响了定位精度。为证明这个结论，固定 N 值，不同速度下的定位误差参见图 5.14。图 5.14 清楚表明单帧的像移量影响了最终定位精度。从前文理论可知，影响定位精度的根本因素是积分时间，固定速度和 N 值，不同积分时间下的 MFAC 和 TDI 定位精度参见图 5.15。图 5.15 表明，随积分时间的增长，定位误差近似与积分时间成反比例关系，接近 CRLB 精度，MFAC 可获得类似 TDI 的精度指标。图 5.13-5.15 表明，像移量较小时，影响定位误差的根本因素是积分时间，而不是 N 数，但较高的 N 数意味着较高的姿态更新率。

为理解空域畸变和时域畸变的影响，固定速度、积分时间和 N 值，分别针对不同加速度进行补偿和不补偿，所得图示参见图 5.16，其中(a)为未校正情形，(b)为校正后结果。图 5.16 显示，要得到接近 CRLB 精度，必须实时补偿运动畸变，获取 FOV 速度分布参数和精确采样时间是必需的一个步骤。

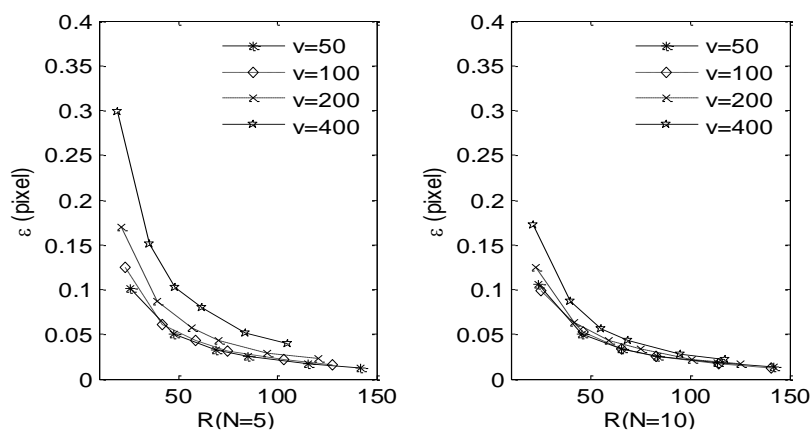

图 5.14 固定 N 值，不同速度下多帧星图定位误差（$\Delta T = 0.1s$）

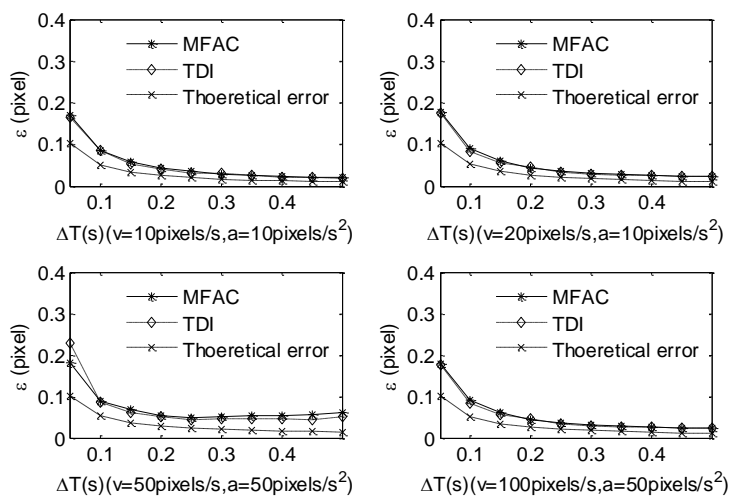

图 5.15 不同速度和积分时间下 MFAC 和 TDI 技术定位误差对比

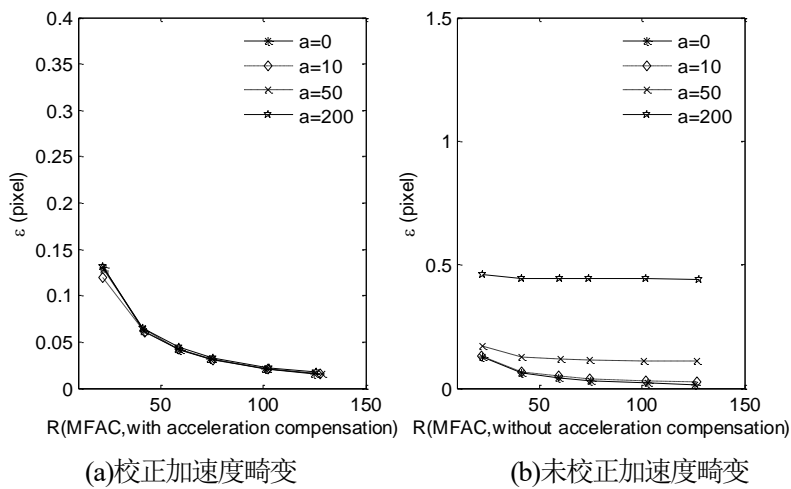

(a)校正加速度畸变　　　　　　　(b)未校正加速度畸变

图 5.16 MFAC 在加速度补偿和不补偿下的定位误差（$\Delta T = 0.1s$）

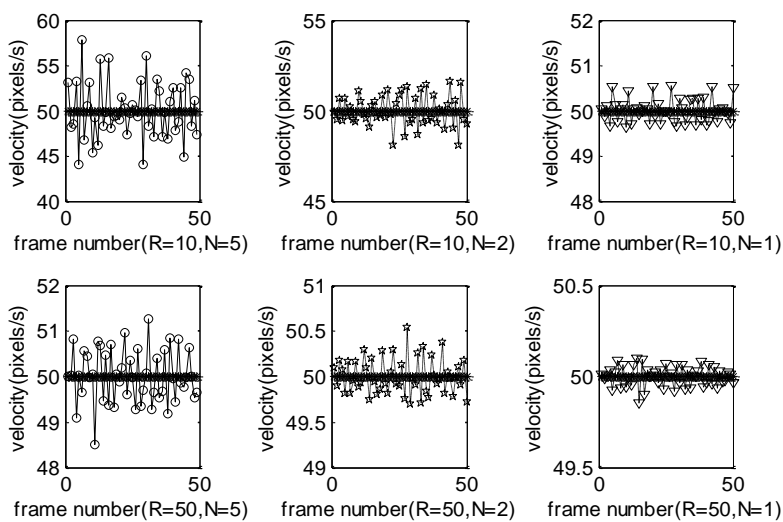

图 5.17 不同信噪比和帧率下速度估计误差（$\Delta T = 0.5s$，$v = 50$）

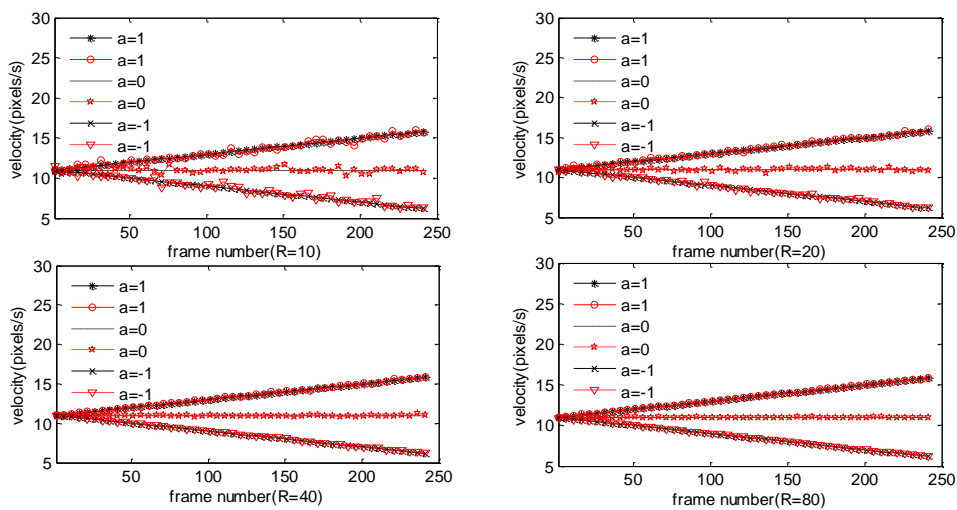

图 5.18 不同加速度下帧差法速度估计误差（$\Delta T = 0.1s$）

前文已经分析了帧差法的性质，为验证这个结论，固定速度，改变信噪比和 N 值，计算结果参见图 5.17。另外一个测试是研究加速度对速度估计得影响，所得结果参见图 5.18。

由图 5.17 可知，不论信噪比如何，高 N 值的速度估计精度较低，而低 N 值的较高，这符合前文理论分析。图 5.18 显示加速度对速度估计误差影响较小。

## 5.3 本章小结

本章通过对星敏感器探测器平面像移精确建模，以 CRLB 理论推导了运动星点的定位精度，由此得到的主要结论如下：

(1)IWCOG 优化算法和差分算法适合甚高精度动态星点提取

与静态星点提取不同的是，由于像移的影响，动态星点存在运动畸变，包括空域运动畸变和时域运动畸变，二者导致的误差在甚高精度星点提取中不可忽略，必须通过建模并实时补偿。文中推导了精确的解析补偿公式，这样在采用 IWCOG 类算法进行星点定位后，精度可达动态情况下的 CRLB 指标，仿真和拟合试验验证了算法精度和补偿公式的正确性。不同视场处的星点运动畸变不一样，帧差分法会导致该帧计算误差积累到后面帧，这是该方法难以有效实现恒星陀螺的基本原因，有必要采用陀螺数据进行运动补偿。由于跟踪环节窗口预测需要的角速率精度较低，帧差分法仍可用于窗口预测。

(2)根据角速率大小调整单帧有效积分时间是降低运动定位误差的根本方法，解析计算运动误差并在轨补偿是实现全精度定位的一种有效方案

匀速运动情况下，当实际像移高于 3.464 倍星点尺度时，运动定位误差是主要误差源，减少单帧积分时间才能根本上降低运动定位误差。为避免精度下降，基于单帧解析补偿技术，提出多帧解析补偿技术解决这一问题。该技术通过增加帧数或降低单帧积分时间，并实时补偿运动畸变误差，既提高了动态性能和姿态输出率，又避免了精度下降。该技术相比 TDI 和 DBA 有一定优势，不存在单帧星点数据缺失导致的精度下降或姿态不能输出问题，无 CTI 效应，除要求相平面速度和精确的时钟信息外，无额外的需求。

# 第六章 在轨校正技术研究

前两章针对静态、动态情况下影响星点高精度定位的因素，如探测器噪声、形貌畸变、运动畸变等提供了有效解决方案，尚未考虑其他因素有：光学畸变、温致焦距漂移、参数均漂效应等。这些因素导致的星点定位误差通常会超过静态和动态下的改进，对这些参数进行实时在轨校正是甚高精度星点定位必不可少的环节。本章针对其中若干因素引入的定位误差，基于第四章探测器理论定位精度公式提出 CRLB 在线误差约束机制，试图将各参数的边界条件在线计算出来，据此判断运行参数处于什么状态，以及相关星点数据是否可用于后续的校正，为地面粗平场校正、在轨误差校正提供依据参考。

## 6.1 在轨校正分析

高级空间任务往往要求星敏感器指向精度达到 0.2"，用常规星敏感器的像元分辨率来表征(10"/pixel)，其测量误差要低于 0.02 像素。但是，大量的重要误差源，如光学镜头畸变、PRNU、温致焦距漂移、像移、速度抖动、热象素、恒星自行等，同样会导致误差高于此值(Chen Y. 1987; Piterman A. et al. 2002; van Bezooijen R. W. H. et al. 2002; Anderson J. et al. 2003; Blarre L. et al. 2008; Schmidt M. et al. 2012)。需要一个在轨校正环节降低这些误差。典型在轨校正的例子有：

(1)热象素点去除。正常星点的定位精度处于亚像元级别，而一个失误星点检测会导致像元级别误差，表现在所有四元数输出姿态同时出现峰值，这在 Herschel 星敏感器中又称为单象素事件。真实星图减掉 CCD 暗帧图像后，会有大量热像素点存在，这些像素点数据不能用于姿态确定。温度降低 7K，热像素能量可降低一半，通过降温可以大幅降低热象素点数量。高于一定星等的耀斑星体同样去除(Schmidt M. et al. 2012)。

(2)光轴扭曲校正。源自焦距变化、更为复杂的空域畸变、光学安装温度场的变化等。机械扭曲源自 CCD 相平面的物理挤压等，采用结构一体化设计可缓轻。对于焦距变化，除温控外，在线实时估计焦距可一定程度上补偿焦距误差；而几何畸变是最大的误差，覆盖整个视场，误差可至几个像元以上，为达到精确

定位，在轨校正要搜集有效的跟踪星点数据，建立合适的光学扭曲补偿模型补偿空域畸变等。国内外文献关于这方面的研究的例子较多，本文对比不展开研究(van Bezooijen R. W. H. et al. 2002; Anderson J. et al. 2003; Schmidt M. et al. 2012)。

(3)星表误差补偿。选择高精度星库，在线补偿恒星自行误差(钟红军 2011)。根据天区，在线优化星数和星等，STR 做了这方面的改进(Schmidt M. et al. 2012)。

尽管国外高精度星敏感器已经通过多项式校正、温度控制、机械部件热解耦设计，高精度的质心算法、可变帧率等技术来降低空域误差、视轴误差和像移影响，但无论是理论还是设计层面上，关于焦距漂移、像移、参数均值漂移等误差源能够降低到何种精度的研究极少。总结起来，存在如下问题需要澄清：(1)一些有规律的大的系统偏差，如光学畸变、焦距、PRNU、CCD 像素亚结构等，星敏感器资料虽提到要采用在轨校正，但能达到何种理论精度几乎没有报道。(2)对于其他无规律的偏差、可以忽略的偏差、方差等，如焦距漂移、运动误差等，设计或运行参数应满足何种条件才满足在轨(或地面)校正要求。这些理论的缺失，一方面导致设计中误差分配不协调，另一方面无法为在轨校正提供实时误差信息，即使实施在轨校正无法取得预期的效果甚至失效。

为收集能够用于在轨校正的星点数据，给在轨校正提供实时误差信息，使得位置误差降低到预期目标，从理论给出了限制误差的全约束公式，获得了在轨校正过程中一些重要误差源的误差边界。首先基于 CRLB 理论给出高斯形貌星点全参数的 CRLB 约束方程(Adorf H. M. 1996; Mendez R. A. et al. 2014)。其次，根据约束方程，为运动、焦距漂移、"均值漂移"效应等重要误差源分别建模，推出该误差源的误差边界及这些参数对定位误差的总体影响。这项研究有助于弄清星敏感器常规设计中许多通用技术的机理，为一些未解决的问题提供了潜在的方案。

## 6.2 CRLB 约束方程

由公式(3.9)可知，单星定位误差由多种重要误差源组成可表述为：

$$\hat{x} = x_{true} + \delta x, \delta x = \sum_i \delta x_i = \delta x_n + \delta x_l + \delta x_v + \delta x_{prnu} + ... + \delta x_f \tag{6.1}$$

其中，$x_{true}$ 为星点真实位置，$\delta x_i$ 分别指器件噪声($\delta x_n$)、光学镜头畸变($\delta x_l$)、CCD 器件的光响应不一致($\delta x_{prnu}$)、温致焦距漂移($\delta x_f$)、像移($\delta x_v$)等误差源带来的位置误差。

其他误差源，如光学畸变、系统装调、运动、焦距漂移等因素互相独立，可以通过工艺改进和设计改善。改善到何种程度，则是由这些没有校正的误差源决定的，这些误差源中，器件噪声是造成定位误差的最根本因素，是第一大误差源，其具有随机性，不可能从信号中去除。统计理论使用 CRLB 方法估计其造成的定位误差大小。

由前面章节可得，噪声导致的定位误差可表述为：

$$\delta x_n = \sqrt{CRLB(x)} = \frac{\sqrt{2}\sigma_s}{R}, R^2 \approx \frac{I_0^2}{2\sqrt{\pi}\sigma_n'\sigma_s} \tag{6.2}$$

其中，$R$ 是信噪比(SNR)。

为有效约束其他误差源，$\delta x_n$ 波动性越小越好。公式(6.2)表明 $\delta x_n$ 正比于 $\sigma_s/R$，由此可知在噪声一定情况下，$\sigma_s$ 和 $I_0$ 越稳定越好。由于

$$\hat{\sigma}_s^2 = \sigma_{s,true}^2 + \delta\sigma_s^2, \delta\sigma_s^2 = \sum_i \delta\sigma_{s,i}^2, \hat{I}_0 = I_{0,true} + \delta I_0, \delta I_0 = \sum_i \delta I_{0,i} \tag{6.3}$$

其中，$\sigma_{s,true}^2$、$I_{0,true}$ 分别为星点真实尺度和亮度；$\delta\sigma_{s,i}^2$、$\delta I_{0,i}$ 为误差源导致的尺度和亮度估计误差。

由前述分析可知，器件噪声特性同样是造成尺度和亮度估计误差的最根本因素。为得到稳定的亮度和尺度，需要计算亮度和尺度的最小方差(Adorf H. M. 1996; Lindegren L. 2008; Mendez R. A. et al. 2014)。重复上述 CRLB 求值过程，关于亮度和尺度的最小误差为：

$$\delta I_{0,n} = \sqrt{CRLB(I_0)} = \frac{I_0}{R}, \delta\sigma_{s,n}^2 = \sqrt{CRLB(\sigma_s^2)} = \frac{4\sigma_s^2}{\sqrt{3}R} \tag{6.4}$$

对于高斯信号，一旦亮度、尺度和位置的估计精度能够达到 CRLB 值，那么即使再好的估计器也不能获得更好的结果。由公式(6.1)和(6.2)可知，任何亮度、尺度和位置误差的增加都会导致最终的定位精度 $\delta x_n$ 低于 CRLB 值。

更详细的 CRLB 估计公式具体参见附录 A，对于非高斯噪声情形，公式(6.2)

和(6.4)同样应做改变。

对于任何参数无偏估计器，一旦达到 CRLB 指标，就不存在更好的无偏估计器(有些有偏估计器的方差可以低于 CRLB 值，但这不是想要的，具体信息参见 Stoica P. 1990)。因此，有些偏置误差，如 $\delta x_l$ 可以通过建模做一定补偿，但更多的误差则只能抑制。对于一些无规律的误差，如果这些量不好实时估计，则可以考虑用工具约束起来，这个工具就是位置的 CRLB，由于位置的 CRLB 同时受亮度估计、尺度估计影响，本章也给出了另外两种 CRLB。既然位置误差指标 $\delta x_n$ 正比于 $\sigma_s / R$，$\delta I_{0,n}$，$\delta x_n$ 和 $\delta \sigma_{s,n}^2$ 可作为全部约束来限制其他重要误差源。因此本章提出以公式(6.2)和(6.4)为基准，约束星敏感器其他重要参数。这样误差约束方程可定义为：

$$\delta \Phi_i \le \eta_{\Phi,i} \delta \Phi_n, \eta_{\Phi} < \infty, \Phi \in \{x, \sigma_s^2, I_0\} \tag{6.5}$$

公式(6.5)可以确定一些参数的取值边界。假定 $\eta_{\Phi,i}$ 小到一定程度，就意味着该因素就对最后总体误差影响较小。那么跟踪的星点数据不仅可以用于输出姿态，也能用于在轨校正。需要注意的是：

(1)不是所有误差源均对位置、亮度、尺度都产生影响，因此主要对产生位置误差的误差源进行限制；

(2)因单星定位精度指标是位置精度，位置约束是最主要的约束，其次为尺度和亮度约束；

(3)公式(6.5)中只有 $\delta \Phi_i$ 未知，需要对误差源 $i$ 建模，推导其产生的 $\delta x_i$，$\delta \sigma_{s,i}^2$，$\delta I_{0,i}$，然后分配相应的 $\eta_{\Phi,i}$，进而得到相应的误差边界，详细内容见第 6.3 节。

对于一些更高端的器件，如果一些设计或运行参数导致的偏差对部件的设计精度已经达到不可忽略的地步，必须考虑针对该参数实施可行的校正技术。

## 6.3 在轨重要误差源的误差边界

对于在轨飞行的星敏感器，除了器件噪声外，平台的运动，焦距的漂移，星等覆盖的范围，星光的颜色等都是影响最终精度的重要误差源(van Bezooijen R.

W. H. et al. 2013)。本节通过建模获得这些参数的误差模型 $\delta\Phi_i$，再使用公式(6.5)约束这些参数，而后进行深入讨论。

## 6.3.1 运动导致的误差

航天器运动是第二种误差源。航天器的角速度会导致星点在积分时间内出现像移，并有长长的拖尾效应，降低了星等探测能力，对星点位置、尺度和亮度估计均造成的误差，可用相应的 CRLB 指标描述。设 $\delta x$、$\delta\sigma_s^2$、$\delta I_0$ 分别表示运动状态下的位置、尺度和亮度最小方差，$\delta x_v$、$\delta\sigma_{s,v}^2$、$\delta I_{0,v}$ 为运动导致的误差，则公式(6.1)和(6.3)可表述为 $\delta x = \delta x_n + \delta x_v$，$\delta\sigma_s^2 = \delta\sigma_{s,n}^2 + \delta\sigma_{s,v}^2$，$\delta I_0 = \delta I_{0,n} + \delta I_{0,v}$。为得到 $\delta x_v$、$\delta\sigma_{s,v}^2$、$\delta I_{0,v}$，需推出运动下的 $\delta x$、$\delta\sigma_s^2$、$\delta I_0$，而获得这三个指标需要对运动状态下的星点建模，将其等效为静态状态下的星点信号，从而导出这三个误差指标。

如第 5 章和前文所述，由公式(6.2)可得 $R'^2 \approx I_0^2/2\sqrt{\pi}\sigma_n'\sigma_s'$，则运动下位置、尺度和亮度最小标准差为：

$$\delta x = \sqrt{2}\sigma_s'/R' \tag{6.6a}$$

$$\delta\sigma_s^2 = \frac{4\sigma_s'^2}{\sqrt{3}R'} \tag{6.6b}$$

$$\delta I_0 = \frac{I_0}{R'} \tag{6.6c}$$

若 $\sigma_r^2 \ll \sigma_s^2$，按公式(6.5)，关于 $\delta x_v$、$\delta\sigma_{s,v}^2$ 和 $\delta I_{0,v}$ 的约束(具体推导过程参见附录 B)可表述为：

$$v^2 T^2 \le 16\eta_{x,v}\sigma_s^2 \tag{6.7a}$$

$$v^2 T^2 \le 48\eta_{\sigma_s^2,v}\sigma_s^2/5 \tag{6.7b}$$

$$v^2 T^2 \le 48\eta_{I,v}\sigma_s^2 \tag{6.7c}$$

举例说明，假设像素尺度 $20''/pixel$，$\sigma_s = 2\,pixel$，$\eta_{x,v} = 0.1$，$T = 0.1s$，$R = 50$，不存在速度畸变和抖动，公式(6.7a) 表明为达到在轨校正要求，最高速度应不大于 $0.14°/s$，在运动补偿后，由速度导致的定位误差是极为有限的，或可以忽略。

在这个速度下，公式(6.7b)和(6.7c)给出 $\eta_{\sigma_s^2,v}=5\eta_{x,v}/3=0.17$，$\eta_{t,v}=\eta_{x,v}/3=0.03$。可见，对于运动参数，亮度约束>位置约束>尺度约束，因星敏感器主要关心位置误差，所以仍然以位置约束为主。

运动导致的定位误差与运动速率、有效积分时间等的关系已经在第五章深入研究，并用案例证明了公式的正确性。

根据 Liebe 公式，减小像元分辨率是提高静态精度的方法。目前像元的尺度多在 $10\mu m$ 以下，等效角为 $10''/pixel$，但是从公式(6.7a)-(6.7c)可知，减小像元分辨率或像元尺度并不是好的思路，原因有以下几点：(1)分辨率越小，保持静态的最高速度越小，而速度不宜控制导致其很容易超过噪声成为主要误差项；(2)在相同焦距下分辨率越小，意味像元尺度越小，以致满井深度太低。这会产生两种不利结果，一是在相同积分时间下信号容易饱和，二是单位像元信噪比太低，根据 CRLB 理论，这意味着静态精度会快速上升；(3)由于制造工艺限制，像元尺度太小可能意味着像素间的光谱响应不均匀加强，由第四章可知，PRNU 本身就相当于噪声，PRNU 变大，静态精度也会随着上升；(4)减少像元分辨率的初始想法是让采样数增加，减少 COG 法的 S 误差，但是由 IWCOG 性质可知，IWCOG 是无偏算法，因此已经不需要减少像元尺度。前 3 点极易抹杀掉减小等效角带来的一点改进，而第 4 条直接否定了原来的想法。纵观甚高精度星敏感器(参见表 1.1-1.2)，AST-301 的像元参数为 $36'',15\mu m$，HAST 的为 $15.46'',15\mu m$，均采纳了较大像元尺度和等效角。只有 DayStar 的像元参数为 $8.8'',6.5\mu m$，但是 DayStar 始终运行在像移模式下，像移量保持在 16-36pixels 之间，更新速度是前两者 5 倍，白天观星滤掉了星光的蓝色波长部分，积分时间为 30ms-70ms，虽然像元尺度只有 $6.5\mu m$，但不可能出现饱和，且其读出噪声不足 3e，信噪比容易保持不低于 5。

## 6.3.2 焦距漂移导致的误差

焦距漂移是第三大误差源(van Bezooijen R. W. H. 2003)。焦距漂移会导致导致 X 和 Y 轴括位置和尺度的不确定性。设 $\delta x_f$、$\delta\sigma_{s,f}$ 为焦距漂移导致位置和尺

度误差$(\delta\sigma_{s,f} \ll \sigma_s)$，则公式(6.1)和(6.3)可表述为$\delta x = \delta x_f$，$\delta\sigma_{s,f}^2 \approx 2\sigma_s \cdot \delta\sigma_{s,f}$。

对于离视场中心一定距离的区域，$\delta x_f$、$\delta\sigma_{s,f}$可表述为：

$$\delta x_f = \frac{k_x d}{f}\delta f \tag{6.8a}$$

$$\delta\sigma_{s,f} = \mp\frac{\delta f}{\Delta f}\sigma_s \tag{6.8b}$$

其中，$k_x$，$d$和$\Delta f$分别为CCD焦平面内此区域的位置，CCD像元尺度(单位为

$''/s$)，离焦距离。

许多因素导致焦距的漂移，如温度呼吸效应，棱镜的色散效应(Jørgensen J. L. et al. 2006)，此时焦距估计不会精确，漂移始终存在。假定温度$\vartheta$和焦距漂移近似满足线性关系，即$f = f_0 + f_{\vartheta_0}(\vartheta - \vartheta_0) + \varsigma, f_\vartheta = \partial f/\partial\vartheta$，$\varsigma$是随机误差，当$\vartheta = \vartheta_0$时满足$E(\varsigma) = 0, \mathrm{var}(\varsigma) = \sigma_\varsigma^2$，$\varsigma$代表由星光谱色散或泰勒高阶误差。当温度随轨道一直波动时，焦距漂移也会导致$\sigma_s$变化。对于离焦的星点，$\sigma_s$可表述为$\sigma_s = md, d = 1, m > 0.5$。根据公式(6.5)和本节的焦距模型，关于$\delta x_f$和$\delta\sigma_{s,f}^2$约束应满足：

$$\frac{k_x}{f}(f_\vartheta^2 \mathrm{var}(\tau) + \sigma_\varsigma^2)^{1/2} \leq \eta_{x,f}\sqrt{2}m/R \tag{6.9a}$$

$$\frac{2\sigma_s^2}{\Delta f}(f_\vartheta^2 \mathrm{var}(\tau) + \sigma_\varsigma^2)^{1/2} \leq \eta_{\sigma_s^2,f}\delta\sigma_{s,n}^2 \tag{6.9b}$$

在静态情况下，设$\sigma_s = 2 pixel$，$\eta_{x,f} = 0.5$，$\eta_{\sigma_s^2,f} = 0.3$，$k = 512, \Delta f = 5\times10^2 \mu m$，$f = 5\times10^4 \mu m$ and $R = 100$，公式(6.9a)给出$\delta f \leq 1.38\mu m$，公式(6.9b)给出$\delta f \leq 1.74\mu m$。同速度误差边界比较后发现，焦距漂移可能是静态情况下的主要误差源。而只有在较高的动态情况下，速度误差才是主要的误差源。加上视场的空域校正残差，该误差在SED36和HYDRA上也被称作低频误差(Blarre L. et al. 2008)。

航天器轨道在远离和趋近太阳时要经受较大的温度梯度变化。由于温度呼吸效应难以补偿，除采用稳定温度，改进机械部件的结构，减少色散误差等减少焦距漂移外，实施分区校正和分段轨道校正是个可行的方法。对于后者，可以用温

度传感器监视探测器焦平面的温度变化，并将温度划分为若干段，收集的跟踪星点数据分别对应每个温度区域，使用同一个温度区域的数据估计在轨校正参数，温度区域之间的在轨数据则可以用插值方法计算。

在采用公式(6.9a)为 X 和 Y 轴建模后，由焦距漂移导致的误差参见图 6.1。图 6.1 显示远离视场中心区域的误差 $\delta x_f$ 与噪声导致的误差属同等量 $\delta x_n = 0.56''$ ($16''/pixel$)。这表明大的像素尺寸和小视场更易于获得低的位置误差，这也证明了区域校正技术的优势。值得一提的是，采用区域校正技术和小视场的主要原因是焦距漂移效应而不是光学畸变效应，因为后者能够实时补偿，而前者只能抑制。另外，为实现时域补偿，分段轨道在轨校正技术应是实现温度补偿的有效方案。

### 6.3.3 均漂效应导致的误差

在 CRLB 推导过程中，暗含假定信号是平均的，星场的工作环境是不变的。但是，许多情形不符合这个规律，如不正确的设计方案，不完美的器件，以及不精确的在线估计等，星敏感器和外环境都会对跟踪星点产生变化(Schmidt M. et al. 2012)。三种情况需要关注。第 1 种是星库中各色星点都可能参与在轨校正和姿态输出，但是这些星点参数并不一致。第 2 种是星点会在视场任何位置参与姿态输出和在轨校正，但是视场不同处的环境是不一致的，第 3 个是温度控制并不完美，会产生色散。这些参数值在均值附近变化，如起伏的噪声方差，光响应不一致性，星谱的色散，不稳定的亮度等是第 4 种误差源。第 4 种误差源会导致亮度、位置、尺度同时发生变化，公式(6.1)和(6.3)可表述为 $\delta x_j = \Delta(\delta x)$，$\delta\sigma_{s,j}^2 = \Delta(\delta\sigma_s^2)$，

$\delta I_j = \Delta(\delta I_0)$。这里考虑星敏感器常工作动态之中，$\delta\sigma_s^2 = \dfrac{4\sigma_s'^2}{\sqrt{3}R'}$，$\delta x = \sqrt{2}\sigma_s'/R'$，

$\delta I_0 = \dfrac{I_0}{R'}$。由 $\sigma_s'^2 \triangleq \sigma_s^2 + \sigma_r^2$、$\dfrac{\Delta f}{f} << \dfrac{\Delta\lambda}{\lambda}$ 可知，$\lambda$ 和 $v$ 引入尺度误差为：

$$\Delta\sigma_s' = \Delta\sigma_{s,\lambda}' + \Delta\sigma_{s,v}' \approx \gamma_1\sigma_s', \gamma_1 = \left|\dfrac{\Delta\lambda}{\lambda}\right|\dfrac{\sigma_s^2}{\sigma_s'^2} + \left|\dfrac{\Delta v}{v}\right|\dfrac{\sigma_r^2}{\sigma_s'^2} \tag{6.10}$$

其中 $\Delta\lambda$ 为恒心中心光谱漂移，$\Delta v$ 为速度抖动误差。

假设均漂误差源互不相关，则当CCD噪声，星点亮度、星点光谱中心波长、速度一直在随平均值变化时，均漂会同时产生亮度、尺度和位置误差。由于亮度误差增加最终会计入位置误差，则"均漂"效应产生的三种误差可分别表述为：

$$\delta\sigma_{s,j}^2 = \eta_{\sigma_s^2,j}\delta\sigma_{s,n}^2 \tag{6.11a}$$

$$\delta I_j = \eta_{I_0,j}\delta I_{0,n} \tag{6.11b}$$

$$\delta x_j = \eta_{x,j}\delta x_n \tag{6.11c}$$

其中，$\left[\left(\dfrac{\Delta\sigma_n'}{\sigma_n'}\right)^2 + \zeta^2 + \dfrac{\gamma_1^2}{4}\right]^{1/2} \cdot \dfrac{R}{R'} \triangleq \eta_{I_0,j}$，$\left[\left(\dfrac{\Delta\sigma_n'}{\sigma_n'}\right)^2 + \dfrac{\eta_{I_0}^2}{R'^2} + \dfrac{25\gamma_1^2}{4}\right]^{1/2} \cdot \dfrac{R}{R'} \triangleq \eta_{\sigma_s^2,j}$，

$\left[\left(\dfrac{\Delta\sigma_n'}{\sigma_n'}\right)^2 + \dfrac{\eta_{I_0}^2}{R'^2} + \dfrac{9\gamma_1^2}{4}\right]^{1/2} \cdot \dfrac{R}{R'} \triangleq \eta_{x,j}$。

设公式(6.11a)-(6.11c)中 $\eta_{\Phi,j}$ 小于某值，这些参数的范围就可以在发射前(设计时)确定。其他影响参数的范围也可以按相同处理方法得到。公式(6.11c)表明为得到更低的定位误差，PRNU、恒星光谱的中心波长和CCD噪声方差、速度的波动性应能满足公式(6.5)的约束要求。公式(6.11c)也表明即使星敏感器处于静态情况下，位置误差也大于0，"均漂"效应引起的定位误差并不能完全根除，例子参见图6.1。

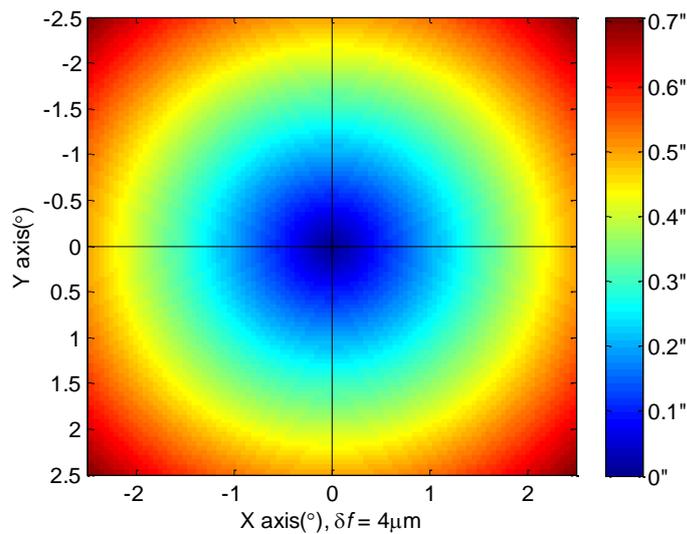

图 6.1 焦距漂移引起的位置残差

为什么要计入这么多的均漂效应，何时去掉一些均漂效应呢？均漂效应实际上是将一些不能实时估计或不能校正的因素计入误差。由于已将未校正的参数漂移计入误差，这里就兼顾了非高精度的星敏感器。更高精度的星敏感器，许多偏差不能忽略，如 PRNU 效应，可用 PRNU MAP 校正(Piterman A. et al. 2012)。

## 6.3.4  全部位置误差和推论

前述已经表明，大量的指标如星点定位精度、动态性能、其他抖动因素等引入的误差均可以用 CRLB 联系起来，它可以辅助设计星敏感器参数、或用于评价星敏感器运行状态。举例如下：

如对于既定的硬件设计、星库恒星参数、探测器尺寸等，公式(6.2)在地面即可确定能够获得大致的姿态输出精度，这同时可以检测测试方法、硬件设计、定位算法等的有效性。公式(6.9a)能够说明既定的定位指标能够容忍的焦距漂移有多大？如公式(6.11a)就可以用于选择恒星参数，不是同一星等的所有亮星均参与校正的，在星库中可以预先去除那些偏红或偏蓝波段的恒星，偏多少不要可以算出来。如公式(6.10)等可确定外界的速度抖动什么情况下可以不用在轨校正，到底多大的抖动是能够接受的应提前计算出来。

表 6.1 不同误差源对位置、亮度、和尺度误差分配

| 误差分配 | 探测器噪声 | 速度 | 焦距漂移 | 均漂效应 |
|---|---|---|---|---|
| 亮度误差 | 1 | $\dfrac{v^2 T^2}{48\sigma_s^2}$ | N/A | $\left[\left(\dfrac{\Delta\sigma_n'}{\sigma_n'}\right)^2 + \zeta^2 + \dfrac{\gamma_1^2}{4}\right]^{1/2} \cdot \dfrac{R}{R'}$ |
| 尺度误差 | 1 | $\dfrac{5v^2 T^2}{48\sigma_s^2}$ | $\dfrac{\sqrt{3}R\delta f}{2\Delta f}$ | $\left[\left(\dfrac{\Delta\sigma_n'}{\sigma_n'}\right)^2 + \dfrac{\eta_{t_0}^2}{R'^2} + \dfrac{25\gamma_1^2}{4}\right]^{1/2} \cdot \dfrac{R}{R'}$ |
| 位置误差 | 1 | $\dfrac{v^2 T^2}{16\sigma_s^2}$ | $\dfrac{k_s R\delta f}{\sqrt{2}fm}$ | $\left[\left(\dfrac{\Delta\sigma_n'}{\sigma_n'}\right)^2 + \dfrac{\eta_{t_0}^2}{R'^2} + \dfrac{9\gamma_1^2}{4}\right]^{1/2} \cdot \dfrac{R}{R'}$ |

所以，CRLB 有助于设计星敏感器。当然，CRLB 也不能做任何事情，星敏感器子部件的性能要由各子部件的特殊测试决定，这不是 CRLB 的事情。

所有关于位置、亮度、尺度的估计误差及量级都总结在表 6.1 中。在信号占主导情况下，除焦距抖动引入的误差分配公式需重新计算外，其他的均保持原值。

根据公式(6.5a), (6.9a)和(6.11c)可得，重要误差源如器件噪声、运动、焦距漂移、抖动效应等引起的全部位置误差可表述为：

$$\delta x^2 = (\delta x_n + \delta x_v)^2 + \delta x_f^2 + \delta x_j^2 = \frac{2\sigma_s'^2}{R'^2} + \frac{k^2 d^2}{f^2}(f_\vartheta^2 \operatorname{var}(\vartheta) + \sigma_\varsigma^2) + \eta_{x,j}^2 \delta x_n^2 \le \eta_x^2 (\delta x_n)^2$$

(6.12)

由于 CRLB 代表了任何方差或偏置误差的边界，假定公式(6.12)中的 $\eta_x$ 值有限，由重要误差源引起的位置误差即得到有效限制。这些有效的误差边界显示了其在系统设计的价值。表 6.1 中和公式(6.12)的一些误差量可以看作对本节两个问题的回答。由于在轨校正需要星点误差越小越好，由此可知，公式(6.12)有如下推论。

**推论 1:** 应根据焦距漂移量实时改变星点在不同处加权系数，提高中心视场处星点对在轨校正的贡献

在速度误差、星点光谱、PRNU 等可以忽略的情况下，即星敏感器达到"准静态"时，焦距漂移成为主要误差源，不同的视场处同一个星点定位误差会不同，中心区域的精度高些，而边缘的最低。当单个星点跑遍整个视场，其提取的精度在整个市场形成一个误差分布图。可以预知，在轨校正应该给予边缘视场一个较低的置信度，而对于视场中心给予一个较高的置信度，或者做分区域校正。

**推论 2:** 应收集亮星数据，根据其 CRLB 值实时计算星点加权值，提高亮星对在轨校正的贡献

由于最终的位置误差正比于 CRLB 值，即信噪比高的单星定位精度高，而信噪比低的单星精度低，因而可以根据星点的星等给予一个反比例与 CRLB 的置信度，这样也可以降低最终误差。整个校正只需判断提取的星点是否符合在轨校正要求，将那些有助于校正的有用信息收集保存起来，然后在做离线或在线解算即可。

按此方式，对多星参与校正情形，就获得了最优加权系数，能够补偿因星点星等不同和所处视场位置不同带来的误差。

本节以 AST-301、HAST 和用于 geoCARB 使用的双头星敏感器(STS)系统为

例(van Bezooijen R. W. H. et al. 2013)，来说明上述整体不确定度对于校正的用处。不确定性和校正的残差参见表 6.2。

表 6.2 典型星敏感器误差分配(1σ, arcsec)

| 型号 | 偏置残差 | 视轴误差 | NEA(噪声、运动、均漂效应) | 角速度抖动 |
|---|---|---|---|---|
| AST-301(X 轴) | 0.103 | ≤0.03 | 0.1248 | 0 |
| HAST | ≤0.11 | ≤0.1 | ≤0.11 | - |
| STS(X 轴) | 1.66 | - | 2.48 | 1.07 |

AST-301 使用一个最优的方差模型和最小二乘拟合法估计和补偿有效焦距的退化和光学畸变。X 轴校正后的偏差为 0.103 角秒，与焦距对 NEA 的贡献一样，偏差会增加 3%达到 0.113 角秒。将这个偏差与 NEA 0.1248 角秒相比较可知这些方法把偏置误差降低大约 10%左右。因此，每个轴在平均星空区域的最终误差可表述为偏置误差和 NEA 的合成，为 0.168 角秒。视轴对齐误差有限，未包括在整个轴的误差当中，因为整个任务时间段内温度变化小于 0.25 摄氏度。

必须指出的是 AST 校正使用了较宽范围的星等数据。如公式(6.12)所示，如果使用足够多的亮星做校正的话，这个校正后的残差可以更小一些。

HAST 获得了很好的性能，并为在轨退化提供了足够的误差余额。尽管剩余偏差与 NEA 都是 0.11arcsec，HAST 为焦距退化留下了足够的空间。另外，STS 星场依赖的偏差为 1.66 角秒也低于 2.48 角秒的 NEA。这个分析表明校正的偏差由 NEA 限制，NEA 可以视为星场校正偏差的上限，这个功能与 CRLB 类似。

这些分析也可以反推 CRLB 可能是估计许多参数引入的实时位置误差的有效工具，因而可用于偏差校正。换句话说，这个研究能够为许多因素建立精确的误差边界以满足在轨校正的需求。这些实用的误差约束显示许多有用特性，在发射前省去大量仿真。另外一个应用是可以使用它优化姿态输出，同样可以获得精度提升，详细参见第七章。在精确补偿空域误差等偏置误差后，在轨校正后的星敏感器姿态精度有望接近 CCD 器件的理论精度，由此得到推论 3。

**推论 3：** 甚高精度星敏感器可定义为接近探测器 CRLB 界的星敏感器

探测器理论精度(而不是光学器件)才是星敏感器的根本瓶颈，这个精度可以通过前述优化理论降低，从而获得更高的精度、动态性能和姿态输出率。

## 6.4 本章小结

本章通过对在轨运行参数误差建模，并对全参数 CRLB 理论进行研究后，得到的主要结论如下：

(1)CRLB 定位精度指标适宜约束若干在轨运行参数或设计参数

通过分析目前在轨校正中存在的问题：如校正数据是否有效，校正到何种程度等，提出 CRLB 误差约束思想，为彻底解决高斯参数关联导致的耦合误差，在此基础上提出 CRLB 误差全约束理论，得到了高斯信号全部的 CRLB 指标，这既包括位置误差，又包括亮度和尺度误差。以此约束速度、焦距漂移、恒星中心光谱变化、速度抖动等设计或在轨运行参数，解析地给出了这些参数的误差边界。这些边界可直接用于星敏感器设计，能够节省一定的仿真试验，避免精度计算错误。

从理论层面弄清了一些常规技术的运行机制，如 TDI 技术、可变帧率技术、小视场技术、区域校正技术等。对于不能补偿的焦距漂移误差和空域畸变误差，可使用分段轨道(温度)校正技术，对于温度区之间的畸变信息可使用插值方法。

另外，分析表明 CRLB 约束相比 NEA 约束指标，更有理论依据和现实基础。星敏感器的理论指向精度受制于探测器而不是光学器件，探测器的 CRLB 界是星敏感器设计精度不可逾越的红线。

(2)对视场和星等加权方法可提高在轨或地面校正精度

将各在轨运行或设计参数所产生的误差加在一起即为总体位置误差，理论研究表明可根据星点星等和所处视场位置不同给予不同的加权系数，加权系数反比于星点总体位置误差。

# 第七章 姿态确定技术优化

前述定位和校正技术研究主要用于降低单星定位误差，而姿态确定算法是以多星为基础的定位算法。由前文研究可知，各星点的单星定位误差并不一致，而现阶段的姿态确定算法却较少考虑这一因素，可以利用这一信息进一步提高指向精度和算法的鲁棒性。基于前期的 CRLB 理论研究，本章提出一种实用的改进方法，能直接加入到现有的程序流程中，相对于目前加权方法，姿态确定精度有不同程度提升。

## 7.1 姿态确定算法分析

目前星敏感器主要使用 QUEST 或类似算法给出姿态信息，用的是星图匹配成功的跟踪星点。这些星点的定位精度决定了输出姿态的精度。已知星点天球参考矢量为 $\mathbf{r}_i$，通过星敏感器测量得到的对应观测矢量为 $\tilde{\mathbf{b}}_i$，卫星姿态方向余弦阵为 $A$，$e_i$ 为测量误差，满足：

$$e_i^T \mathbf{b}_i = 0, \mathrm{E}(e_i) = 0, \mathrm{var}(e_i) = \sigma_i^2 \tag{7.1}$$

目前的姿态确定算法可表述为：

$$\tilde{\mathbf{b}}_i = A\mathbf{r}_i + e_i = \mathbf{b}_i + e_i \tag{7.2}$$

该问题为 Wahba 问题，需求解正交矩阵 $A$，使得指标函数：

$$J(A) = \sum_i \frac{1}{2} \beta_i (A\mathbf{r}_i - \mathbf{b}_i)(A\mathbf{r}_i - \mathbf{b}_i)^T \beta_i, \tag{7.3}$$

达到最小，其中 $\beta_i$ 为加权因子，满足：$\beta_i^2 = \sigma_i^{-2} / \sum_i \sigma_i^{-2}$。

目前姿态估计时域内采用 KF 算法抑制的测量误差，该滤波算法是个时域滤波器，并未针对空域的多个星点进行优化，暗含假定 $\sigma_i^2 = \sigma_j^2$，即 $\beta_i^2 = \beta_j^2$ (Markley F. L. 2000; Crassidis J. L. 2002; Chen Z. 2003)。由上可知，目前姿态确定算法空域采用的是平均加权，即不论星点在视场何处提取，不同星点星等如何，提取精度应相同。但第六章推论 1 和 2 表明，在探测器阵列噪声和响应曲线不变情况下，亮星定位误差小，暗星定位误差大(图 7.1 右)，视场中心视轴偏差小，边缘大(图 7.1 左)，所以，上述假设并不成立。由此可设想，如果能精确地估计星点提取精

度信息并将其引入 QUEST 算法，相应的姿态确定精度定会提升。本节主要考虑 3 种大的难以校正的因素，即运动、噪声和温致焦距漂移导致的误差。

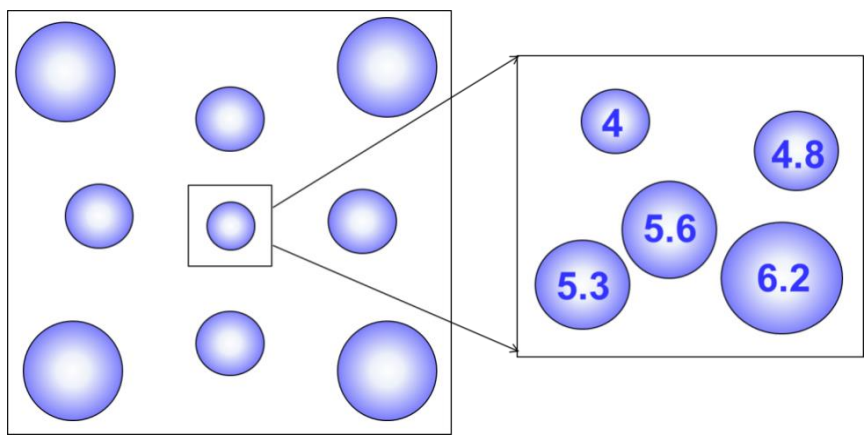

图 7.1 焦距漂移误差(左图)和星点误差示意图(右图，数字单位为星等)

## 7.2 优化方案

由第五章可知，角速度导致误差与噪声是耦合在一起的，该项可表述为：

$$\sigma_v = \alpha \frac{\sigma'_{sv}}{R'}, R' = \frac{I_0}{\sqrt{K}\sigma'_n} \tag{7.4}$$

设 $K$ 星点提取窗口大小(Liebe C. C. 2002)。由文献可知，本节假定 $I_0 \ll K\sigma'^2_n$，即背景和读出噪声占主导，，$\alpha = \sqrt{2}, R'^2 \approx \frac{I_0^2}{2\sqrt{\pi}\sigma'^2_n\sigma'_{s,a}}$。为简化，方差仍然保留在星点信号中。得到星点定位误差精度对于估计加权系数十分重要，在补偿运动误差后，星点定位精度可达CRLB值。

许多因素导致焦距的漂移，由第六章知，该焦距漂移量导致的误差为：

$$\sigma_f = \frac{kd}{f}\delta f \tag{7.5}$$

其中，$d$ 为像元尺度，$f$ 为有效焦距，$\delta f$ 为焦距漂移量。

由此可知，两种误差导致的总和为：

$$\sigma^2 = \sigma_v^2 + \sigma_f^2 = \frac{2\sigma'^2_s}{R'^2} + \frac{k^2 d^2}{f^2}\delta f^2 \tag{7.6}$$

即在速度和焦距漂移两种误差下，星点定位精度应为二者之和(Zhang J. et al 2016)。仍然利用第六章结论，可以得到如下优化方案。

由 $\beta_i^2$ 定义，可得，

$$\beta_i^2 = \frac{1/(\sigma_{v,i}^2 + \sigma_{f,i}^2)}{\sum_i 1/(\sigma_{v,i}^2 + \sigma_{f,i}^2)} \tag{7.7}$$

其中， $\sigma_{v,i} = \alpha_i \dfrac{\sigma'_{s,i}}{R'}$ ， $\sigma_{f,i} = \dfrac{k_i d}{f} \delta f$ 。

若探测器本底噪声、星点尺度一致， $\alpha_i = \alpha_j$ ，不考虑焦距漂移，静态，则 $\beta_i$

可简化为： $\beta_i^2 = \dfrac{I_{0,i}^2}{\sum_i I_{0,i}^2}$ ，此处的 $\beta_i$ 与亮度成一定关系。由于视场内多星参与姿态确定，弱星占比大，而亮星占比小，中心视场区域小，而边缘面积大。因而，该技术能够较好的解决原有的 QUEST 算法各星等同等加权和视场同等加权问题，改进的算法通过提高视场中心和亮星的权值，姿态精度进一步提高。实施上，软件需要作出如下关键步骤修改：(1)星点提取环节，除计算星点位置外，实时计算多颗星点亮度和视轴误差等数据并保存；(2)在星图匹配成功后，利用上述加权系数法改进 QUEST 算法或其他姿态确定方法，再利用 QUEST 算法输出姿态；由此改进的 QUEST 算法提高了姿态精度，适用于所有星敏感器产品。

## 7.3 仿真分析

为验证方法的正确性，做了一系列仿真试验，该试验测定各个算法对姿态确定精度的影响。紧跟其后的是姿态改进试验，证明该策略的有效性。

所有仿真测试中的星点，其星等 $m$ 服从 $6.58e^{1.08m}$ 分布(Liebe C. C. 2002)，加入高斯噪声和焦距漂移误差。然后用 COG 和 IWCOG 算法计算质心，再确定每个星的位置。得到位置和星等信息后，可用常规 QUEST 算法输出姿态，该算法采用平等加权策略，也可用新的算法输出姿态，加权系数可由前文公式得出。整个试验重复 M 次以获得稳定的统计信息。所有仿真参数和取值范围参见表 7.1。

如不加注释，采用四元数误差作为评估加权策略和定位算法的根本手段。

表 7.1 仿真参数和值

| 参数 | 值 | 参数 | 值 |
|---|---|---|---|
| 星等(Mv) | 4.5-7.5 | 高斯噪声 | $N(0,4)$ |
| 焦距和焦距漂移($\mu m$) | 50000,2 | 像素尺度 | $20'',10\mu m$ |
| 单次积分时间 | 0.1 | 星跟踪数 | 26 |
| M | 400 | 有效视场(pixel) | $8^\circ \times 8^\circ (1024 \times 1024)$ |

表 7.2 姿态四元数改进

| 四元数 | (1) | (2) | (a) | (b) | (c) | (d) |
|---|---|---|---|---|---|---|
| $q_1$ | 0.8695 | 0.8384 | 0.8707 | 0.2179 | 0.8525 | 0.2121 |
| $q_2$ | 0.9194 | 0.9010 | 0.8713 | 0.2263 | 0.8527 | 0.2211 |
| $q_3$ | 0.8902 | 0.8542 | 0.8715 | 0.2258 | 0.8537 | 0.2208 |
| $q_4$ | 0.8891 | 0.8546 | 0.8694 | 0.2254 | 0.8513 | 0.2202 |

(1) 和 (2) 表示图 7.1 姿态改进均值, (a)-(d) 表示图 7.2 同样数据

假设方法 1 表示目前用的等值加权策略，方法 2 为 CRLB 加权策略，方法 3 为 CRLB 加上 FOV 加权策略。图 7.2 显示了 3 种算法估计的四元数误差，其中左图比较方法 1 和方法 2，右图比较方法 2 和方法 3。

设第一个下标 c 表示 COG 算法，i 表示 IWCOG 算法；第 2 个下标 u 表示方法 1，r 为方法 2，f 为方法 3，则 $\sigma_{cu}$ 表示利用 COG 做质心算法、方法 1 加权策略所得的标准误差。为量化新加权策略性能，做了以下四种比较： (a) $\eta = \sigma_{ir}/\sigma_{iu}$ ; (b) $\eta = \sigma_{ir}/\sigma_{cu}$ ; (c) $\eta = \sigma_{if}/\sigma_{iu}$ ; (d) $\eta = \sigma_{if}/\sigma_{cu}$ .

重复第 1 个方针所有例程，其中每次仿真生成的星点均具有不同的星等和位置，姿态改进参见图 7.3。 该改进的均值显示在表 7.2 中，其中 (1) 和 (2) 表示图 7.2 中的姿态精度改进，(a)-(d) 表示图 7.3 中的姿态改进。

从表格 7.2 中 (1) 、(2) 和图 7.2 可知，方法 2 和方法 3 显示比常规算法更有优势。考虑 FOV 加权后，姿态又改进 3% 左右。最后，该方法能在现有软件少许改动基础上降低姿态误差约 15%。

图 7.3 显示了与图 7.2 相同的趋势。尽管有些不好的星点会增加姿态误差，大部分案例都显示改进的效果，该方法能获得约 14% 的姿态精度增长。另外，所有样本显示 IWCOG 算法能够得到更好的效果，想比于 COG 法，这个改进在 78% 左右。

表 7.3 不同点高位算法和加权策略的精度改进

| Sky zone | A(pixel) | B(pixel) | C(pixel) | B/A | C/B | C/A | C/A limit |
|---|---|---|---|---|---|---|---|
| sky1 | 0.061 | 0.031 | 0.027 | 50.81% | 87.1% | 44.26% | 37.1% |
| sky2 | 0.058 | 0.031 | 0.026 | 53.45% | 83.87% | 44.83% | 37.2% |
| sky3 | 0.065 | 0.042 | 0.033 | 64.62% | 78.57% | 50.77% | 38.6% |
| sky4 | 0.060 | 0.032 | 0.026 | 53.33% | 81.25% | 43.33% | 36.1% |

A:COG+方法 1, B:IWCOG+ 方法 1,C:IWCOG+ 方法 2.

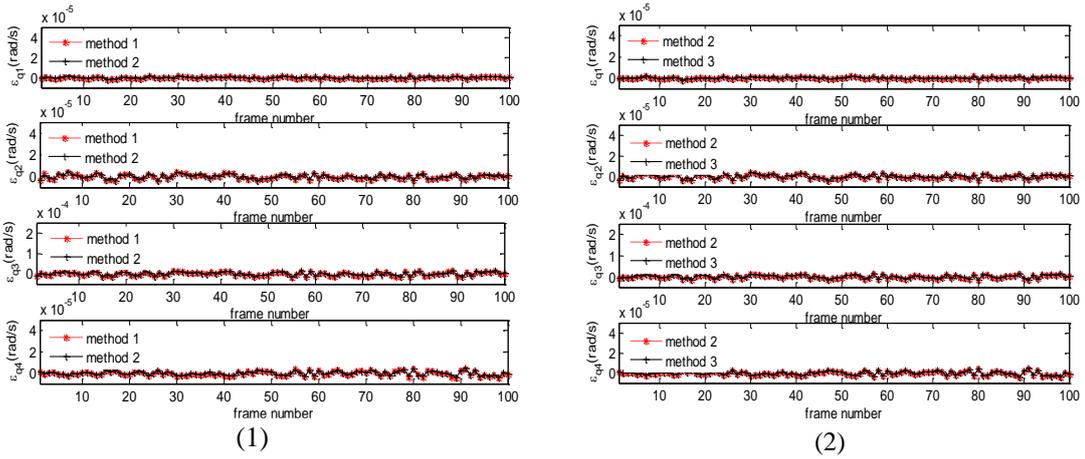

图 7.2 不同加权策略四元数估计误差

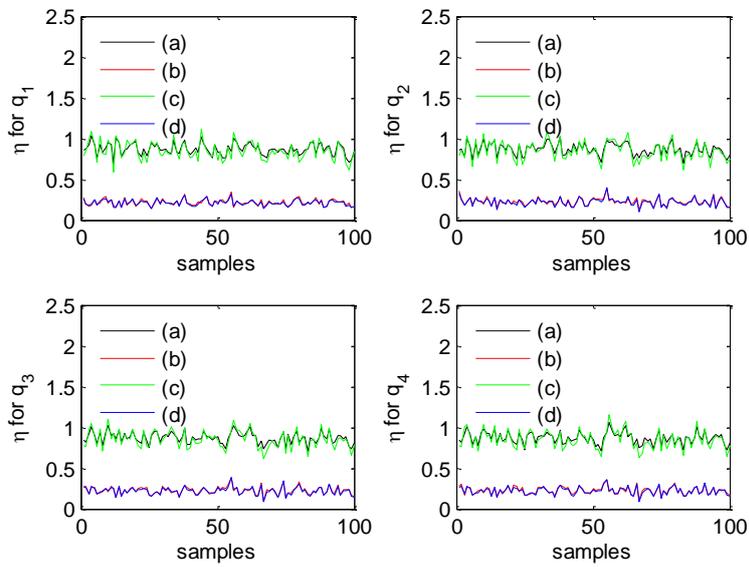

图 7.3 不同加权策略和定位技术的改进比较

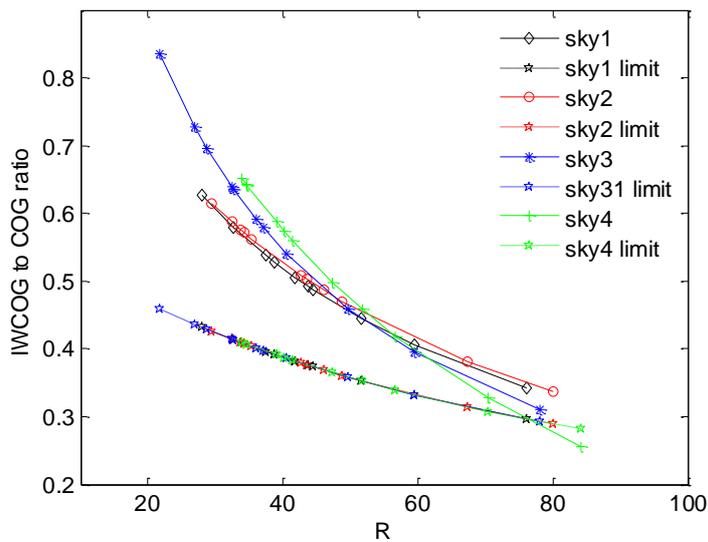

图 7.4 四个天区不同定位技术下的姿态改进

以此类推，仿真了地面可能的测试结果。星图内同时捕获多颗星，均参与姿态输出。同时计算在这些参数下应该可获得的理论改进值。不同天区星点的测试误差参见图 7.4。可以看出，信噪比高的星点性能接近 CRLB 理论值。由前文可知，这是因为，亮星施加了更大的加权系数。因此，QUEST 算法同时验证了方法 1 和 2。每个算法的姿态改进参见表格 7.3。

从表 7.3 可知，CRLB 加权法降低误差 18% 左右，低于表 7.2 的估计值 22%。考虑到 IWCOG 算法贡献 55% 左右姿态改进，新的 QUEST 算法降低误差约 44%。尽管这个值低于 37% 理论值，考虑到实际仿真和加权策略问题，这个改进是可以接受的。改进的 QUEST 算法适用于所有星敏感器产品。地面测试受平台振动和大气视宁度影响，改进可能会比仿真值低一点。

## 7.4 本章小结

通过对当前 QUEST 算法空域滤波分析和仿真研究，可知对视场和星等加权能进一步提高姿态确定精度。

QUEST 姿态确定算法存在星等同等加权和视场同等加权问题，应用第六章的部分推论，改进的算法可实时计算各个星点加权系数，提高视场中心和亮星的权值，加大了亮星和视场中心星点对最终姿态确定算法精度的贡献，最终提升的精度在 10-20%。由于跟踪环节也涉及星点定位，这个加权策略也适用于局部天区星图跟踪识别环节。

# 第八章 总结和展望

## 8.1 研究成果总结

由于高端星敏感器产品尤其是甚高精度星点定位一直是悬而未决的问题。本课题基于自然科学基金(批准号 61174004）和国家重大仪器发展专项项目背景,以实现甚高精度星点定位为目标,开展了关于星敏感器产品定位理论和校正技术方面的基础研究,试图找到提高现有甚高精度星敏感器产品精度的新途径。论文涉及星点提取、星点追踪和姿态确定等决定姿态精度的主要技术,主要研究成果有:

(1)基于自然基金"高精度星敏感器星像分布对精度的影响研究"(批准号61174004)思路提出一种根本性的星点误差降低方法。首先论证了频域分解法只适合解决非星敏感器导致的低频误差,而不是降低星敏感器自身误差的主要方式。对与星敏感器自身相关的误差源,包括算法、器件噪声、光学镜头畸变、CCD 器件的光响应不一致、温致焦距漂移、像移、速度畸变和抖动等,提出对各误差源所产生的单星定位误差建模的方法,为研究根本性地降低误差方法提供了统一的思路,第四、五、六章成功地应用该思路计算了各种因素导致的定位误差。

(2)针对算法导致的星点提取误差,获得了迭代加权质心定位算法(IWCOG)优化算法和差分 IWCOG 算法。在理想高斯形貌下算法导致的 S 误差可以忽略,经验证是目前为止最好的算法,可用于甚高精度星点提取。

(3)针对探测器噪声导致的星点提取误差,通过 CRLB 理论获得了探测器的理论定位精度,揭示了 IWCOG 算法在高斯噪声下的最优性和泊松噪声下的近优性。针对非高斯星点形貌畸变导致的定位误差,通过研究高采样 PSF 数据匹配算法获得了相应的理论精度。这些理论定位精度将为星敏感器器件选型、参数设计、模型畸变校正等提供直接理论依据。

(4)针对运动和速度抖动引入的定位误差,获得了动态情况下一般性解析补偿公式,能够完全补偿由速度、视场、加速度引入的运动畸变,解决了单帧运动补偿问题。基于上述解析补偿公式提出一种新的全精度高动态技术,能够将精度、

动态、姿态更新率三个指标解耦，可实现三个指标同时提高，为未来更先进的星敏感器研制提供了理论基础和实现方案。

(5)针对焦距漂移、星点光谱中心变化等引入的定位误差，基于探测器理论定位精度公式提出 CRLB 在线误差约束机制，能够在线计算各参数的边界条件，据此判断运行参数处于什么状态，以及相关星点数据是否有效并用于后续的校正，为地面粗平场校正、在轨误差校正提供了理论依据和实现策略。这项研究同时表明探测器理论精度(而不是光学器件)才是星敏感器的根本瓶颈，为甚高精度星敏感器设计提供了可能的改进方向。针对 QUEST 算法未能有效空域滤波问题，基于前述研究提出一种对星等、视场加权的姿态确定优化算法，实现了对视场和星等引入误差的在线补偿。

围绕定位和校正技术，上述研究包含 2 个主要创新点：

(1)获得了可用于甚高精度星点提取的 IWCOG 优化算法及差分 IWCOG 算法

证实 IWCOG 算法和差分 IWCOG 算法在近高斯形貌、高斯噪声下属于无偏估计算法，在综合要求效率和精度两个指标场合下，比最小二乘高斯拟合法、高斯拟合法、质心法及改进方法、频率补偿法等有很大优势，是目前为止最优异算法，将解决静态、动态很多情况下噪声、像移等导致的星点难以精确提取问题。

(2)将 CRLB 理论应用于星点静态、动态定位和在轨校正环节，导出相应环节下星点定位理论精度，实现了其在星敏感器领域的应用

应用 CRLB 理论获得的阶段性研究进展有：

(a)CRLB 理论可作为定位算法和一些动态技术优越性的判断标准。如获得的星点静态定位理论精度证明了 IWCOG 类算法的最优性，获得的星点动态定位理论精度证明 IWCOG 类算法在静态、动态等情况下具有其他算法难以比拟的技术优势，指出了单帧解析补偿技术和多帧解析补偿全精度技术应能达到的精度。

(b)提出的 CRLB 误差全约束理论给出了判断星点数据是否能用于在轨校正的基本标准，指出了在轨焦距校正和基于多项式拟合的几何畸变校正理应达到的定位精度。该约束理论能实时计算许多运行参数的在线边界，可应用于在轨和地面校正环节。

(c)CRLB 理论能够实时计算校正、跟踪和姿态确定等多星参与环节的多星加权系数。在姿态确定方程中，用 CRLB 理论实时计算中心视场星点和亮星的加权

系数，突出了这两个因素对定位精度的贡献，仿真显示该技术直接提高指向精度10%-20%。

## 8.2 展望

本文将 CRLB 理论引入到星敏感器的动态定位和校正环节，导出了星敏感器设计和在轨运行参数与星点定位精度的定量关系，这些方面虽是关于甚高精度星敏感器星点定位的基础理论研究，但其应用并不限于本领域。在其他领域，如 Shack-Hartmann 传感器波前定位、热核聚变激光点火、无线通讯定位等需要精确点定位的场合，本文研究的算法和思路可作参考。

鉴于 IWCOG 类算法和 CRLB 理论的出色表现，以及尚未深入的问题，未来可在以下几个方面展开进一步研究：

在星点定位算法上，无论是 IWCOG 类算法、质心法、最小二乘高斯拟合算法等均以星点形貌近似符合高斯对称分布为假设前提，实际星点受光学像差等因素影响，星点 PSF 多偏离高斯对称分布，由此导致的质心偏差已不可忽略，需要研究相关算法获得精确的质心偏差。由 CRLB 理论可知，如何实现质心偏差的精确提取同时达到 CRLB 定位精度是算法设计者应努力的目标，本文分析的 PSF 高采样数据匹配法在提取偏差上有优势，但精度上有所下降，可知要获得目标算法仍需要进一步努力。

在动态技术上，提出的多帧解析补偿动态技术属于一种新的全精度技术，国外没有进行过相关报道，该技术是否真实可行需要进一步试验验证。

在相关畸变在轨校正上，目前国内对于温致低频误差、空域低频误差的校正取得了部分成果，部分参数的校正精度有待进一步验证。本文研究虽为后续甚高精度在轨校正工作提供了一些理论基础和可能的潜在技术，但该理论是否有助于将空域畸变等校正到甚高精度，是否存在更好的理论也需要国内人员进一步探索。

# 附录 A

附录 A 计算出高斯信号全部参数在各种噪声下 CRLB 误差。

星点采样样本为 $\{s_m\}$，$\boldsymbol{\theta} = \{\theta_j\}$ 为模型参数，$\lambda_m(\boldsymbol{\theta})$ 为模型估计得参数值，每个像元 $m$ 收集的光电子 $s_m$ 包括信号 $\lambda_m(\boldsymbol{\theta})$ 和噪声 $e_m$ 两部分。$\lambda_m$ 服从高斯形貌分布 $I_0 g(x_c)$，噪声包括泊松噪声 $P(\lambda_n, \lambda_g)$ 和高斯读出噪声 $N(0, \sigma_n^2)$，由此可得，

$$\mathrm{E}(s_m) = \lambda_m(\boldsymbol{\theta}) + \lambda_n = \lambda_m', \mathrm{var}(s_m) = \lambda_m' + \sigma_n^2 \tag{A1}$$

公式(A1)显示 $\{s_m\}$ 既非高斯也非泊松类型分布，为能计算 CRLB 值，需要改进该模型，使其符合高斯或泊松类型分布，并满足公式(A1)。

## A.1: 泊松似然函数信号模型

这种情形下，公式(A1)可改写为：

$$\mathrm{E}(s_m + \sigma_n^2) = \lambda_m' + \sigma_n^2, \mathrm{var}(s_m + \sigma_n^2) = \lambda_m' + \sigma_n^2 \tag{A2}$$

由公式(A2)可得，$s_m + \sigma_n^2$ 项值均值和方差相等，符合泊松分布定义。由此可得似然分布函数为：

$$p(\boldsymbol{\theta} | \{s_m\}) = \prod_{m=1}^{M} p(s_m | \boldsymbol{\theta}, \sigma_n), p(s_m | \boldsymbol{\theta}, \sigma_n) = \frac{e^{-(\lambda_m' + \sigma_n^2)}(\lambda_m' + \sigma_n^2)^{(s_m + \sigma_n^2)}}{(s_m + \sigma_n^2)!} \tag{A3}$$

从数学上讲，可以改为 log 似然函数形式，于是可得，

$$\begin{aligned} l(\boldsymbol{\theta} | \{s_m\}) &= \ln p(\boldsymbol{\theta} | \{s_m\}) \\ &= \sum_m \left[ -(\lambda_m' + \sigma_n^2) + (s_m + \sigma_n^2)\ln(\lambda_m' + \sigma_n^2) - \ln(s_m + \sigma_n^2)! \right] \end{aligned} \tag{A4}$$

最大似然函数求解矢量 $\boldsymbol{\theta}$ 应满足：$\partial l(\boldsymbol{\theta} | \{s_m\}) / \partial \boldsymbol{\theta} = 0$。利用公式(A4)，可化简为：

$$\sum_m \frac{s_m - \lambda_m'}{\lambda_m' + \sigma_n^2} \frac{\partial \lambda_m}{\partial \boldsymbol{\theta}} = 0 \tag{A5}$$

由 Fisher 信息矩阵定义，可得，

$$F_{jj'} = -\mathrm{E}\left(\frac{\partial^2 l}{\partial \theta_j \partial \theta_{j'}}\right) = -\mathrm{E}\left(\frac{\partial}{\partial \theta_{j'}} \sum_m \frac{s_m - \lambda_m'}{\lambda_m' + \sigma_n^2} \frac{\partial \lambda_m}{\partial \theta_j}\right) = \tag{A6}$$

$$-\mathrm{E}\left(\sum_m\left[\left(\frac{-1}{\lambda'_m+\sigma_n^2}-\frac{s_m-\lambda'_m}{(\lambda'_m+\sigma_n^2)^2}\right)\frac{\partial\lambda_m}{\partial\theta_{j'}}\frac{\partial\lambda_m}{\partial\theta_j}+\frac{s_m-\lambda'_m}{\lambda'_m+\sigma_n^2}\frac{\partial^2\lambda_m}{\partial\theta_j\partial\theta_j}\right]\right)$$

将(A1)带入(A6)，$F_{jj'}$ 可化简为：

$$F_{jj'}=\sum_m\frac{1}{\lambda'_m+\sigma_n^2}\frac{\partial\lambda_m}{\partial\theta_{j'}}\frac{\partial\lambda_m}{\partial\theta_j}=\sum_m\frac{1}{\lambda_m+\sigma_n'^2}\frac{\partial\lambda_m}{\partial\theta_{j'}}\frac{\partial\lambda_m}{\partial\theta_j},\sigma_n'^2=\sigma_n^2+\lambda_n \qquad (A7)$$

**A.2: 高斯似然函数信号模型**

这种情形下，公式 (A1) 可改写为：

$$\mathrm{E}(s_m-\lambda'_m)=\mathrm{E}(e_m-\lambda_n)=0,\mathrm{var}(s_m-\lambda'_m)=\mathrm{var}(e_m-\lambda_n)=\lambda_n+\sigma_n^2\triangleq\sigma_n'^2 \qquad (A8)$$

由公式(A8)可得，$s_m-\lambda'_m$ 项均值为 0 和方差 $\sigma_n'^2$，可以假设符合高斯分布定义。

由此可得似然分布函数为：

$$p(\boldsymbol{\theta}\,|\,\{s_m\})=\prod_{m=1}^M p(s_m\,|\,\boldsymbol{\theta},\sigma'_n),p(s_m\,|\,\boldsymbol{\theta},\sigma'_n)=\frac{1}{\sqrt{2\pi}\sigma'_n}\exp\left[-\frac{(s_m-\lambda'_m)^2}{2\sigma_n'^2}\right] \qquad (A9)$$

log 似然函数为：

$$l(\boldsymbol{\theta}\,|\,\{s_m\})=\ln p(\boldsymbol{\theta}\,|\,\{s_m\})=\sum_m\left[-\ln\sqrt{2\pi}\sigma'_n-(s_m-\lambda'_m)^2/2\sigma_n'^2\right] \qquad (A10)$$

利用 $\partial l(\boldsymbol{\theta}\,|\,\{s_m\})/\partial\boldsymbol{\theta}=0$，关于矢量 $\boldsymbol{\theta}$ 求解为：

$$\sum_m\frac{s_m-\lambda'_m}{\sigma_n'^2}\frac{\partial\lambda_m}{\partial\boldsymbol{\theta}}=0 \qquad (A11)$$

Fisher 信息矩阵是，

$$F_{jj'}=-\mathrm{E}\left(\frac{\partial^2 l}{\partial\theta_{j'}\partial\theta_j}\right)=-\mathrm{E}\left(\sum_m\left[\frac{-1}{\sigma_n'^2}\frac{\partial\lambda_m}{\partial\theta_{j'}}\frac{\partial\lambda_m}{\partial\theta_j}+\frac{s_m-\lambda'_m}{\sigma_n'^2}\frac{\partial^2\lambda_m}{\partial\theta_j\partial\theta_j}\right]\right) \qquad (A12)$$

将(A1)带入(A12)，$F_{jj'}$ 化为：

$$F_{jj'}=\sum_m\frac{1}{\sigma_n'^2}\frac{\partial\lambda_m}{\partial\theta_{j'}}\frac{\partial\lambda_m}{\partial\theta_j} \qquad (A13)$$

比较公式(A13)和(A7)可得，高斯似然函数的 Fisher 信息矩阵是泊松似然函数的特殊情形，具体是哪种取决于信号还是背景占优势（$\lambda_m\ll\sigma_n'^2$ 或 $I_0\ll M\sigma_n'^2$).

**A.3: Fisher 矩阵的解析形式**

可令

$$\boldsymbol{\theta}=\{\theta_j\}=\begin{bmatrix}I_0 & x_c & \sigma_s^2\end{bmatrix}' \qquad (A14)$$

$$F = F' = \begin{bmatrix} F_{I_0,I_0} & F_{I_0,x_c} & F_{I_0,\sigma_s^2} \\ F_{x_c,I_0} & F_{x_c,x_c} & F_{x_c,\sigma_s^2} \\ F_{\sigma_s^2,I_0} & F_{\sigma_s^2,x_c} & F_{\sigma_s^2,\sigma_s^2} \end{bmatrix} \tag{A15}$$

$$\frac{\partial \lambda_m}{\partial I_0} = g_m, \frac{\partial \lambda_m}{\partial x_c} = \frac{I_0 \varphi_m}{\sigma_s^2} g_m = p_x g_m, \frac{\partial \lambda_m}{\partial \sigma_s^2} = \frac{I_0(\varphi_m^2 - \sigma_s^2)}{2\sigma_s^4} g_m = p_\sigma g_m, \varphi_m = x_m - x_c \tag{A16}$$

为进一步推导 CRLBs，计算分为以下两种情形：(a) 信号主导过程，

$\lambda_m \gg \sigma_n'^2$；(b) 背景主导过程，$\lambda_m \ll \sigma_n'^2$。

(a) 信号主导过程

为方便，$1/(\lambda_m + \sigma_n'^2)$ 可泰勒展开为 0 阶项和 1 阶项：

$$\frac{1}{\lambda_m + \sigma_n'^2} \approx \frac{1}{\lambda_m} - \frac{\sigma_n'^2}{\lambda_m^2} = c_{0m} + c_{1m} \tag{A17}$$

由公式(A17)，公式(A15) 可化为：

$$F \approx F_{c0} + F_{c1}, F_{ck} = \begin{bmatrix} \sum_m c_{km} g_m^2 & \sum_m c_{km} p_x g_m^2 & \sum_m c_{km} p_\sigma g_m^2 \\ \sum_m c_{km} p_x g_m^2 & \sum_m c_{km} p_x^2 g_m^2 & \sum_m c_{km} p_x p_\sigma g_m^2 \\ \sum_m c_{km} p_\sigma g_m^2 & \sum_m c_{km} p_x p_\sigma g_m^2 & \sum_m c_{km} p_\sigma^2 g_m^2 \end{bmatrix}, k = 0,1 \tag{A18}$$

化简后，$F$ 矩阵元素可写为解析形式，这方便推导 CRLBs。由此可得，

$$F_{c0} = \begin{bmatrix} \dfrac{1}{I_0} & 0 & 0; 0 & \dfrac{I_0}{\sigma_s^2} & 0; 0 & 0 & \dfrac{I_0}{2\sigma_s^4} \end{bmatrix} \tag{A19}$$

和

$$F_{c1} = \begin{bmatrix} \dfrac{-M\sigma_n'^2}{I_0^2} & -\dfrac{\sigma_n'^2}{I_0\sigma_s^2}\sum_m \varphi_m & -\dfrac{\sigma_n'^2}{2I_0\sigma_s^4}\sum_m(\varphi_m^2 - \sigma_s^2) \\ -\dfrac{\sigma_n'^2}{I_0\sigma_s^2}\sum_m \varphi_m & -\dfrac{\sigma_n'^2}{\sigma_s^4}\sum_m \varphi_m^2 & -\dfrac{\sigma_n'^2}{2\sigma_s^6}\sum_m(\varphi_m^3 - \varphi_m\sigma_s^2) \\ -\dfrac{\sigma_n'^2}{2I_0\sigma_s^4}\sum_m(\varphi_m^2 - \sigma_s^2) & -\dfrac{\sigma_n'^2}{2\sigma_s^6}\sum_m(\varphi_m^3 - \varphi_m\sigma_s^2) & -\dfrac{\sigma_n'^2}{4\sigma_s^8}\sum_m(\varphi_m^2 - \sigma_s^2)^2 \end{bmatrix} \tag{A20}$$

公式(A19)可以改写为信噪比的函数。信噪比按公式(2.16)定义，可表述为：

$R^2 = \sum_m \lambda_m^2 / (\lambda_m + \sigma_n'^2)$。由 $\lambda_m \gg \sigma_n'^2$，可得，

$$R^2 = \sum_m \lambda_m^2 / (\lambda_m + \sigma_n'^2) = \int \lambda_m = I_0 \tag{A21}$$

$F_{c0}$ 即化为：

$$F_{c0} = \left[ \begin{array}{ccccccc} \dfrac{R^2}{I_0^2} & 0 & 0; 0 & \dfrac{R^2}{\sigma_s^2} & 0; 0 & 0 & \dfrac{R^2}{2\sigma_s^4} \end{array} \right] \tag{A22}$$

公式(A20) 显示 $F_{c1}$ 矩阵交叉项要么是奇函数，要么趋近与中心矩。 对于以下情形： $\lambda_m \gg \sigma_n'^2$ ， $F_{c1} \propto 0 (\ll F_{c0})$ 。CRLBs 即为公式(A22)的反比，即：

$CRLB = F_{c0}^{-1}$ 。

(b) 背景主导过程

同上， $1/(\lambda_m + \sigma_n'^2)$ 可泰勒展开为：

$$\frac{1}{\lambda_m + \sigma_n'^2} \approx \frac{1}{\sigma_n'^2} - \frac{\lambda_m}{\sigma_n'^4} = c_{0m} + c_{1m} \tag{A23}$$

公式(A18)仍然成立。化简可得，

$$F_{c0} = \left[ \begin{array}{ccccccc} \dfrac{1}{2\sqrt{\pi}\sigma_s\sigma_n'^2} & 0 & U_0; 0 & \dfrac{I_0^2}{4\sqrt{\pi}\sigma_s^3\sigma_n'^2} & 0; U_0 & 0 & \dfrac{3I_0^2}{32\sqrt{\pi}\sigma_s^5\sigma_n'^2} \end{array} \right] \tag{A24}$$

和

$$F_{c1} = \left[ \begin{array}{ccc} \dfrac{-I_0}{2\sqrt{3}\pi\sigma_s^2\sigma_n'^4} & 0 & \dfrac{I_0^2}{6\sqrt{3}\pi\sigma_s^4\sigma_n'^4} \\[3mm] 0 & -\dfrac{I_0^3}{6\sqrt{3}\pi\sigma_s^4\sigma_n'^4} & 0 \\[3mm] \dfrac{I_0^2}{6\sqrt{3}\pi\sigma_s^4\sigma_n'^4} & 0 & -\dfrac{I_0^3}{12\sqrt{3}\pi\sigma_s^6\sigma_n'^4} \end{array} \right] \tag{A25}$$

其中，

$$U_0 = \frac{I_0}{2\sigma_s^4\sigma_n'^2}\sum_m (\varphi_m^2 - \sigma_s^2)g_m^2 = 0 \quad or \quad \frac{-I_0}{8\sqrt{\pi}\sigma_s^3\sigma_n'^2} \tag{A26}$$

当 $\lambda_m \ll \sigma_n'^2$ ，信噪比按公式(2.17)-(2.18)定义，可表述为：

$R^2 = \sum_m \dfrac{\lambda_m^2}{\lambda_m + \sigma_n'^2} = \int \dfrac{\lambda_m^2}{\sigma_n'^2} = \dfrac{I_0^2}{2\sqrt{\pi}\sigma_s\sigma_n'^2}$ 。 公式(A24)可表述为：

$$F_{c0} = \left[ \begin{array}{ccccccc} \dfrac{R^2}{I_0^2} & 0 & 0; 0 & \dfrac{R^2}{2\sigma_s^2} & 0; 0 & 0 & \dfrac{3R^2}{16\sigma_s^4} \end{array} \right], U_0 = 0 \quad or \quad \dfrac{-R^2}{4I_0\sigma_s^2} \tag{A27}$$

当 $U_0 \neq 0$ ，交叉项对星点亮度和尺度估计产生较大影响(22%误差增幅)。正常情形下，CRLB 与 $F_{c0}$ 成反比： $CRLB = F_{c0}^{-1}$ ，即为第六章公式(6.2)和(6.4)。

**A.4: CRLB 总结**

由上述分析可知:

(a) $\delta I_0 = \alpha_1 \dfrac{I_0}{R}, \alpha_1 = 1$ (信号或背景主导);

(b) $\delta x_c = \alpha_2 \dfrac{\sigma_s}{R}, \alpha_2 = 1$ (信号主导), $\alpha_2 = \sqrt{2}$ (背景主导);

(c) $\delta \sigma_s^2 = \alpha_3 \dfrac{\sigma_s^2}{R}, \alpha_3 = \sqrt{2}$ (信号主导), $\alpha_3 = 4/\sqrt{3}$ ( 背景主导)。

这些结果在探测器为线性器件、像元内像素响应和电荷转移无效可以忽略、信号是过采样的(高分辨率)的情形下成立。对于 2D 信号,采样窗口多在 3×3pixel 以上,可以认为是完全采样的。离散情况下的 CRLB 值应接近本文推导结果,此时可用离散结果替换公式(6.2)和(6.4)来约束其他参数。

# 附录 B

附录 B 给出第 6.3 节所有公式。

(1) 公式(6.7a)-(6.7c)

由公式(6.5)的定义，速度误差源导致的位置误差、亮度和尺度误差可限制为：

$$\delta x_v = \delta x - \delta x_n \le \eta_{x,v} \delta x_n \tag{B2}$$

$$\delta I_{0,v} = \delta I - \delta I_{0,n} \le \eta_{I,v} \delta I_{0,n}$$

$$\delta \sigma_{s,v}^2 = \delta \sigma_s^2 - \delta \sigma_{s,n}^2 \le \eta_{\sigma_s^2,v} \delta \sigma_{s,n}^2$$

例如，关于速度约束的可计算如下：

$$\delta x - \delta x_n = \frac{\sqrt{2}\sigma_s'}{R'} - \frac{\sqrt{2}\sigma_s}{R} \le \eta_{x,v} \frac{\sqrt{2}\sigma_s}{R}, R = \frac{I_0^2}{2\sqrt{\pi}\sigma_s \sigma_n'^2}$$

由此可得，$\left(\dfrac{\sigma_s'^2}{\sigma_s^2}\right)^{3/4} - 1 \approx \dfrac{3\sigma_r^2}{4\sigma_s^2} = \dfrac{v^2 T^2}{16\sigma_s^2} \le \eta_{x,v}$。

公式(6.7b)和(6.7c)可同样推得。

(2) 公式(6.8a)-(6.8b), (6.9a)-(6.9b)

星敏感器探测器焦平面主导星的方向可定义为：

$$\omega_i = \frac{1}{\sqrt{(\vec{r}_i - \vec{r}_0)^2 + f^2}} \begin{bmatrix} \vec{r}_i - \vec{r}_0 \\ -f \end{bmatrix}, \vec{r} = (x, y) \tag{B3}$$

由于焦距不能实时准确估计出来，当焦距漂移发生时，主导星的方向是不动的，可定义为：

$$\frac{\omega_{i,r}}{\omega_{i,f}} = \frac{\vec{r}_i - \vec{r}_0}{-f} = \frac{\vec{r}_i - \vec{r}_0 + \delta \vec{r}_f}{-(f + \delta f)} \tag{B4}$$

设 $\vec{r}_i - \vec{r}_0 = (k_x d, k_y d)$，公式(B4) 即推得，

$$\delta \vec{r}_f = \frac{(k_x, k_y)d}{f} \delta f, \delta \vec{r}_f = (\delta x_f, \delta y_f) \tag{B5}$$

同样，星敏感器采用离焦技术使得星点尺度与像元大小相当。可知，在某段时间内，对于一定的离焦距离，在其他参数不变情况下，其星点尺度也是一定的，即，

$$\frac{\Delta f}{\sigma_s} = \frac{(\Delta f \pm \delta f)}{\sigma_s + \delta \sigma_{s,f}} \tag{B6}$$

于是，

$$\delta \sigma_{s,f} = \mp \frac{\sigma_s}{\Delta f} \delta f \tag{B7}$$

其中±主要取决于离焦方案在焦距左边（−偏向透镜）还是右边（+偏离透镜）。

注意误差没有正负号概念。由 (B5), (B7) 和 $\delta \sigma_{s,f}^2 = (\sigma_s + \delta \sigma_{s,f})^2 - \sigma_s^2 \approx 2\sigma_s \cdot \delta \sigma_{s,f}$，可推出公式(6.9a)-(6.9b)。

(3) 公式(6.10)

对 $\sigma_s'^2 = \sigma_s^2 + \sigma_r^2$ 微分得：$2\sigma_s' \Delta \sigma_s' = 2\sigma_s \Delta \sigma_s + 2\sigma_r \Delta \sigma_r$。由 $\sigma_s = \beta \frac{\lambda f}{\pi D}, \sigma_r = \frac{vT}{\sqrt{12}}$，

该项化简为：

$$\sigma_s' = \frac{\sigma_s}{\sigma_s'} \Delta \sigma_{s,\lambda} + \frac{\sigma_r}{\sigma_s'} \Delta \sigma_r = \frac{\sigma_s}{\sigma_s'} \left| \frac{\Delta \lambda}{\lambda} \right| \sigma_s + \frac{\sigma_r}{\sigma_s'} \left| \frac{\Delta v}{v} \right| \sigma_r \tag{B8}$$

$$= \gamma_1 \sigma_s', \gamma_1 = \frac{\sigma_s^2}{\sigma_s'^2} \left| \frac{\Delta \lambda}{\lambda} \right| + \frac{\sigma_r^2}{\sigma_s'^2} \left| \frac{\Delta v}{v} \right|$$

(4) 公式(6.11a)-(6.11c)

由 $R' = \frac{I_0}{\sqrt{2\sqrt{\pi} \sigma_s' \sigma_n'}}$，可得，

$$\frac{\partial R'}{\partial \sigma_n'} = \frac{-R'}{\sigma_n'}, \frac{\partial R'}{\partial I_0} = \frac{R'}{I_0}, \frac{\partial R'}{\partial \sigma_s'} = \frac{-R'}{2\sigma_s'} \tag{B9}$$

(a) 公式(6.11a)

假设 PRNU 引入的亮度误差为：$\Delta I_0 = \zeta I_0, |\zeta| \ll 1$。

利用 $\delta I_j = \Delta \left( \frac{I_0}{R'} \right) = \Delta \left( \frac{1}{\sqrt{2\sqrt{\pi} \sigma_s' \sigma_n'}} \right)$，$\frac{\delta I_{0,n}}{\delta I_0} = \frac{R'}{R}$，可得，

$$\delta I_j = \Delta\left(\frac{I_0}{R'}\right) = -\frac{I_0}{R'^2}\left(\frac{\partial R'}{\partial \sigma'_n}\Delta\sigma'_n + \frac{\partial R'}{\partial \sigma'_s}\Delta\sigma'_s\right) + \frac{\Delta I_0}{R'} = \eta_{I_0}\delta I_0 = \eta_{I_0,j}\delta I_{0,n} \tag{B10}$$

于是，
$$\eta_{I_0} = \left[\left(\frac{\Delta\sigma'_n}{\sigma'_n}\right)^2 + \zeta^2 + \frac{\gamma_1^2}{4}\right]^{1/2}, \eta_{I_0,j} = \eta_{I_0}\frac{R}{R'}$$

(b) 公式(6.11b)

利用 $\delta x_j = \Delta\left(\dfrac{\sqrt{2}\sigma'_s}{R'}\right) = \Delta\left(\dfrac{2\pi^{1/4}\sqrt{\sigma'_s}\sigma'_s\sigma'_n}{I_0}\right), \dfrac{\delta x_n}{\delta x} = \dfrac{R'}{R}$，可得，

$$\delta x_j = -\frac{\sqrt{2}\sigma'_s}{R'^2}\left(\frac{\partial R'}{\partial \sigma'_n}\Delta\sigma'_n + \frac{\partial R'}{\partial I_0}\delta I_j + \frac{\partial R'}{\partial \sigma'_s}\Delta\sigma'_s\right) + \frac{\sqrt{2}\Delta\sigma'_s}{R'} = \eta_{x,j}\delta x_n. \tag{B11}$$

于是，
$$\eta_{x,j} = \left[\left(\frac{\Delta\sigma'_n}{\sigma'_n}\right)^2 + \frac{\eta_{I_0}^2}{R'^2} + \frac{9\gamma_1^2}{4}\right]^{1/2}\frac{R}{R'}$$

(c) 公式(6.11c)

利用 $\delta\sigma'^2_{s,j} = \Delta\left(\dfrac{4\sigma'^2_s}{\sqrt{3}R'}\right) = \Delta\left(\dfrac{4\sqrt{2}\pi^{1/4}\sqrt{\sigma'_s}\sigma'^2_s\sigma'_n}{\sqrt{3}I_0}\right), \dfrac{\delta\sigma^2_{s,n}}{\delta\sigma^2_s} = \dfrac{R'}{R}$，可得，

$$\delta\sigma^2_{s,j} = -\frac{4\sigma'^2_s}{\sqrt{3}R'^2}\left(\frac{\partial R'}{\partial \sigma'_n}\Delta\sigma'_n + \frac{\partial R'}{\partial I_0}\delta I_j + \frac{\partial R'}{\partial \sigma'_s}\Delta\sigma'_s\right) + \frac{8\sigma'_s\Delta\sigma'_s}{\sqrt{3}R'} = \eta_{\sigma^2_s,j}\delta\sigma^2_{s,n}. \tag{B12}$$

于是，
$$\eta_{\sigma^2_s,j} = \left[\left(\frac{\Delta\sigma'_n}{\sigma'_n}\right)^2 + + \frac{\eta_{I_0}^2}{R'^2} + + \frac{25\gamma_1^2}{4}\right]^{1/2}\frac{R}{R'}$$

# 参考文献

# 博士在读期间论文和专利发表情况

# 致 谢

匆匆又过 4 年，在 502 所的学习经历是令人难忘的，"天行键，君子以自强不息"，或许将星敏感器在向甚高精度推进上需要这样做，博学之、审问之、慎思之、明辨之、笃行之。博士经历对我来说是一次重要考验，在论文完成之际，向多年来关心和帮助过我的老师、同学和朋友表示诚挚地感谢。

首先要感谢的是我的导师郝云彩研究员。博士课题是在郝老师悉心指导下完成的，在论文选题、撰写、修改和相关试验上，郝老师都有精准的把握并做了细致的指导。在论文研究之初，课题进展很慢，是老师及时点拨才很快进入主题。在科学论文写作上，郝老师不厌其烦地一次次纠正语言表述上的问题，文章可读性、思想性大大加强，在此对老师的辛苦工作和耐心指导表示衷心的感谢。

其次要感谢 502 所和成像事业部的领导和同事们。502 所和事业部为作者独立展开探索研究提供了宽松的学习环境和学术氛围。吴宏鑫院士、卢欣研究员等老师在论文开题、课题阶段性评估中给予了不少的指导和建议。武延鹏研究员在课题定位算法和试验方面提供了不少帮助和支持。王立、梅志武、赵春晖、钟红军等专家对本人课题研究深入给予了相关指导意见。张春明、周建涛、刘达等同事提供了不少论文上帮助。在此对各位老师在论文进行过程中的建议和指导表示诚挚感谢。

再要感谢北京控制工程研究所人事处邱孜、宋歌、孙昕等老师对我工作和生活上的无微不至的关怀，谢谢你们！同时要感谢 502 所和中国空间技术研究院为本人提供帮助的各位同学和朋友。

最后感谢岳父母、父母对攻读博士学位的鼎力支持。感谢老婆和莽儿，博士顺利毕业得益于老婆你的理解和宽容。

谨在此感谢所有关心、帮助过我的人们！

张 俊

3/15/2016，于北京控制工程研究所

# 博士学位论文原创性声明

本人郑重声明：此处所提交的博士学位论文《甚高精度星敏感器星点定位和校正技术研究》，是本人在导师指导下，在中国空间技术研究院北京控制工程研究所攻读博士学位期间独立进行研究工作所取得的成果。据本人所知，论文中除已注明部分外不包含他人已发表或撰写过的研究成果。对本文的研究工作做出重要贡献的个人和集体，均已在文中以明确方式注明。本声明的法律结果将完全由本人承担。

签字            日期：    年   月   日

# 博士学位论文使用授权书

《甚高精度星敏感器星点定位和校正技术研究》系本人在中国空间技术研究院北京控制工程研究所攻读博士学位期间在导师指导下完成的博士学位论文。本论文的研究成果归中国空间技术研究院所有，本论文的研究内容不得以其它单位的名义发表。本人同意院方保留并向有关部门送交论文的复印件和电子版本，允许论文被查阅和借阅。本人授权中国空间技术研究院，可以采用影印、缩印或其他复制手段保存论文，可以公布论文的全部或部分内容。

学生签名：        日期：    年   月   日

导师签名：        日期：    年   月   日